\documentclass[preprint,12pt]{elsarticle}
\usepackage{graphicx}
\graphicspath{{./}}
\usepackage{float}
\usepackage{caption}
\usepackage{subcaption}
\usepackage{tabularx}
\usepackage[utf8]{inputenc}  
\DeclareUnicodeCharacter{02BC}{'}
\usepackage{amsmath,amssymb}
\providecommand{\sech}{\operatorname{sech}}
\providecommand{\csch}{\operatorname{csch}}
\usepackage{booktabs}
\usepackage{hyperref}

\journal{Communications in Nonlinear Science and Numerical Simulation}
\biboptions{sort&compress}

\begin{document}
\begin{frontmatter}

\title{Effective-potential classification of kink-antikink collision channels in the $\phi^8$ scalar field theory}

\author[aff1]{Yun Feng}
\ead{202200700090@mail.sdu.edu.cn}
\author[aff2]{Yunguo Jiang\corref{cor1}}
\ead{jiangyg@sdu.edu.cn}
\cortext[cor1]{Corresponding author}
\affiliation[aff1]{organization={SDU-ANU Joint Science College, Shandong University},
            city={Weihai},
            postcode={264209},
            country={Peopleʼs Republic of China}}
\affiliation[aff2]{organization={Shandong Provincial Key Laboratory of Optical Astronomy and Solar-Terrestrial Environment, Institute of Space Sciences, Shandong University},
            city={Weihai},
            postcode={264209},
            country={Peopleʼs Republic of China}}

\begin{abstract}
We study kink-antikink collisions in a $(1+1)$-dimensional $\phi^8$ scalar field theory with multiple degenerate vacua. We derive soliton solutions for different vacuum structures labeled by $n=p_2/p_1$, and focus on the cases $n=2$ and $n=3$. We perform numerical simulations in all topological sectors and for both kink-antikink ($K\bar K$) and $\bar K K$ orderings. In the $(-1/2,1/2)$ sector, the kink-antikink pair annihilates for all initial velocities. To the best of our knowledge, this full-velocity annihilation regime has not been reported  in $\phi^8$ kink collisions. We also find fractal multi-bounce windows in the $(-1,-1/2)$, $(-1,-1/3)$, and $(-1/3,1/3)$ sectors. Our main result indicates an effective-potential classification of these collision outcomes. We show that the shape of the effective potential is closely related to the final channel. It determines whether the pair escapes, forms a bion, annihilates, or changes sector. When the solitons pass through each other, the effective potential can change suddenly. This gives a possible mechanism for annihilation and sector change. Our results establish connections among topological structure, spectrum and effective potentials in higher-order scalar field theories.
\end{abstract}

\begin{keyword}
Kink-antikink collisions; Phi8 field theory; Resonance windows;
Effective potential; Topological solitons
\end{keyword}
\end{frontmatter}

\section{The introduction of kink dynamics} \label{sec:0}
Topological defects are important non-perturbative objects in scalar field theories. In higher-dimensional theories, kink-like configurations are related to domain walls, while in $(1+1)$ dimensions they appear as particle-like topological excitations. Their collisions provide a simple real-time setting for studying how localized field configurations can be produced, annihilated, or transformed in multi-vacuum scalar field theories. Solitons, also known as solitary waves, are a special type of structure that exist extensively in nonlinear wave equations. A kink is a soliton in $(1+1)$ dimensions \cite{ref1}. Kink dynamics is not merely a theoretical mathematical construct, it finds practical applications in multiple physical fields, such as condensed matter physics, particle physics, nonlinear optics, biophysics and so on \cite{ref2,ref3,ref4,ref5}.

The study of localized, particle-like solutions in nonlinear field theories, known as solitons, has been a cornerstone of theoretical physics for decades. In (1+1) dimensions, kink solitons—topological defects connecting disparate vacua of a scalar field—serve as a powerful theoretical laboratory for probing fundamental nonlinear phenomena. Among the canonical models, the $\phi^4$ theory has been instrumental in elucidating complex dynamical processes. Extensive research has revealed a rich phenomenology, including the formation of bound states (bions), the existence of a critical velocity separating capture from escape, and a characteristic fractal structure of resonance windows in the velocity-impact parameter space \cite{ref6,ref7}. To approximate these intricate dynamics analytically, the collective coordinate (or covariant coordinate) method has emerged as a leading effective approach. This method, which treats the kink's position as a dynamical variable, has successfully captured the resonant energy exchange mechanism between translational kinetic energy and internal vibrational modes, accurately predicting the locations of resonance windows and critical velocities in $\phi^4$ theory \cite{ref7,ref8,ref9}. Similar investigations have extended these findings to the $\phi^6$ model, revealing parallel yet distinct dynamical features \cite{ref10,ref11}.

While the physics of lower-order polynomial models is well-established, the frontier of soliton dynamics is increasingly moving toward more complex potentials. The $\phi^8$ theory, characterized by an eighth-degree polynomial potential, represents a particularly compelling and challenging system. Unlike its simpler counterparts, the $\phi^8$ potential supports four degenerate vacua, enabling a richer taxonomy of kinks—including elementary and nested configurations—and more intricate interaction landscapes. This complexity makes it an ideal candidate for exploring phenomena that are inaccessible in simpler models. Moreover, although collective coordinate method successfully forecasts the bounce windows, the strong nonlinearity of the $\phi^8$ system causes it to break down. Recent studies have begun to unravel its unique properties, from the excitation spectra of kinks with exponential asymptotics \cite{ref12} to the scattering dynamics of kinks exhibiting power-law asymptotics \cite{ref13}. A crucial insight from this work is the discovery of a collective vibrational mode supported by the kink-antikink pair as a composite system. Even in the absence of an individual soliton shape mode, this pair-dependent mode facilitates energy transfer during collisions, a mechanism reminiscent of the resonance observed in the $\phi^6$ theory \cite{ref10,ref13}. Conversely, the presence of vibrational modes does not necessarily guarantee the occurrence of two-bounce resonance windows: their complete suppression has been observed in a deformed $\phi^4$ model even when vibrational modes are present \cite{ref14}. Furthermore, progress has been made in constructing both implicit and explicit kink solutions under specific parameter regimes \cite{ref15}, and in applying robust numerical techniques like the pseudospectral method to solve the challenging boundary value problems associated with these solitons \cite{ref16,ref17}. 


Despite these significant advances, a profound and conspicuous gap in our understanding persists. The majority of previous research has focused on individual soliton properties or two-body scattering at specific energies. A systematic, large-scale investigation into the global, long-term dynamics of kink-antikink collisions in the $\phi^8$ model—particularly the emergent fractal structures as a function of both initial velocity and potential shape parameters—is conspicuously absent. This paper is dedicated to filling this gap. Building upon the methodological foundations laid by previous work, we conduct a systematic numerical study of kink-antikink collisions across different topological sectors of the $\phi^8$ theory. Our primary goal is to find a systematic framework to classify all possible collision channels in the $\phi^8$ model. In this line of thought, we do provide a potential-based classification principle to explain channels like escape, bion formation, annihilation, and topological-sector change in higher-order scalar field theories. By doing so, we  not only extend the current knowledge of the $\phi^8$ model but also  provide deeper insights into the universal principles governing soliton interactions in higher-order nonlinear field theories.

This paper is structured as follows. Section \ref{sec:0} is the introduction of kink solutions. Section \ref{sec:1} introduces the $\phi^8$ model, detailing both the implicit soliton solutions and the explicit forms derived under special conditions. Section \ref{sec:2} aims to work out boundary conditions. Section \ref{sec:3} outlines the numerical methods employed and presents a detailed analysis of our simulation results, focusing on the fractal structures and dynamical phenomena observed for different parameter sets. Finally, Section \ref{sec:4} summarizes our findings and discusses their implications.

\section{The $\phi^8$ theory}\label{sec:1}
Considering the (1 + 1) dimension scalar field $\phi(x, t)$, the Lagrangian density can be denoted by \cite{ref1}
\begin{equation}\label{2-1}
\mathcal{L} = \frac{1}{2}\partial_\mu \phi \partial^\mu \phi - V(\phi), \tag{2-1}
\end{equation}
where $V(\phi)$ is the potential, which can be depicted in different types for different theories. Considering the $\phi^8$ theory, its potential $V(\phi)$ can be denoted by 
\begin{equation}\label{2-2}
V (\phi)= \frac{1}{2}(\phi^2 - p_1^2)^2(\phi^2 - p_2^2)^2, \quad (p_1 < p_2) \tag{2-2}
\end{equation}
where $p_1$ and $p_2$ determine the vacuum values. There are four vacuum solutions, i.e., $\phi =-p_2, -p_1, p_1$, and $p_2$. Its form of Euler-Lagrange equation is
\begin{equation}\label{2-3}
\partial_\mu \partial^\mu \phi + \frac{dV}{d\phi} = 0. \tag{2-3}
\end{equation}
Ignoring the kinetic term, the total energy of the system can be given by
\begin{equation}\label{2-4} 
E_{\text{static}}[\phi] = E_{\text{BPS}}[\phi] + \frac{1}{2}\int \left( \frac{d\phi}{dx} \pm\frac{dW}{d\phi}\right)^2 dx, \tag{2-4}
\end{equation}
where $E_{\text{BPS}} = W[\phi(+\infty)] - W[\phi(-\infty)]$ is the bounded energy and the mass term for the kink and antikink configurations, and $W(\phi)$ is the super potential. 
For $\phi^8$ theory, the super-potential can be denoted by
\begin{equation}\label{2-5} 
W(\phi) = \frac{1}{5}\phi^5 - \frac{1}{3}(p_1^2 + p_2^2)\phi^3 + p_1^2 p_2^2 \phi. \tag{2-5}
\end{equation}
The Bogomolynyi-Prasad-Sommerfield (BPS) equation \cite{ref18,ref19} is obtained by minimizing the total energy, which reads
\begin{equation}\label{2-6} 
\frac{d\phi}{dx} = \pm \frac{dW}{d\phi}. \tag{2-6}
\end{equation}
It is a first-order equation which can be used to work out the specific form of kink solutions. All the masses of kinks and antikinks \cite{ref15} can be denoted by
\begin{equation}\label{2-7}
\begin{split}   
M(p_1, p_2) &= M(-p_2, -p_1) = \frac{2(p_2 - p_1)^3(p_1^2 + 3p_1 p_2 + p_2^2)}{15},\\
M(-p_1, p_1) &= \frac{4p_1^3(5p_2^2 - p_1^2)}{15}. 
\end{split}
\tag{2-7}
\end{equation}

\begin{figure} [h] 
        \centering
        \includegraphics[width=0.5\linewidth]{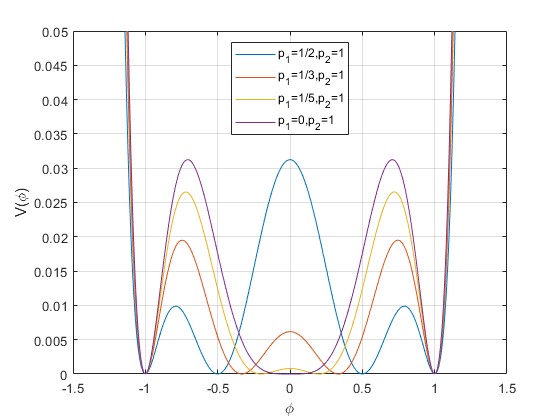}
        \caption{Different potentials in the $\phi^8$ theory}
        \label{Fig:1}
    \end{figure}

The kink solutions of the $\phi^8$ theory can be worked out by the BPS equation, and they interpolate between the adjacent vacuums. There are six kink solutions in the $\phi^8$ theory. Because of the intricate high-level nonlinear potential field, there is no universal explicit solution. When $p_2/p_1$ takes some unique values, there are explicit solutions ($\phi = \phi(x)$) for the $\phi^8$ theory. These explicit potentials are presented in Figure \ref{Fig:1}. 
    
There are still some implicit kink solutions. 
First, we consider the Topological sectors $(p_1, p_2)$ and $(-p_2, -p_1)$. By integrating the BPS equation under the condition $0 < a < |\phi| < b$, we can obtain an implicit solution \cite{ref15}
\begin{equation}\label{2-8} 
x = \frac{1}{2(p_2^2 - p_1^2)} \ln\left[ \left( \frac{\phi - p_1}{\phi + p_1} \right)^{\frac{1}{p_1}} \left( \frac{p_2 + \phi}{p_2 - \phi} \right)^{\frac{1}{p_2}} \right]. \tag{2-8}
\end{equation}
And Equation \ref{2-8} can be transformed into
\begin{equation}\label{2-9} 
\left( \frac{\phi - p_1}{\phi + p_1} \right)^{\frac{p_2}{p_1}} \left( \frac{p_2 + \phi}{p_2 - \phi} \right) = \exp\left[ 2p_2(p_2^2 - p_1^2)x \right]. \tag{2-9}
\end{equation}
Then denoting $p_2/p_1 = n$ and seting $p_2 = 1$, Equation \ref{2-9} becomes
\begin{equation}\label{2-10}
\left( \frac{n\phi - 1}{n\phi + 1} \right)^n \left( \frac{1 + \phi}{1 - \phi} \right) = \exp\left[ 2\left( 1 - \frac{1}{n^2} \right)x \right]. \tag{2-10}
\end{equation}
We will present the solution for different  $n$ values below.

\subsection{$n = p_2/p_1 = 2$}  \label{sec:3.1}
\begin{figure} [h]
\centering
\includegraphics[width=0.5\textwidth]{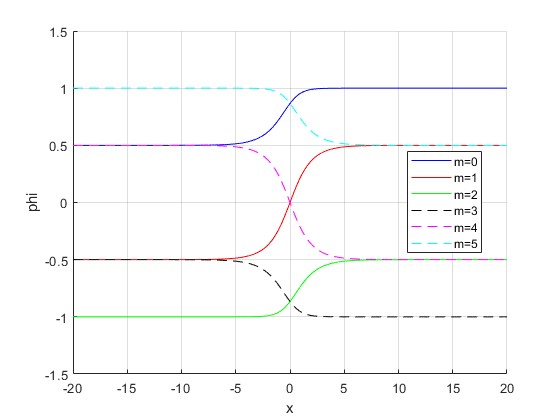} 
\caption{Kinks and antikinks when $n=2$ in the $\phi^8$ theory for $p_1=0.5$ and $p_2=1$. }
\label{Fig:2}
\end{figure}

When $n=2$, by solving the Equation \ref{2-10}, the kink solutions can be described by the universal form
\begin{equation} \label{2-11}
\phi_m(x) = \cos\left\{ \frac{1}{3}\arccos\left[\tanh ( \frac{3}{4}x ) \right] + \frac{\pi}{3}m \right\}, \tag{2-11}
\end{equation}
where $m=0,1,2,3,4,5$.
In this formula, the subscript $m$ of $\phi$ illustrates the different types of kinks. As a prime instance, $m=1$ represents that kink is between the vacuums $\phi(-\infty) = -p_1$ and $\phi(+\infty) = p_1$.   There are six different types of kinks which are plotted in the Figure  \ref{Fig:2}. The corresponding soliton mass can be derived from Equation \ref{2-7}, i.e.,
\begin{equation}\label{2-12}
M_{\left( \frac{1}{2}, 1 \right)} = M_{\left( -1, -\frac{1}{2} \right)} = \frac{11}{240}, \quad M_{\left( -\frac{1}{2}, \frac{1}{2} \right)} = \frac{19}{240}. \tag{2-12}
\end{equation}

\subsection{$n = p_2/p_1 = 3$}
\begin{figure} [h] 
\centering
\includegraphics[width=0.5\textwidth]{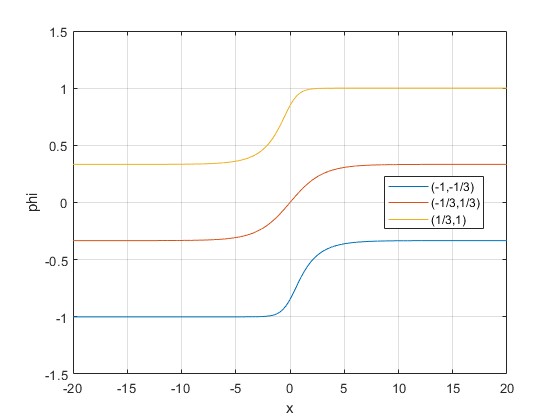} 
\caption{Kinks (no antikinks) when $n=3$ in the $\phi^8$ theory}
\label{Fig:3}
\end{figure}

When $n=3$ ($p_2=1$ and $p_1=1/3$), the kink solutions can be given in three intervals.
For the sector $(-1, -1/3)$, the kink solution reads
\begin{equation}\label{2-13}
\phi_K(x) = 
\begin{cases}
&\frac{1}{3}\left [{-\sqrt{1-{{(\sech{\frac{8}{9}x)}}^{\frac{2}{3}}}}-\sqrt{2+{{(\sech{\frac{8}{9}x)}}^{\frac{2}{3}}}-\frac{2\tanh{\frac{8}{9}x}}{\sqrt{1-{{(\sech{\frac{8}{9}x)}}^{\frac{2}{3}}}}}}}\right ], \quad x<0. \\
&\frac{1}{3}\left [{\sqrt{1-{{(\sech{\frac{8}{9}x)}}^{\frac{2}{3}}}}-\sqrt{2+{{(\sech{\frac{8}{9}x)}}^{\frac{2}{3}}}+\frac{2\tanh{\frac{8}{9}x}}{\sqrt{1-{{(\sech{\frac{8}{9}x)}}^{\frac{2}{3}}}}}}}\right ], \quad x>0. \tag{2-13}
\end{cases}
\end{equation}
Its limit is
\begin{equation}\label{2-14}
\lim_{x\to-0}\phi_K(x) = \lim_{x\to+0}\phi_K(x) = -\frac{\sqrt{3 + 2\sqrt{3}}}{3}. \tag{2-14}
\end{equation}
For the sector $(1/3, 1)$, the kink solution reads
\begin{equation}\label{2-15}
\phi_K(x) = 
\begin{cases}
&\frac{1}{3}\left [{-\sqrt{1-{{(\sech{\frac{8}{9}x)}}^{\frac{2}{3}}}}+\sqrt{2+{{(\sech{\frac{8}{9}x)}}^{\frac{2}{3}}}-\frac{2\tanh{\frac{8}{9}x}}{\sqrt{1-{{(\sech{\frac{8}{9}x)}}^{\frac{2}{3}}}}}}}\right ], \quad x<0\\
&\frac{1}{3}\left [{\sqrt{1-{{(\sech{\frac{8}{9}x)}}^{\frac{2}{3}}}}+\sqrt{2+{{(\sech{\frac{8}{9}x)}}^{\frac{2}{3}}}+\frac{2\tanh{\frac{8}{9}x}}{\sqrt{1-{{(\sech{\frac{8}{9}x)}}^{\frac{2}{3}}}}}}}\right ], \quad x>0. \tag{2-15}
\end{cases}
\end{equation}
Its limit is
\begin{equation}\label{2-16}
\lim_{x\to-0}\phi_K(x) = \lim_{x\to+0}\phi_K(x) = \frac{\sqrt{3 + 2\sqrt{3}}}{3}. \tag{2-16}
\end{equation}
For sector $(-1/3, 1/3)$, the kink solution reads
\begin{equation}\label{2-17}
\phi_K(x) = 
\begin{cases}
&\frac{1}{3}\left [{\sqrt{1+{{(\csch{\frac{8}{9}x)}}^{\frac{2}{3}}}}-\sqrt{2-{{(\csch{\frac{8}{9}x)}}^{\frac{2}{3}}}-\frac{2\coth\frac{8}{9}x}{\sqrt{1+{{(\csch{\frac{8}{9}x)}}^{\frac{2}{3}}}}}}}\right ], \quad x<0\\ 
&\frac{1}{3}\left [{{-\sqrt{1+{{(\csch{\frac{8}{9}x)}}^{\frac{2}{3}}}}+}\sqrt{{2-{{(\csch{\frac{8}{9}x)}}^{\frac{2}{3}}}+}\frac{{2}{\coth{\frac{8}{9}x}}}{{\sqrt{1+{{(\csch{\frac{8}{9}x)}}^{\frac{2}{3}}}}}}}}\right ],\quad x>0. \tag{2-17}
\end{cases}
\end{equation}
Its limit is
\begin{equation}\label{2-18}
\lim_{x\to-0}\phi_K(x) = \lim_{x\to+0}\phi_K(x) = 0. \tag{2-18}
\end{equation}
The anti-kink solution $\phi_{\bar{K}}$ could be obtained by replacing $x\to -x$ in $\phi_K$.
The corresponding kink mass can be derived from Equations \ref{2-7} 
\begin{equation}\label{2-19}
M_{\left( \frac{1}{3}, 1 \right)} = M_{\left( -1, -\frac{1}{3} \right)} = \frac{304}{3645}, \quad M_{\left( -\frac{1}{3}, \frac{1}{3} \right)} = \frac{176}{3645}. \tag{2-19}
\end{equation}
These three types of kinks  are depicted in the Figure  \ref{Fig:3}.

Based on the explicit formula of the $n=2$ and $n=3$ kinks, we can speculate that if there are explicit kink solutions in higher $n$, they will be more complicated.
\subsection{Other $n$}
According to the kink solution $\phi^8$ theory \cite{ref15}, when $n$ is a rational value, its asymptotics of kink solutions should be discussed for different sectors. We consider the asymmetric kink $(1/n, 1)$ and symmetric kink $(-1/n, 1/n)$ which are given below
\paragraph{(i) asymmetric kink $(1/n, 1)$}\mbox{}\\
\begin{equation}\label{2-20}
\phi_K(x) = \frac{1}{n} + \frac{2}{n}\left( \frac{n-1}{n+1} \right)^{1/n} \exp\left[ \frac{2}{n}\left( 1 - \frac{1}{n^2} \right)x \right], \quad \text{when } x\to-\infty. \tag{2-20}
\end{equation}
\begin{equation}\label{2-21}
\phi_K(x) = 1 - 2\left( \frac{n-1}{n+1} \right)^n \exp\left[ -2\left( 1 - \frac{1}{n^2} \right)x \right], \quad \text{when } x\to+\infty. \tag{2-21}
\end{equation}
\paragraph{(ii) symmetric kink $(-1/n, 1/n)$}\mbox{}\\
\begin{equation}\label{2-22}
\phi_K(x) = -\frac{1}{n} + \frac{2}{n}\left( \frac{n-1}{n+1} \right)^{1/n} \exp\left[ \frac{2}{n}\left( 1 - \frac{1}{n^2} \right)x \right], \quad \text{when } x\to-\infty. \tag{2-22}
\end{equation}
\begin{equation}\label{2-23}
\phi_K(x) = \frac{1}{n} - \frac{2}{n}\left( \frac{n-1}{n+1} \right)^{1/n} \exp\left[ -\frac{2}{n}\left( 1 - \frac{1}{n^2} \right)x \right], \quad \text{when } x\to+\infty. \tag{2-23}
\end{equation}
Based on Equations above, the asymptotic form can be used to derive the boundary conditions for a general $n$.

\section{The kink-antikink collision in the $\phi^8$ Theory} \label{sec:2}
In the last section they are all static kink solutions. A Lorentz transformation applied to the static solution yields a relativistic dynamics for the kink-antikink collisions \cite{ref20}.  Here, we consider a pair of kink-antikink moving towards each other with the same velocity $v_{{in}}$. Initially, the kink and antikink are separated by a distance of $2a$, and the general form of the field can be expressed as
\begin{equation}\label{3-1}
\phi_{K\bar{K}}(x, t) = \phi_K\left[ \gamma(x - v_{{in}}t + a) \right] + \phi_{\bar{K}}\left[ \gamma(x + v_{{in}}t - a) \right] + p, \tag{3-1}
\end{equation}
where $\gamma$ is Lorentz factor, which can be expressed as
\begin{equation}\label{3-2}
\gamma = \frac{1}{\sqrt{1 - v_{{in}}^2}}. \tag{3-2}
\end{equation}
Here, $v_{{in}}$ denotes the kink velocity normalized by the speed of light $c=1$, and $a$ represents the initial position of the kink.
The $p$ in Equation \ref{3-1} is a constant which makes the configuration of kink-antikink satisfy the boundary condition. There are three different $p$ for the corresponding sectors:
(i) When $(-p_2,-p_1)=(-1,-1/n)$, $p = p_1$;
(ii) When $(-p_1,p_1)=(-1/n,1/n)$, $p = -p_1$;
(iii) When $(p_1,p_2)=(1/n,1)$, $p = -p_2$.
In the previous research on $\phi^8$ theory, due to the strong nonlinearity of the potential field as well as the complexity and globle internal modes, it is extremely challenging to apply the covariant coordinate method and make effective approximations. In $\phi^8$ theory, for a potential parameter $n = p_2/p_1$, there exist six solitons with distinct configurations. The collision channels of soliton pairs vary with their different configurations and quantities \cite{ref15,ref16}.
This paper will focus on the kink dynamics for Topological sectors $n=2$ and $n=3$. We also present several cases where $n$ takes other values.

\subsection{Setup for $\phi_{K\bar{K}}$}
Based on Equation \ref{3-1}, the relativistic formulation of the $\phi_{K\bar{K}}$ for a kink-antikink system moving with an initial velocity $v_{{in}}$ and an initial separation $2a$ is considered. The Lorentz factor $\gamma$ is determined solely by $v_{{in}}$ through Equation \ref{3-2} and is independent of $n$.
\paragraph{(1) $n = p_2/p_1 = 2$}\mbox{}\\
Based on the single soliton in Equation \ref{2-11}, we can get that  $m=0$ and $m=5$ correspond to the kink and antikink, respectively. Together with its topological distribution sector $(1/2,1)$, its relativistically covariant form can be written as
\begin{equation}\label{3-3}
\phi_{0,5}(x) = \phi_0\left( \gamma(x - v_{{in}}t + a) \right) + \phi_5\left( \gamma(x + v_{{in}}t - a) \right) - 1. \tag{3-3}
\end{equation}
Correspondingly, for the kink-antikink pair with $m=1$ and $m=4$ over sector $(-1/2,1/2)$, and the other kink-antikink pair with $m=2$  and $m=3$ over sector $(-1,-1/2)$, the relativistically covariant forms of $\phi$ can be written as
\begin{equation}\label{3-4}
\phi_{1,4}(x) = \phi_1\left( \gamma(x - v_{{in}}t + a) \right) + \phi_4\left( \gamma(x + v_{{in}}t - a) \right) - \frac{1}{2}, \tag{3-4}
\end{equation}
\begin{equation}\label{3-5}
\phi_{2,3}(x) = \phi_2\left( \gamma(x - v_{{in}}t + a) \right) + \phi_3\left( \gamma(x + v_{{in}}t - a) \right) + \frac{1}{2}, \tag{3-5}
\end{equation}
here $\phi_m(x)$ is in Equation \ref{2-11}.
\paragraph{(2) $n = p_2/p_1 = 3$}\mbox{}\\
For the case of $n=3$, the potential field expression of a single soliton is rather complex by itself. Moreover, it also splits into two separate expressions in the potential field, and the corresponding kink-antikink expressions are complicated as well. For the sake of convenience in our discussion, we first describe the scenario where $v_{{in}} \cdot t < a$.
When the kink is in the sector $(-1,-1/3)$, the $\phi_{K\bar{K}}$ is written as\\
(i) $x < -a + v_{{in}}t$, 
\begin{equation*}\label{3-6}
\begin{split}
\phi_{K\bar{K}}(x) = &-\frac{1}{3}\left \{{-\sqrt{1-{{\left[\sech(y_1)\right]}^{\frac{2}{3}}}}+\sqrt{2+{{\left[\sech(y_1)\right ]}^{\frac{2}{3}}}-\frac{2\tanh{\left(y_1\right)}}{\sqrt{1-{{\left[\sech(y_1)\right]}^{\frac{2}{3}}}}}}}\right\}\\
&-\frac{1}{3}\left \{\sqrt{1-{{\left[\sech(y_2)\right]}^{\frac{2}{3}}}}+\sqrt{2+{{\left[\sech(y_2)\right]}^{\frac{2}{3}}}+\frac{2\tanh{\left(y_2\right)}}{\sqrt{1-{{\left[\sech(y_2)\right]}^{\frac{2}{3}}}}}}\right\}+\frac{1}{3}.\\
\end{split}
\tag{3-6}
\end{equation*}
(ii) $-a + v_{{in}}t < x < a - v_{{in}}t$
\begin{equation*}\label{3-7}
\begin{split}
\phi_{K\bar{K}}(x) = &-\frac{1}{3}\left \{-\sqrt{1-{{\left[\sech(y_1)\right]}^{\frac{2}{3}}}}+\sqrt{2+{{\left[\sech(y_1)\right]}^{\frac{2}{3}}}-\frac{2\tanh{\left(y_1\right)}}{\sqrt{1-{{\left[\sech(y_1)\right]}^{\frac{2}{3}}}}}}\right \}\\
&-\frac{1}{3}\left \{-\sqrt{1-{{\left[\sech(y_2)\right]}^{\frac{2}{3}}}}+\sqrt{2+{{\left[\sech(y_2)\right]}^{\frac{2}{3}}}-\frac{2\tanh{\left(y_2\right)}}{\sqrt{1-{{\left[\sech(y_2)\right]}^{\frac{2}{3}}}}}}\right \}+\frac{1}{3}.
\end{split}
\tag{3-7}
\end{equation*}
(iii) $x > a - v_{{in}}t$
\begin{equation*}\label{3-8}
\begin{split}
\phi_{K\bar{K}}(x) = &-\frac{1}{3}\left \{\sqrt{1-{{\left[\sech(y_1)\right]}^{\frac{2}{3}}}}+\sqrt{2+{{\left[\sech(y_1)\right]}^{\frac{2}{3}}}+\frac{2\tanh{\left(y_1\right)}}{\sqrt{1-{{\left[\sech(y_1)\right]}^{\frac{2}{3}}}}}}\right \}\\
&-\frac{1}{3}\left \{-\sqrt{1-{{\left[\sech(y_2)\right]}^{\frac{2}{3}}}}+\sqrt{2+{{\left[\sech(y_2)\right]}^{\frac{2}{3}}}-\frac{2\tanh{\left (y_2\right)}}{\sqrt{1-{{\left[\sech(y_2)\right]}^{\frac{2}{3}}}}}}\right \}+\frac{1}{3},
\end{split}
\tag{3-8}
\end{equation*}
When the potential field lies in the interval $(-1/3,1/3)$, the $\phi_{K\bar{K}}$ is written as\\
(i) $x < -a + v_{{in}}t$
\begin{equation*}\label{3-9}
\begin{split}
\phi_{K\bar{K}}(x) = &-\frac{1}{3}\left \{\sqrt{1+{{\left [\csch(y_1)\right]}^{\frac{2}{3}}}}-\sqrt{2-{{\left [\csch(y_1)\right]}^{\frac{2}{3}}}-\frac{2\coth{\left(y_1\right)}}{\sqrt{1+{{\left [\csch(y_1)\right]}^{\frac{2}{3}}}}}}\right\}\\
&-\frac{1}{3}\left \{-\sqrt{1+{{\left [\csch(y_2)\right]}^{\frac{2}{3}}}}+\sqrt{2-{{\left [\csch(y_2)\right]}^{\frac{2}{3}}}+\frac{2\coth{\left(y_2\right)}}{\sqrt{1+{{\left [\csch(y_2)\right]}^{\frac{2}{3}}}}}}\right\}-\frac{1}{3}.
\end{split}
\tag{3-9}
\end{equation*}
(ii) $-a + v_{{in}}t < x < a - v_{{in}}t$
\begin{equation*}\label{3-10}
\begin{split}
\phi_{K\bar{K}}(x) = &-\frac{1}{3}\left \{\sqrt{1+{{\left [\csch(y_1)\right]}^{\frac{2}{3}}}}\mathrm{-}\sqrt{\mathrm{2-}{{\left [\csch(y_1)\right]}^{\mathrm{\frac{2}{3}}}}\mathrm{-}\frac{\mathrm{2}\coth{\left (y_1\right )}}{\sqrt{\mathrm{1+}{{\left [\csch(y_1)\right]}^{\mathrm{\frac{2}{3}}}}}}}\right\}\\
&-\frac{1}{3}\left \{\sqrt{\mathrm{1+}{{\left [\csch(y_2)\right]}^{\mathrm{\frac{2}{3}}}}}\mathrm{-}\sqrt{\mathrm{2-}{{\left [\csch(y_2)\right]}^{\mathrm{\frac{2}{3}}}}\mathrm{-}\frac{\mathrm{2}\coth{\left(y_2\right)}}{\sqrt{{1+}{{\left [\csch(y_2)\right]}^{\mathrm{\frac{2}{3}}}}}}}\right\}-\frac{1}{3}.
\end{split}
\tag{3-10}
\end{equation*}
(iii) $x > a - v_{{in}}t$
\begin{equation*}\label{3-11}
\begin{split}
\phi_{K\bar{K}}(x) = &-\frac{1}{3}\left \{-\sqrt{1+{{\left [\csch(y_1)\right]}^{\frac{2}{3}}}}\mathrm{+}\sqrt{\mathrm{2-}{{\left [\csch(y_1)\right]}^{\mathrm{\frac{2}{3}}}}\mathrm{+}\frac{\mathrm{2}\coth{\left(y_1\right)}}{\sqrt{\mathrm{1+}{{\left [\csch(y_1)\right]}^{\mathrm{\frac{2}{3}}}}}}}\right\}\\
&-\frac{1}{3}\left\{\sqrt{\mathrm{1+}{{\left [\csch(y_2)\right]}^{\mathrm{\frac{2}{3}}}}}\mathrm{-}\sqrt{\mathrm{2-}{{\left [\csch(y_2)\right]}^{\mathrm{\frac{2}{3}}}}\mathrm{-}\frac{\mathrm{2}\coth{\left(y_2\right)}}{\sqrt{\mathrm{1+}{{\left [\csch(y_2)\right]}^{\mathrm{\frac{2}{3}}}}}}}\right\}-\frac{1}{3}.
\end{split}
\tag{3-11}
\end{equation*}
When the potential field lies in the interval $(1/3,1)$, the $\phi_{K\bar{K}}$ is written as\\
(i) $x < -a + v_{{in}}t$
\begin{equation*}\label{3-12}
\begin{split}
\phi_{K\bar{K}}(x) =&-\frac{1}{3}\left \{-\sqrt{\mathrm{1-}{{\left[\sech(y_1)\right]}^{\mathrm{\frac{2}{3}}}}}-\sqrt{2+{{\left[\sech(y_1)\right]}^{\frac{2}{3}}}-\frac{2\tanh{\left(y_1\right)}}{\sqrt{1-{{\left[\sech(y_1)\right]}^{\frac{2}{3}}}}}}\right\}\\
&-\frac{1}{3}\left \{\sqrt{1-{{\left[\sech(y_2)\right]}^{\frac{2}{3}}}}-\sqrt{2+{{\left[\sech(y_2)\right]}^{\frac{2}{3}}}+\frac{2\tanh{\left(y_2\right)}}{\sqrt{1-{{\left[\sech(y_2)\right]}^{\frac{2}{3}}}}}}\right\}-1.
\end{split}
\tag{3-12}
\end{equation*}
(ii) $-a + v_{{in}}t < x < a - v_{{in}}t$
\begin{equation*}\label{3-13}
\begin{split}
\phi_{K\bar{K}}(x) = &-\frac{1}{3}\left \{-\sqrt{1-{{\left[\sech(y_1)\right]}^{\frac{2}{3}}}}-\sqrt{2+{{\left[\sech(y_1)\right]}^{\frac{2}{3}}}-\frac{2\tanh{\left(y_1\right)}}{\sqrt{1-{{\left[\sech(y_1)\right]}^{\frac{2}{3}}}}}}\right \}\\
&-\frac{1}{3}\left \{-\sqrt{1-{{\left[\sech(y_2)\right]}^{\frac{2}{3}}}}-\sqrt{2+{{\left[\sech(y_2)\right]}^{\frac{2}{3}}}-\frac{2\tanh{\left (y_2\right)}}{\sqrt{1-{{\left[\sech(y_2)\right]}^{\frac{2}{3}}}}}}\right\}-1.
\end{split}
\tag{3-13}
\end{equation*}
(iii) $x > a - v_{{in}}t$
\begin{equation*}\label{3-14}
\begin{split}
\phi_{K\bar{K}}(x) = &-\frac{1}{3}\left \{\sqrt{1-{{\left[\sech(y_1)\right]}^{\frac{2}{3}}}}-\sqrt{2+{{\left[\sech(y_1)\right]}^{\frac{2}{3}}}+\frac{2\tanh{\left(y_1\right)}}{\sqrt{1-{{\left[\sech(y_1)\right]}^{\frac{2}{3}}}}}}\right\}\\
&-\frac{1}{3}\left \{-\sqrt{1-{{\left[\sech(y_2)\right]}^{\frac{2}{3}}}}-\sqrt{2+{{\left[\sech(y_2)\right]}^{\frac{2}{3}}}-\frac{2\tanh{\left(y_2\right)}}{\sqrt{1-{{\left[\sech(y_2)\right]}^{\frac{2}{3}}}}}}\right\}-1.
\end{split}
\tag{3-14}
\end{equation*}
Here $y_1 ={{\frac{8}{9}\gamma(x-a+{v_{in}}t)}}$, $y_2 = {{ {\frac{8}{9}\gamma}(-x-a+{v_{in}}t)}}$.
\paragraph{(3) Other $n$}\mbox{}\\
For a general value of $n$, an explicit expression for the potential field is not straightly available except for the cases of $n=2$ and $n=3$. The general solution can only be derived by combining mathematical or numerical approximations of the boundary conditions with numerical methods by now. This content will be presented on in the Section 3.2 and will not be discussed in excessive detail here.

\subsection{Setup of the boundary conditions}  \label{Sec:bondary}
In the numerical calculation, the boundary conditions are different  for the explicit and implicit cases. We present them in the following.
\paragraph{(1) Explicit cases}\mbox{}\\
To do numerical simulations, we present the initial boundary conditions. If an explicit analytical solution or an explicit approximate expression of kink is available, the initial boundary condition can be directly presented via the finite difference method. In general, the boundary conditions required for numerical simulations can be obtained by combining the derived 
$\phi_{K\bar{K}}(x, t)$ with the first-order  finite  difference  method, whose general expression is given by
\begin{equation}\label{3-15}
\frac{\partial\phi_{K\bar{K}}}{\partial{x}}|_{\Delta{x},\Delta{t}\to0,t=0}\approx\frac{\phi_{K\bar{K}}(x+\Delta{x},0)-\phi_{K\bar{K}}(x,0)}{\Delta{x}}. \tag{3-15}
\end{equation}
According to Equation \ref{3-1}, the initial time derivative can be approximated by
\begin{equation}\label{3-16}
\begin{split}
\left.\frac{\partial\phi_{K\bar{K}}}{\partial{t}}\right|_{t=0}\approx
\frac{1}{\Delta{t}}\Big\{&
\phi_K[\gamma(x-v_{{in}}\Delta{t}+a)]
{}+\phi_{\bar{K}}[\gamma(x+v_{{in}}\Delta{t}-a)]\\
&-\phi_K[\gamma(x+a)]
-\phi_{\bar{K}}[\gamma(x-a)]
\Big\}.
\end{split}
\tag{3-16}
\end{equation}

\paragraph{(2) Implicit cases}\mbox{}\\
In the interval $0 < p_1 < |\phi| < p_2$, $x$ can be expressed as an explicit function of $\phi$, denoted as $x = f(\phi_K)$. For a kink, the corresponding $\phi_K$ can be solved by the nonlinear equation $f(\phi_K)-\gamma x=0$ over the interval $[-x_{max},x_{max}]$ ($x_{max}$ is the boundary), which gives the Dirichlet boundary condition. Based on Equation \ref{3-16}, the Neumann boundary condition can be derived by solving the equation $f(\phi_K)-\gamma(x-v_{{in}}k)=0$ (here $k$ denotes the time bin $\Delta t$) in combination with the Dirichlet boundary condition.
For a soliton pair with the kink initially at $-a$ and the antikink at $a$, the general Dirichlet boundary condition for the kink is given by $\gamma(x+a)=f(\phi_K)$, and that for the antikink by $\gamma(x-a)=-f(\phi_{\bar{K}})$; the general Neumann boundary condition for the kink is $\gamma(x+a-v_{{in}}k)=f(\phi_K)$, and that for the antikink is $\gamma(x-a+v_{{in}}k)=-f(\phi_{\bar{K}})$. Based on the above analysis, the following formulas are obtained
\begin{equation}\label{3-17}
\frac{1}{2(p_2^2 - p_1^2)} \ln\left[ \left( \frac{\phi_K(x) - p_1}{\phi_K(x) + p_1} \right)^{1/p_1} \cdot \left( \frac{\phi_K(x) + p_2}{p_2 - \phi_K(x)} \right)^{1/p_2}\right] - \gamma(x+a) = 0, \tag{3-17}
\end{equation}
\begin{equation}\label{3-18}
-\frac{1}{2(p_2^2 - p_1^2)} \ln\left[ \left( \frac{\phi_{\bar{K}}(x) - p_1}{\phi_{\bar{K}}(x) + p_1} \right)^{1/p_1} \cdot \left( \frac{\phi_{\bar{K}}(x) + p_2}{p_2 - \phi_{\bar{K}}(x)} \right)^{1/p_2} \right] - \gamma(x-a) = 0, \tag{3-18}
\end{equation}
\begin{equation}\label{3-19}
\frac{1}{2(p_2^2 - p_1^2)} \ln\left[ \left( \frac{\phi_K(\bar{x}_K)- p_1}{\phi_K(\bar{x}_K) + p_1} \right)^{1/p_1} \cdot \left( \frac{\phi_K(\bar{x}_K) + p_2}{p_2 - \phi_K(\bar{x}_K)} \right)^{1/p_2} \right] - \gamma(\bar{x}_K+a ) = 0, \tag{3-19}
\end{equation}
\begin{equation}\label{3-20}
-\frac{1}{2(p_2^2 - p_1^2)} \ln\left[ \left( \frac{\phi_{\bar{K}}(\bar{x}_{\bar K}) - p_1}{\phi_{\bar{K}}(\bar{x}_{\bar K}) + p_1} \right)^{1/p_1} \cdot \left( \frac{\phi_{\bar{K}}(\bar{x}_{\bar K}) + p_2}{p_2 - \phi_{\bar{K}}(\bar{x}_{\bar K})} \right)^{1/p_2} \right] - \gamma(\bar{x}_{\bar K}-a ) = 0. \tag{3-20}
\end{equation}
Here $\bar{x}_K=x-v_{{in}}k$ in Equation \ref{3-19} and $\bar{x}_{\bar K}=x+v_{{in}}k$ in Equation \ref{3-20}. The Newton's method or the pseudospectral method \cite{ref16,ref17} can be adopted for the numerical solution. In this work, we adopt the Newton's method.  The Dirichlet and Neumann boundary conditions can be derived by combining with the first-order finite difference method
\begin{equation}\label{3-21}
\left. \phi_{K\bar{K}} \right|_{\Delta x, \Delta t \to 0,t=0} = \phi_K(x) + \phi_{\bar{K}}(x) + p, \tag{3-21}
\end{equation}
\begin{equation}\label{3-22}
\left. \frac{\partial \phi_{K\bar{K}}}{\partial t} \right|_{t=0} \approx \frac{\left[\phi_K(\bar{x}_K) + \phi_{\bar{K}}(\bar{x}_{\bar K})\right] - \left[\phi_K(x) + \phi_{\bar{K}}(x)\right]}{k}. \tag{3-22}
\end{equation}

\section{Numerical results} \label{sec:3}
\subsection{The numerical illustration of collisions}   \label{sec:num}
In this paper, a numerical scheme combining the finite difference method introduced above with the Runge-Kutta method \cite{ref21} is adopted to solve the time-dependent potential field. The finite difference method discretizes the spatial derivatives of the wave equation, thus yielding a set of coupled ordinary differential equations with time as the independent variable. Standard methods (e.g., the Runge-Kutta method) are used to solve these coupled ordinary differential equations. Based on the former analysis, the $\phi^8$ theory can be sorted by different $n$ values.  
Here we use the Dirichlet boundary condition, and set that the spatial boundary is far from the initial kink and antikink pair. We make some preliminary numerical tests, and find that $x_{max} = 200$ for $v<0.7$ and $x_{max} = 300$ for  $v>0.7$ are proper. Thus, we consider $x \in [-200,200]$ and $t \in [0,300]$ for $v<0.7$, and $x \in [-300,300]$ and $t \in [0,400]$ for $v>0.7$ in the numerical simulations. With these settings,  the boundary perturbations have almost no effect on the core region of interest. To ensure the computational feasibility, the step bins are set as  $h=\Delta x=k=\Delta t=0.05$.

To manifest the kink and anti-kink collisions, we consider three cases of $n$ with different topological sectors. The initial velocity $v$ is varied to find the special bounce states.

\paragraph{(1) $n = p_2/p_1 = 2$}\mbox{}\\
 Here we consider that $\phi_{K\bar{K}}$ is inclined to the vacuum boundary at $|x|=200$, and  the initial distance from the kink to the center is set as $a=10$. As presented in Section  \ref{sec:3.1}, there are three fundamental topological sectors, $(-1,-1/2)$, $(-1/2,1/2)$, and $(1/2, 1)$. The sector  $(-1/2,1/2)$ was named as inner sector, while the other two sectors were named as the outer sector \cite{ref16}. We will study these sectors in the follows.
 
\subparagraph{(i) Topological sector $(-1, -1/2)$}\mbox{}\\
\begin{figure} [h]
\centering
\subcaptionbox{$v=0.06$}{\includegraphics[height=0.24\textwidth]{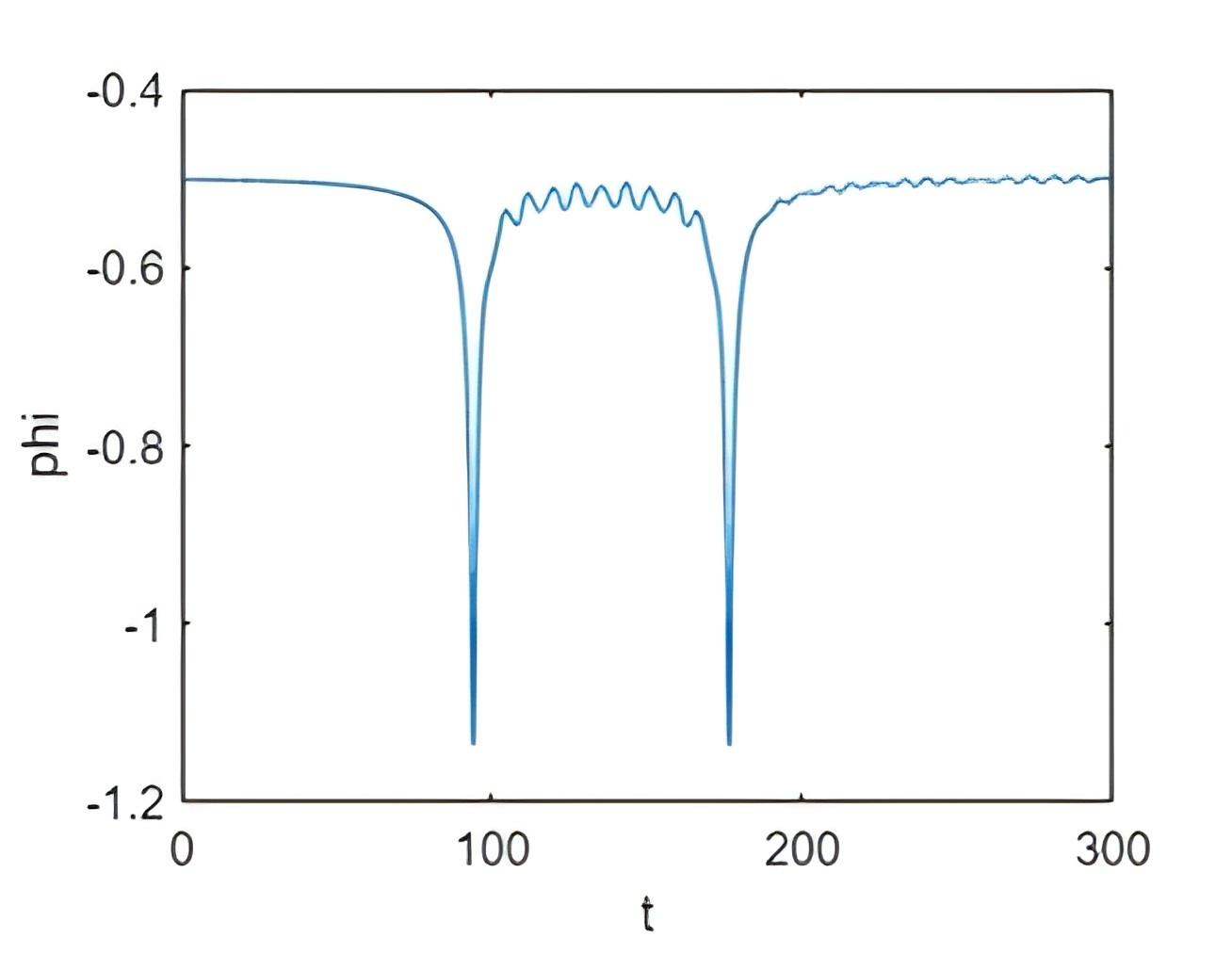}}
\subcaptionbox{$v=0.0762$}{\includegraphics[height=0.24\textwidth]{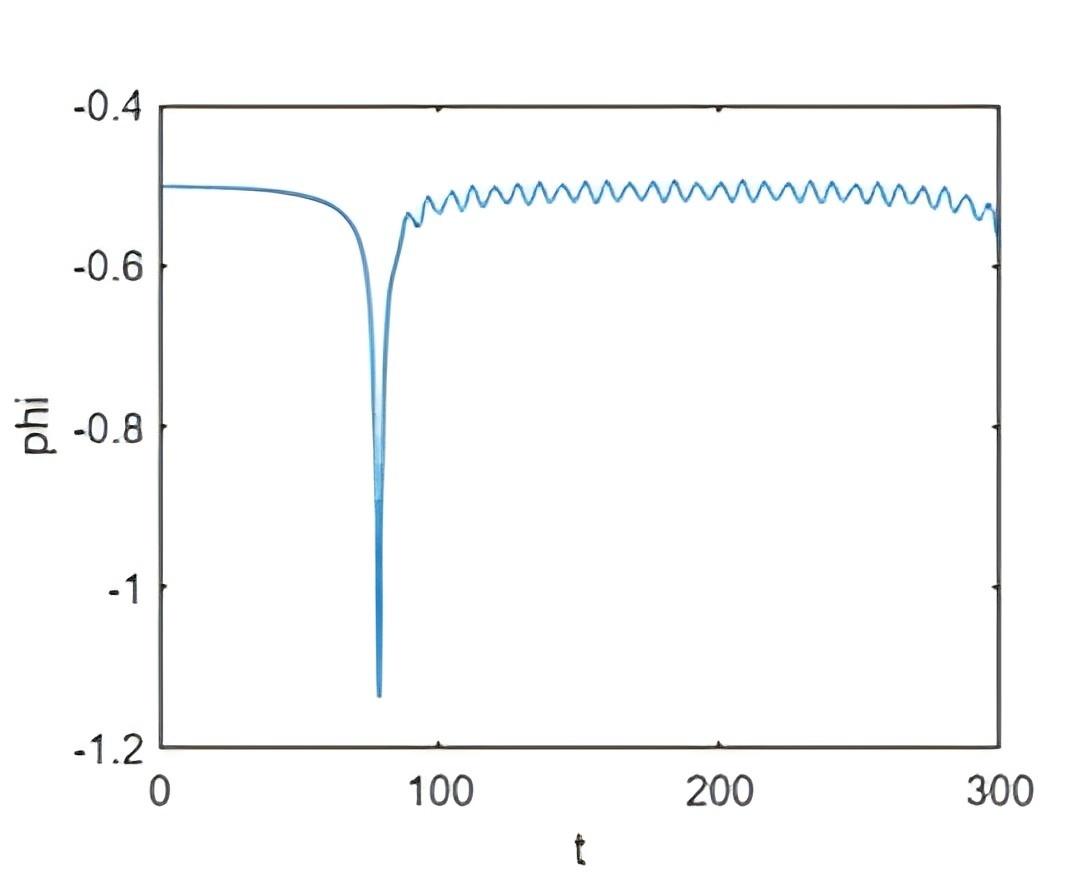}}
\subcaptionbox{$v=0.04$}{\includegraphics[height=0.24\textwidth]{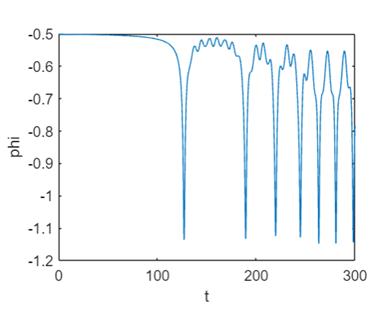}}
\caption{2D plot of the $\phi_{K\bar{K}}$ at $x=0$. From left to right panels, the initial velocities  are $v=0.06$, $v=0.0762$ and $v=0.04$, respectively. }
\label{Fig:4}
\end{figure}

\begin{figure} [h]
\centering
\subcaptionbox{$v=0.04$}{\includegraphics[height=0.21\textwidth]{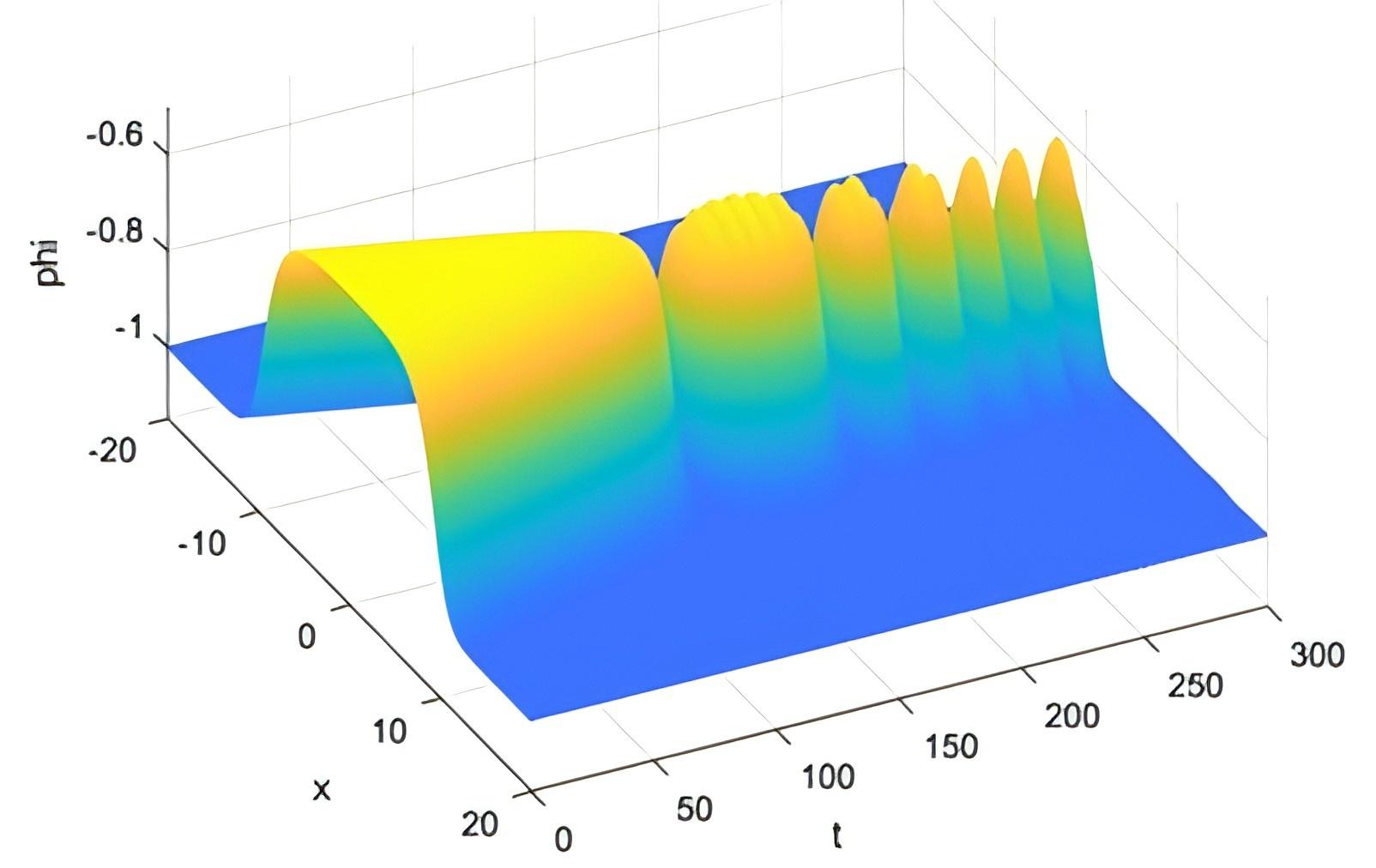}}
\subcaptionbox{$v=0.06$}{\includegraphics[height=0.21\textwidth]{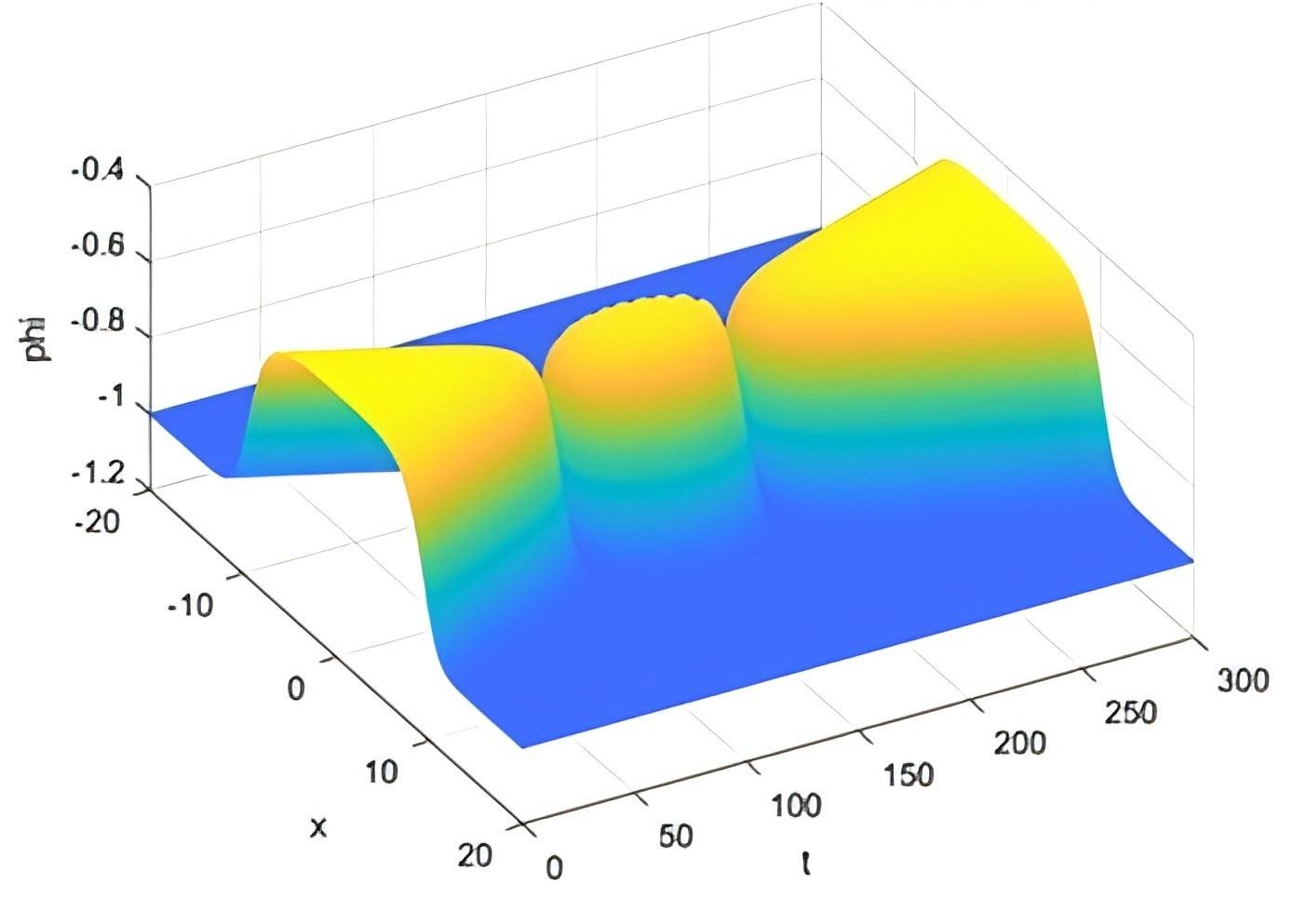}}
\subcaptionbox{$v=0.0762$}{\includegraphics[height=0.21\textwidth]{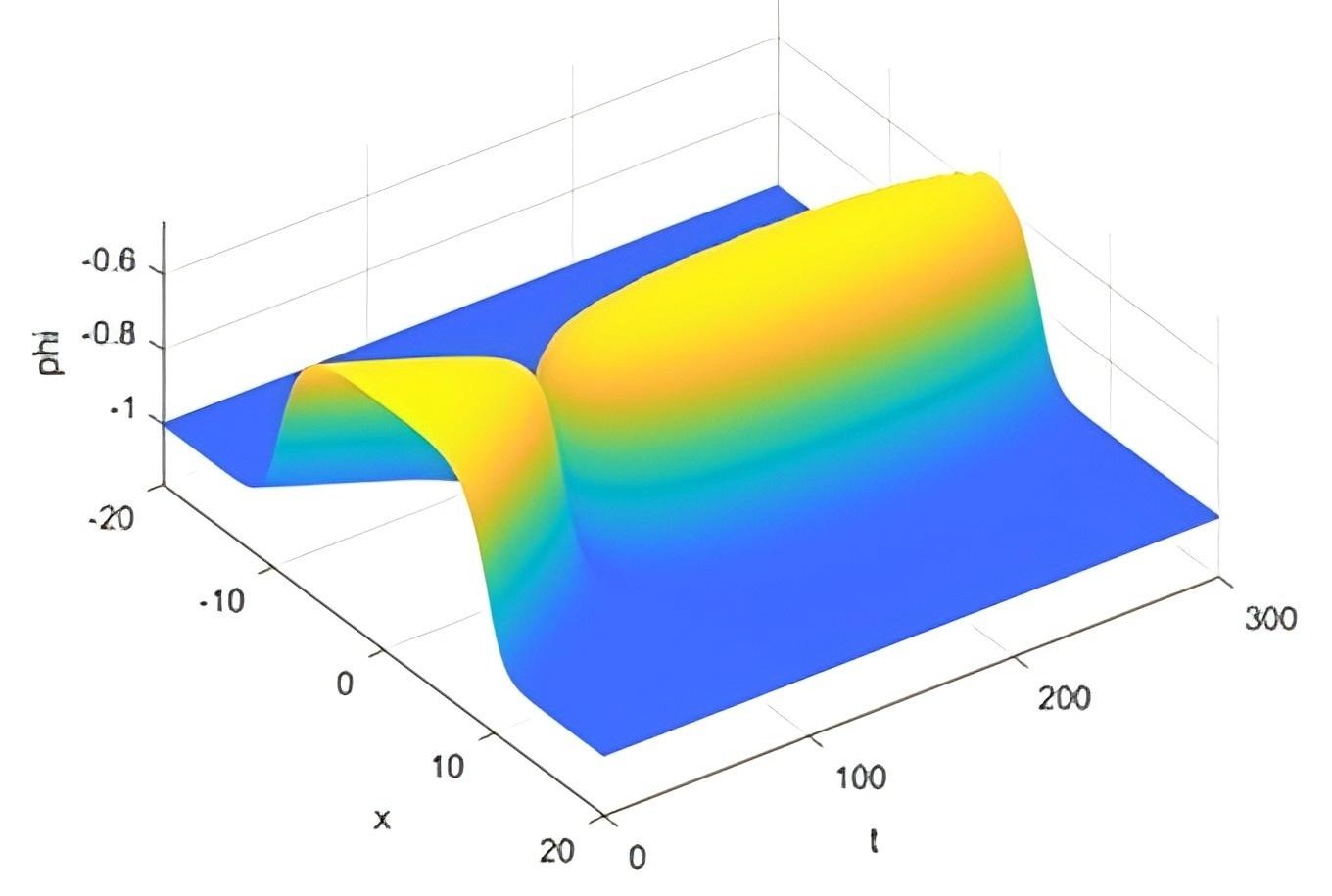}}
\caption{3D plot of $\phi_{K\bar{K}}$ with the initial velocities $v=0.04$, $v=0.06$, and $v=0.0762$, respectively.}
\label{Fig:5}
\end{figure}

Figure \ref{Fig:4} illustrates the evolutions of $\phi_{K\bar{K}}$ at $x=0$ for $t \in [0,300]$. It can be seen that the number of bounces are 2, 1 and 7 at the left, middle and right panels, respectively.  The corresponding 3D plots  are presented in Figure \ref{Fig:5}.
For $v=0.04$, the soliton pair forms a long-lived bound oscillatory state, known as the bion.  In this case, the initial kinetic energy is partly converted to both the vibration energy, and the remaining kinetic energy is insufficient to overcome the long-range attractive potential after each collision
For $v=0.06$, the kink-antikink pair initially separates after the first collision, turning to a transient vibrational state. They subsequently re-approach, collide for a second time, and then escape to infinity. This is a case of escaping state. 
For $v=0.0762$, the soliton pair separates after the initial collision, but exhibits a turning back tendency for a long-term time duration $\Delta t\sim 200$. Such long-term oscillation is common in our $\phi^8$ simulations.

\subparagraph{(ii) Topological sector $(-1/2, 1/2)$}\mbox{}\\  \label{sec:m1212}
\begin{figure} [h]
\centering
\subcaptionbox{$v=0.02$}{\includegraphics[width=0.3\textwidth]{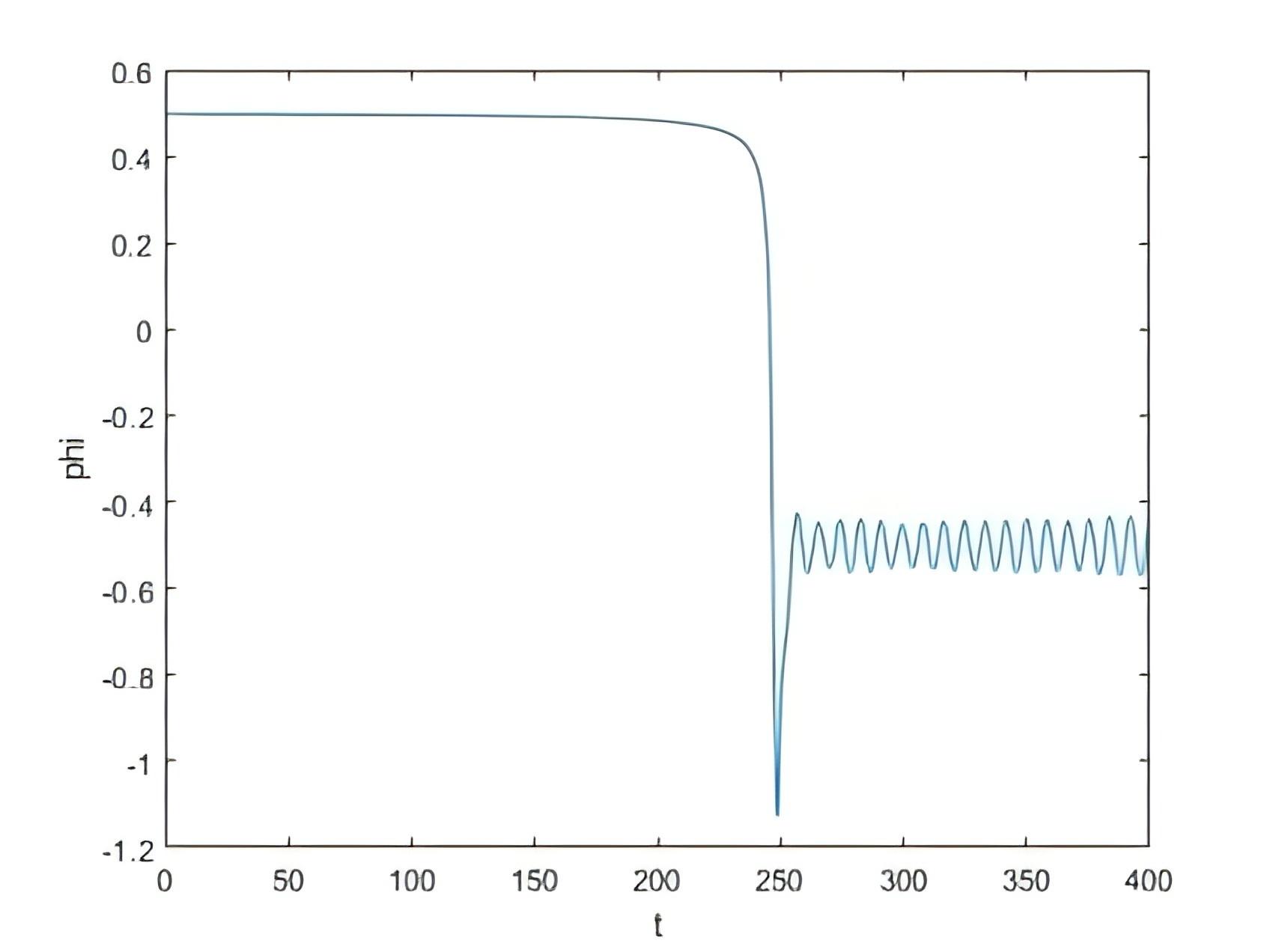}}
\subcaptionbox{$v=0.2$}{\includegraphics[width=0.3\textwidth]{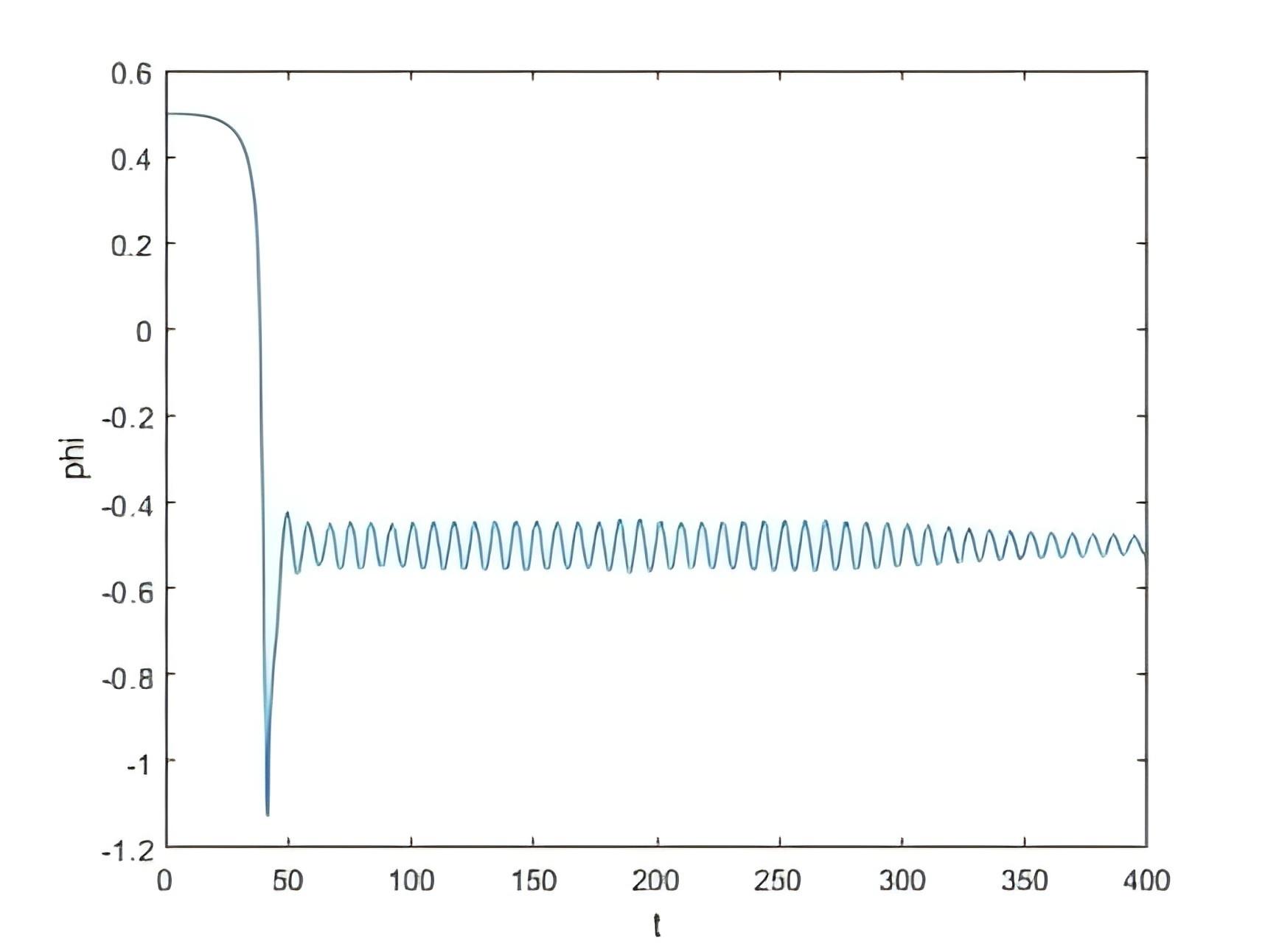}}
\subcaptionbox{$v=0.8$}{\includegraphics[width=0.3\textwidth]{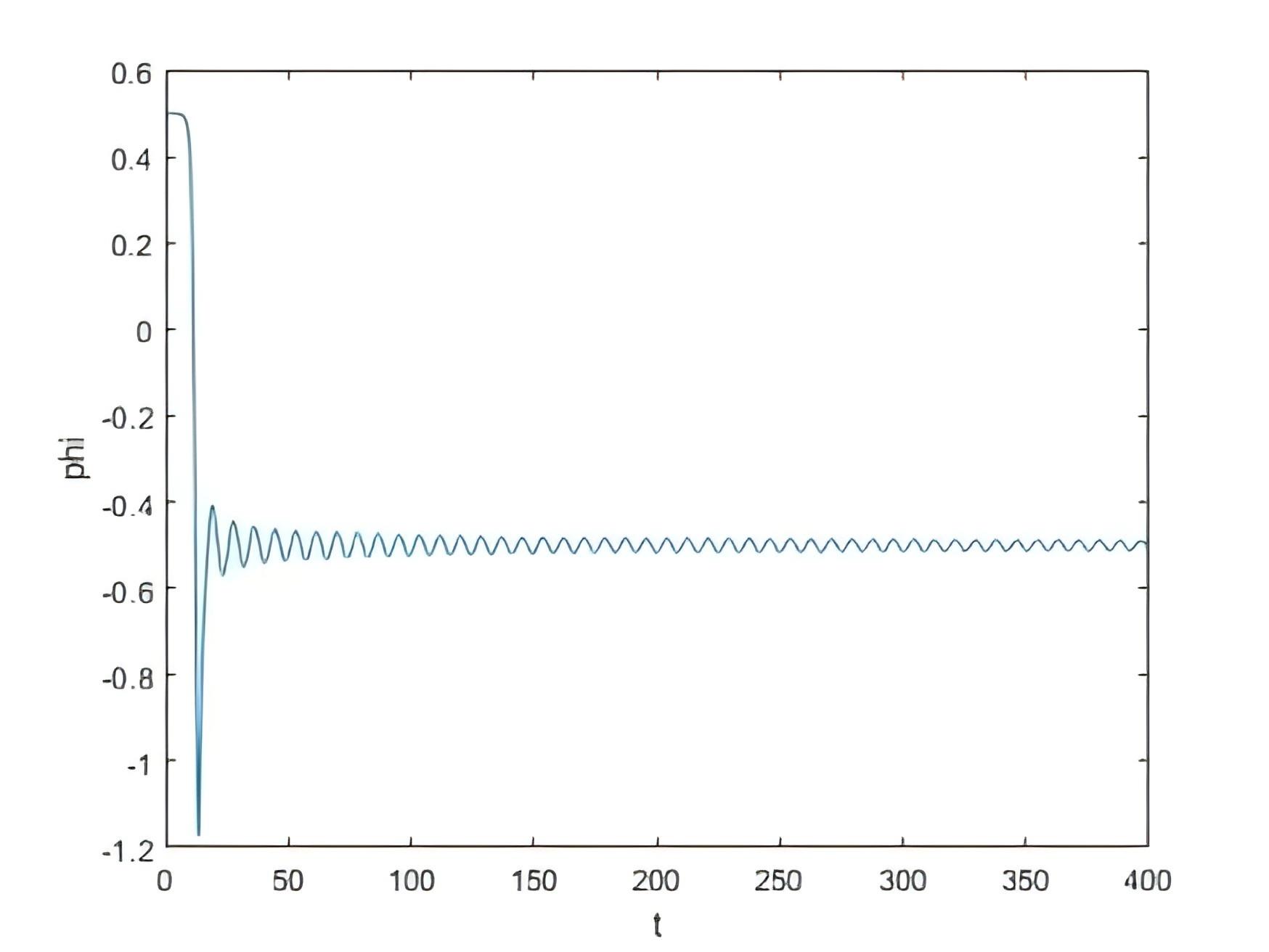}}
\caption{2D plot of the $\phi_{K\bar{K}}$ at $x=0$. From left to right panels, the velocities are $v=0.02$, $v=0.2$ and $v=0.8$, respectively.}
\label{Fig:6}
\end{figure}

\begin{figure} [h]
\centering
\subcaptionbox{$v=0.02$}{\includegraphics[width=0.3\textwidth]{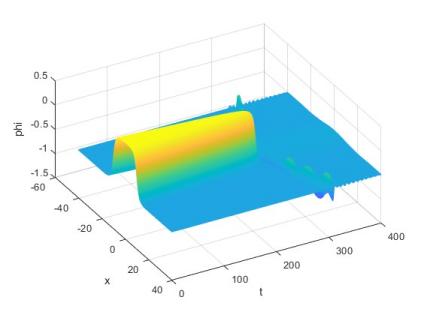}}
\subcaptionbox{$v=0.2$}{\includegraphics[width=0.3\textwidth]{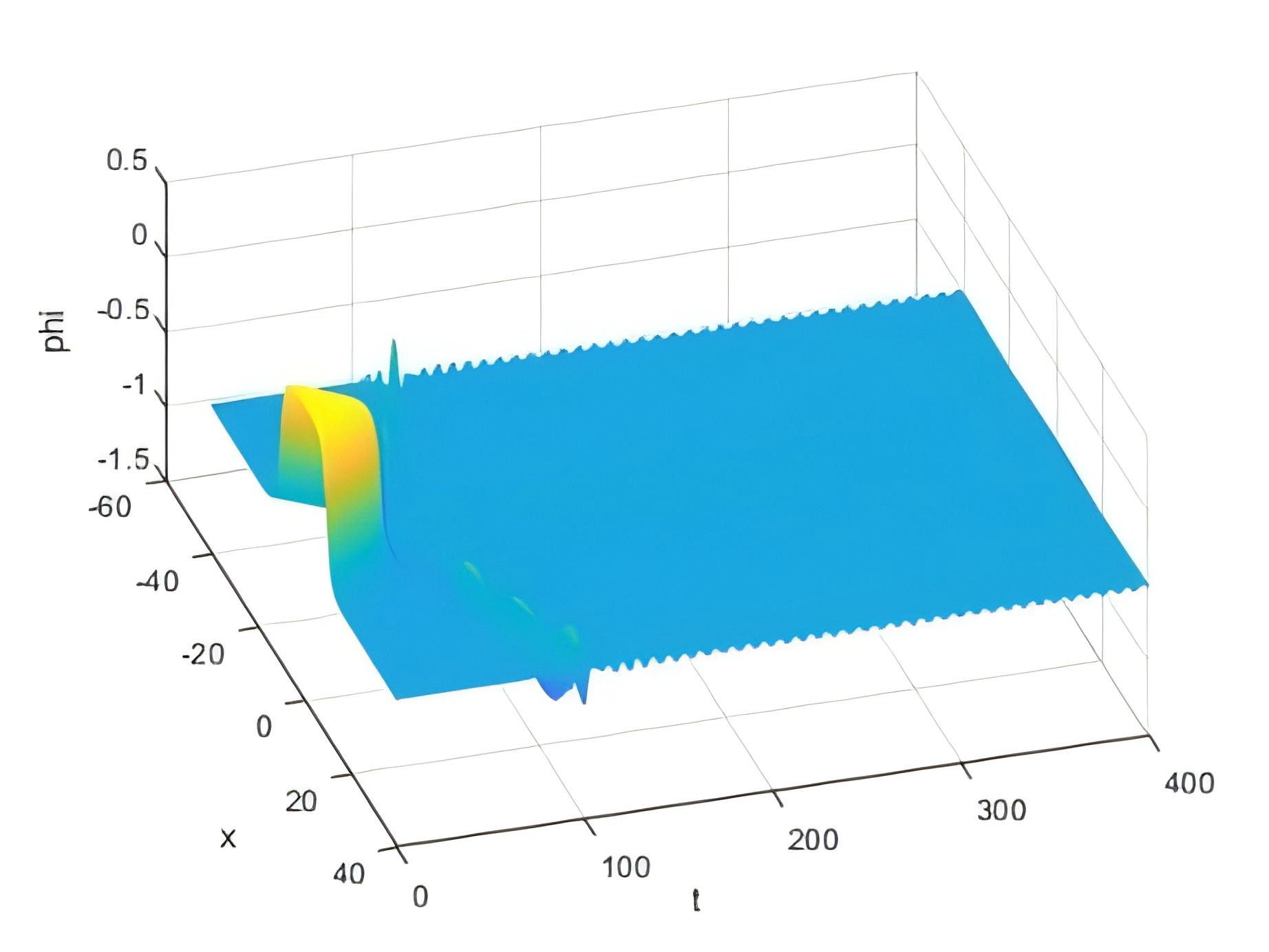}}
\subcaptionbox{$v=0.8$}{\includegraphics[width=0.3\textwidth]{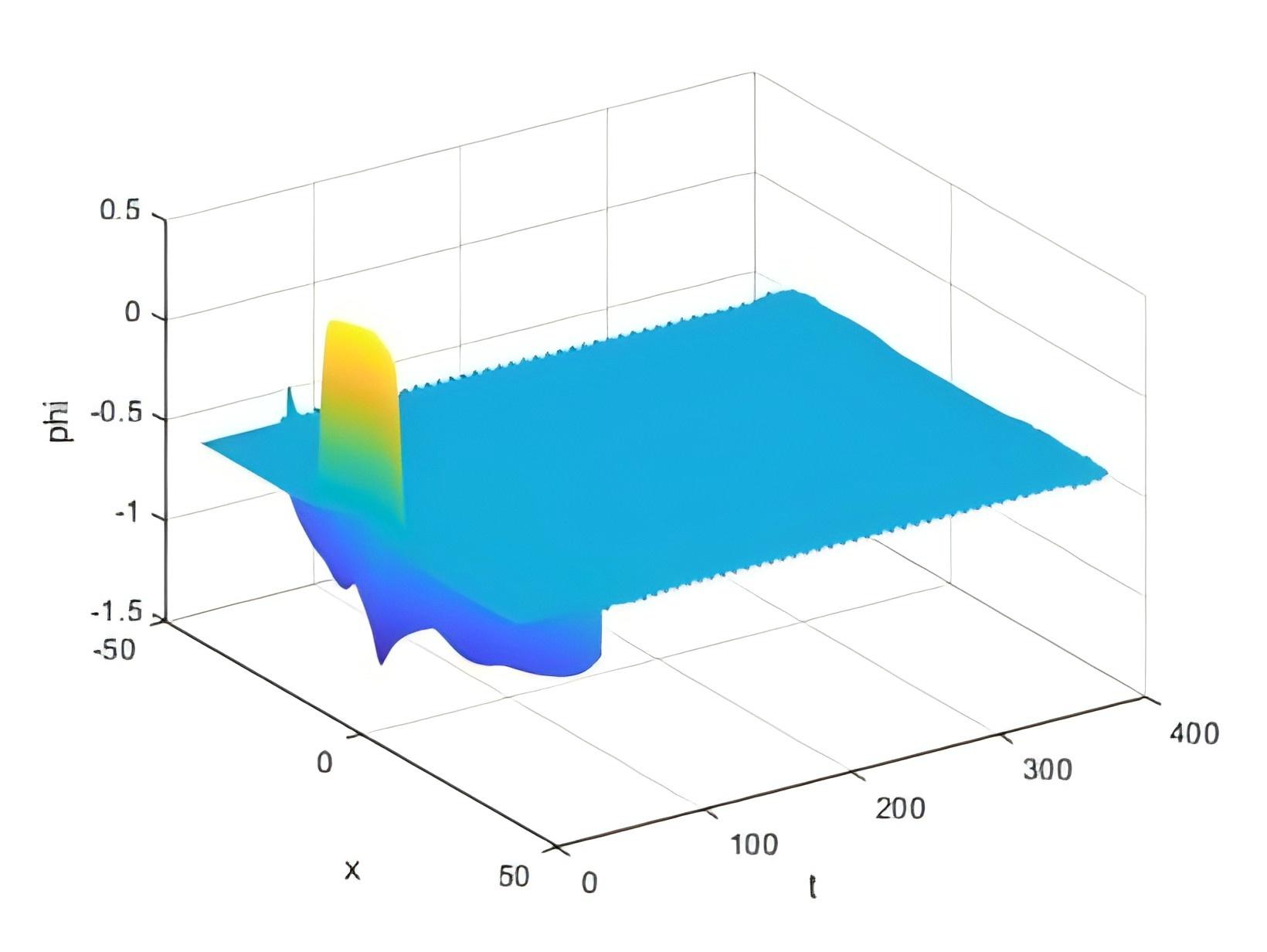}}
\caption{3D plot of $\phi_{K\bar{K}}$ with the initial velocities $v=0.02$, $v=0.2$ and $v=0.8$, respectively.}
\label{Fig:7}
\end{figure}

Figure \ref{Fig:6} depicts the time evolution of $\phi_{K\bar{K}}$ at $x=0$ for the sector $(-1/2,1/2)$. We present three cases with $v=0.02, 0.2$, and $0.8$, respectively.   The corresponding 3D plots are presented in Figure \ref{Fig:7}.
In all three cases shown, the kink-antikink pair undergoes direct annihilation after the initial collision, with the field relaxing to the vacuum  at the spatial boundary. No escaping or bion state is observed. Some radiation escapes to the boundary with an outgoing velocity exceeding the incident velocity. This annihilation behavior is found to be the unique outcome for the full range of initial velocities within this topological sector. Bazeia et al. have reported annihilation in the inner sector $(-0.82,0.82)$ of the $\phi^8$ theory at $v=0.2$ \cite{ref16}. However, the authors did not study the full velocity range.
From the view of energy, all the initial kinetic energy and configuration energy are turned into the radiation and vibration mode of the vacuum. To understand this phenomenon, we also presented its energy density plot.
\begin{figure} [h]
\centering
\subcaptionbox{$v=0.02$}{\includegraphics[width=0.3\textwidth]{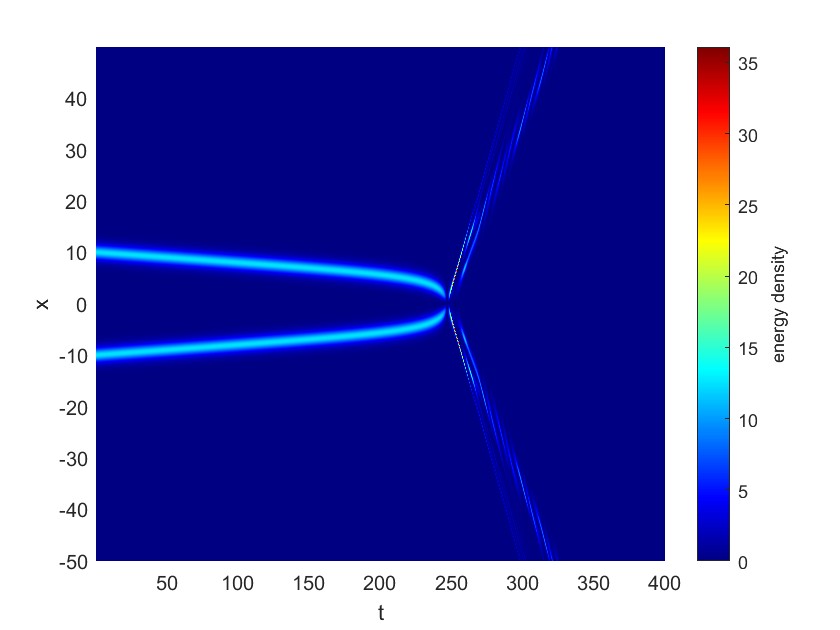}}
\subcaptionbox{$v=0.2$}{\includegraphics[width=0.3\textwidth]{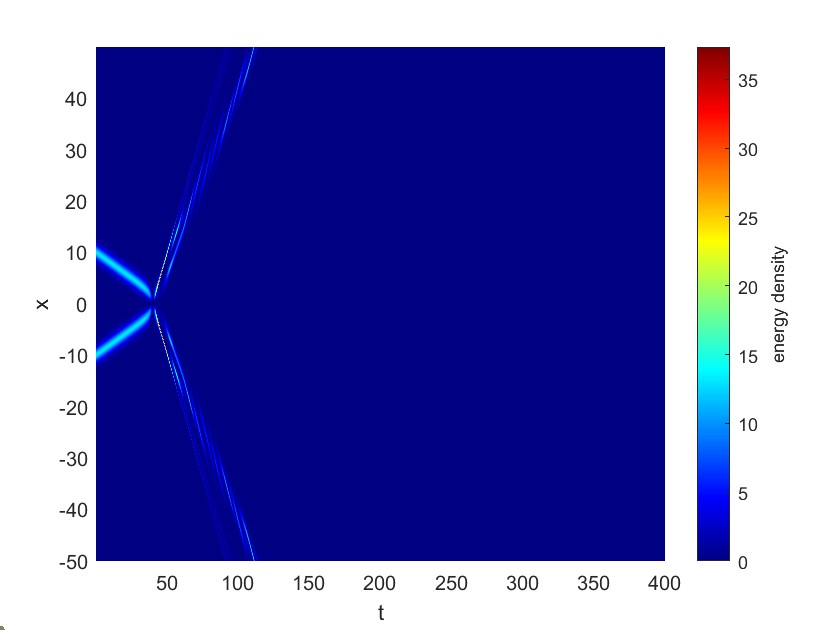}}
\subcaptionbox{$v=0.8$}{\includegraphics[width=0.3\textwidth]{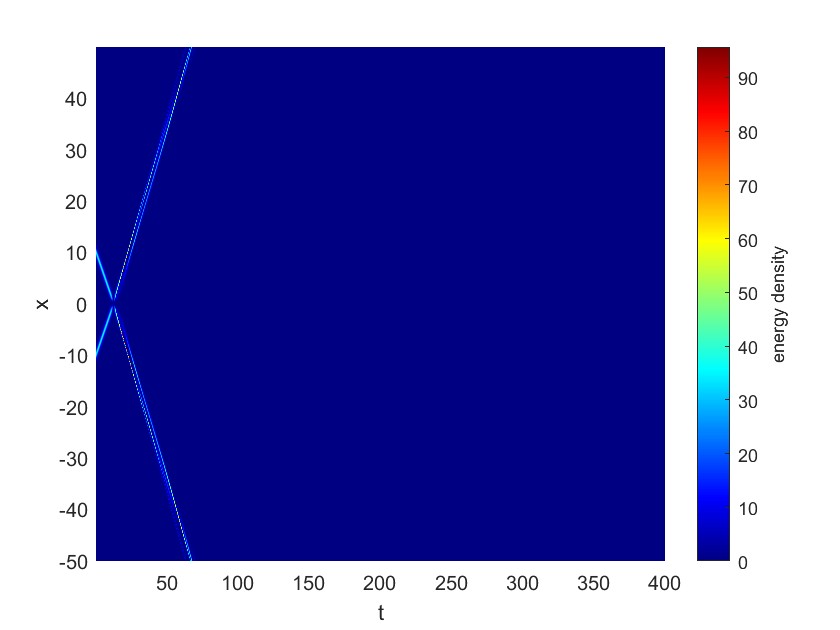}}
\caption{energy density plots of $\phi_{K\bar{K}}$ with the initial velocities $v=0.02$, $v=0.2$ and $v=0.8$, respectively.}
\label{Fig:A}
\end{figure}
In Figure \ref{Fig:A}, the originally concentrated energy density can no longer maintain the localized soliton configuration and collapses abruptly. The localized energy is fully radiated toward both boundary sides in the form of dispersive wave modes. 

\subparagraph{(iii) Topological sector $(1/2, 1)$}\mbox{}\\
\begin{figure} [h] 
\centering
\subcaptionbox{$v=0.1$}{\includegraphics[width=0.3\textwidth]{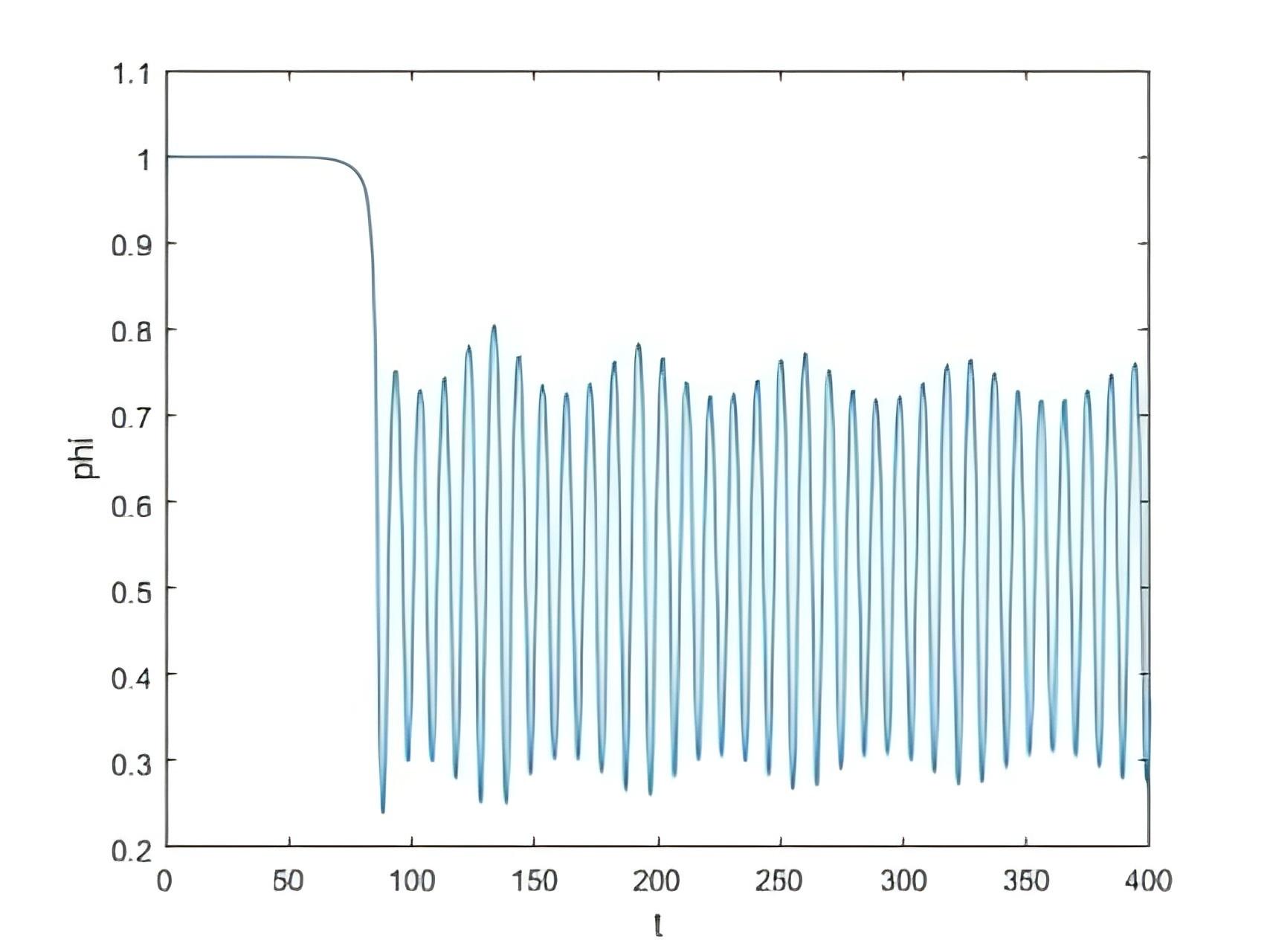}}
\subcaptionbox{$v=0.75$}{\includegraphics[width=0.3\textwidth]{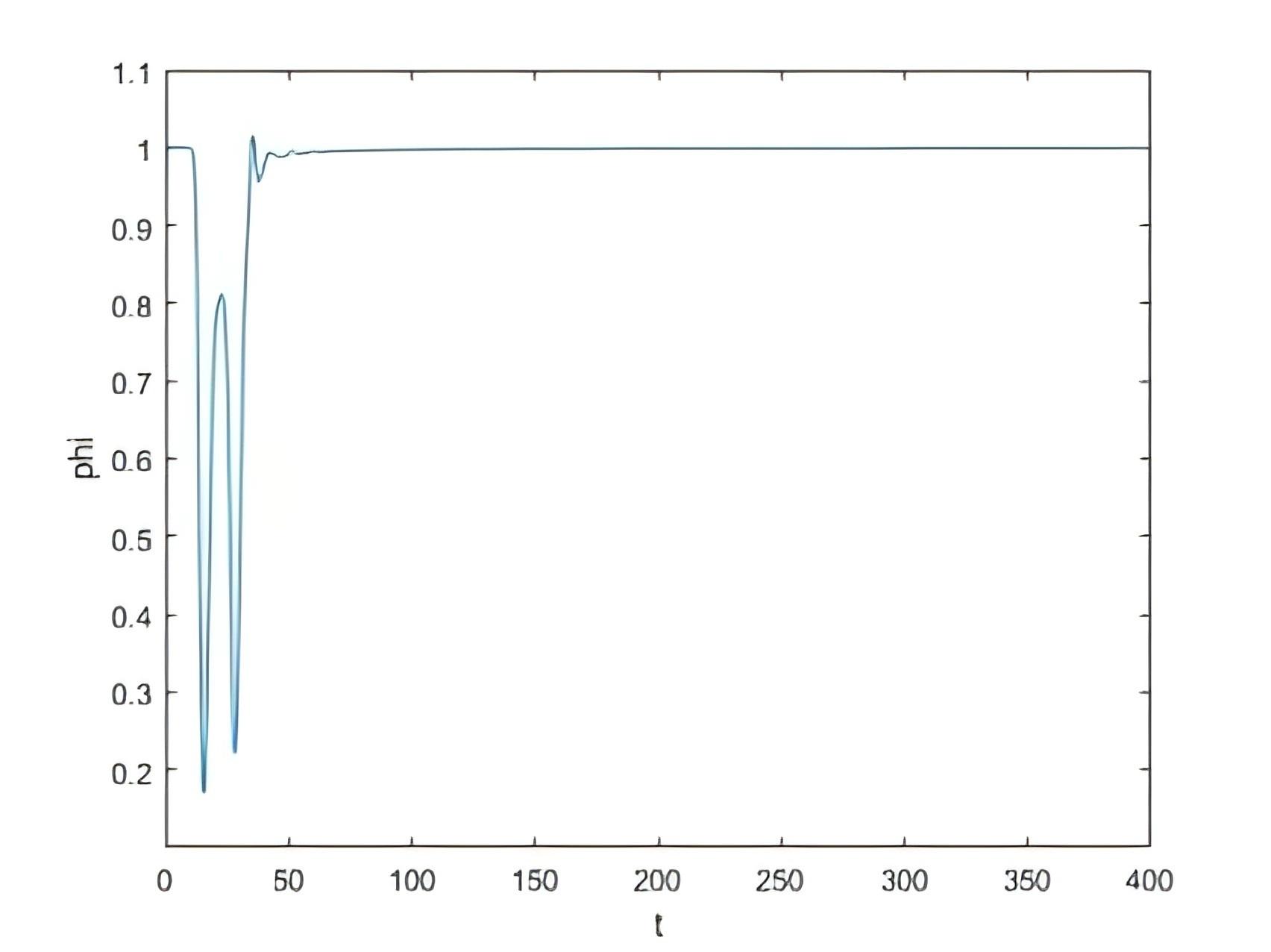}}
\subcaptionbox{$v=0.85$}{\includegraphics[width=0.3\textwidth]{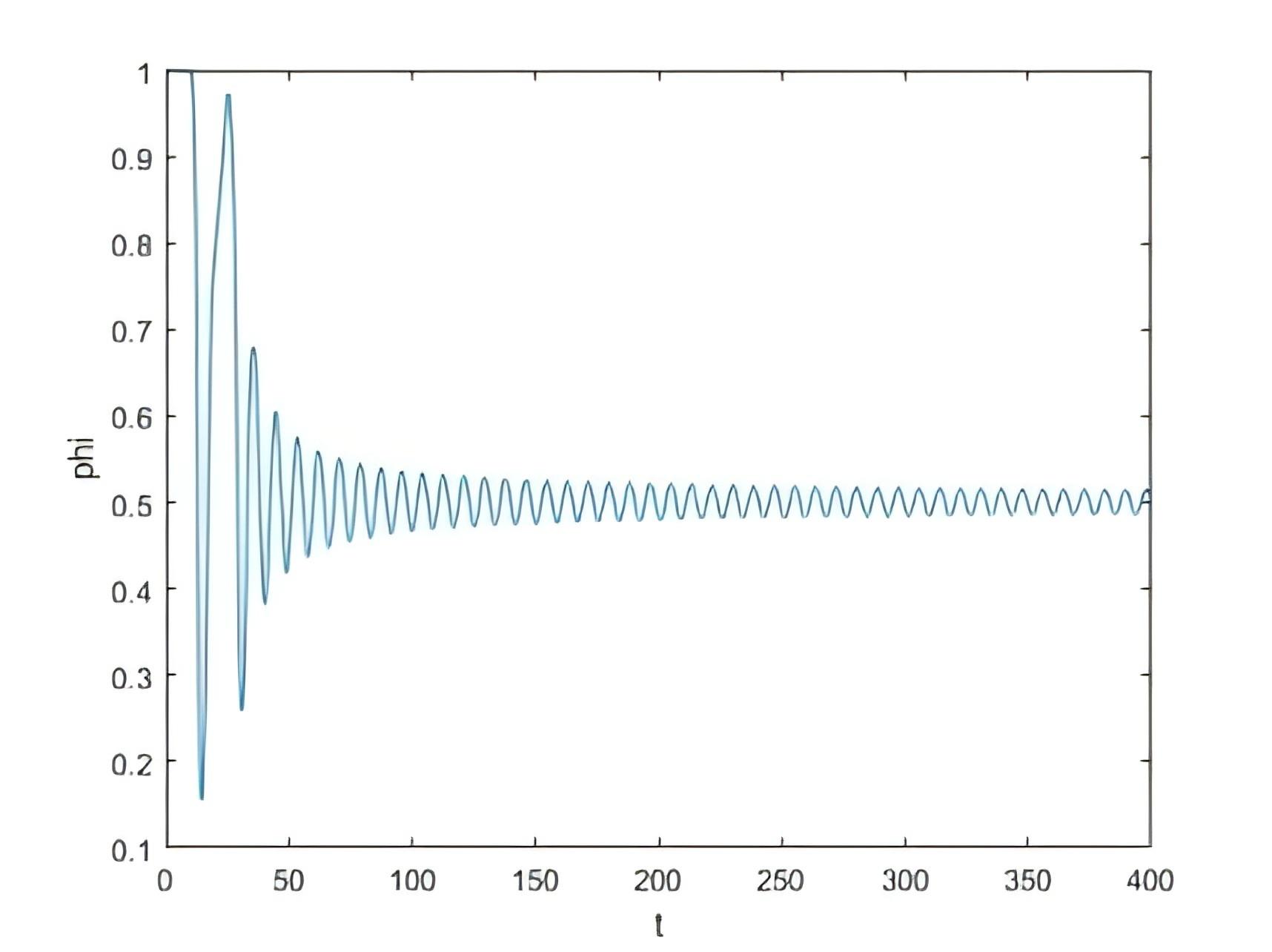}}
\caption{2D plot of the $\phi_{K\bar{K}}$ at $x=0$. From left to right panels, incident velocities are $v=0.1$, $v=0.75$ and $v=0.85$, respectively.}
\label{Fig:8}
\end{figure}

Figure \ref{Fig:8} demonstrates the time evolution of $\phi_{K\bar{K}}$ at $x=0$ for the topological sector $(1/2, 1)$. In the left panel,  the kink-antikink soliton pair oscillates near the vacuum after collision over the range $t \in [0,400]$. In the middle panel, the number of bounces is two, and the pair reflects back. In the right panel, different from the panel (a), the kink-antikink soliton pair first bounces one time, and then quickly annihilates to vacuum in a manner similar to the phenomenon in topological sector $(-1/2,1/2)$. The 3D plots in conjunction with the velocities in Figure \ref{Fig:8} is presented below. \\
\begin{figure} [h]
\centering
\subcaptionbox{$v=0.1$}{\includegraphics[width=0.3\textwidth]{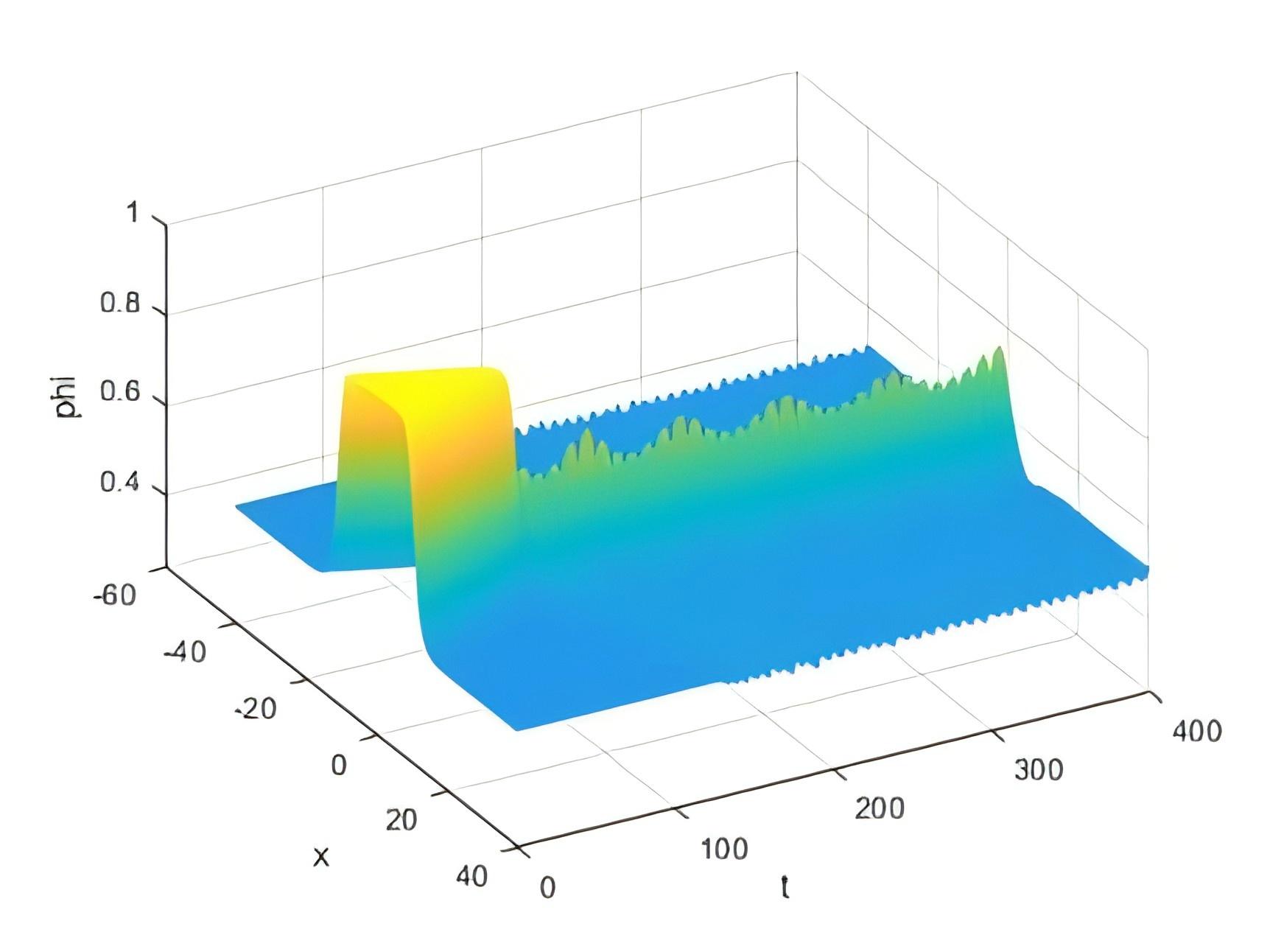}}
\subcaptionbox{$v=0.75$}{\includegraphics[width=0.3\textwidth]{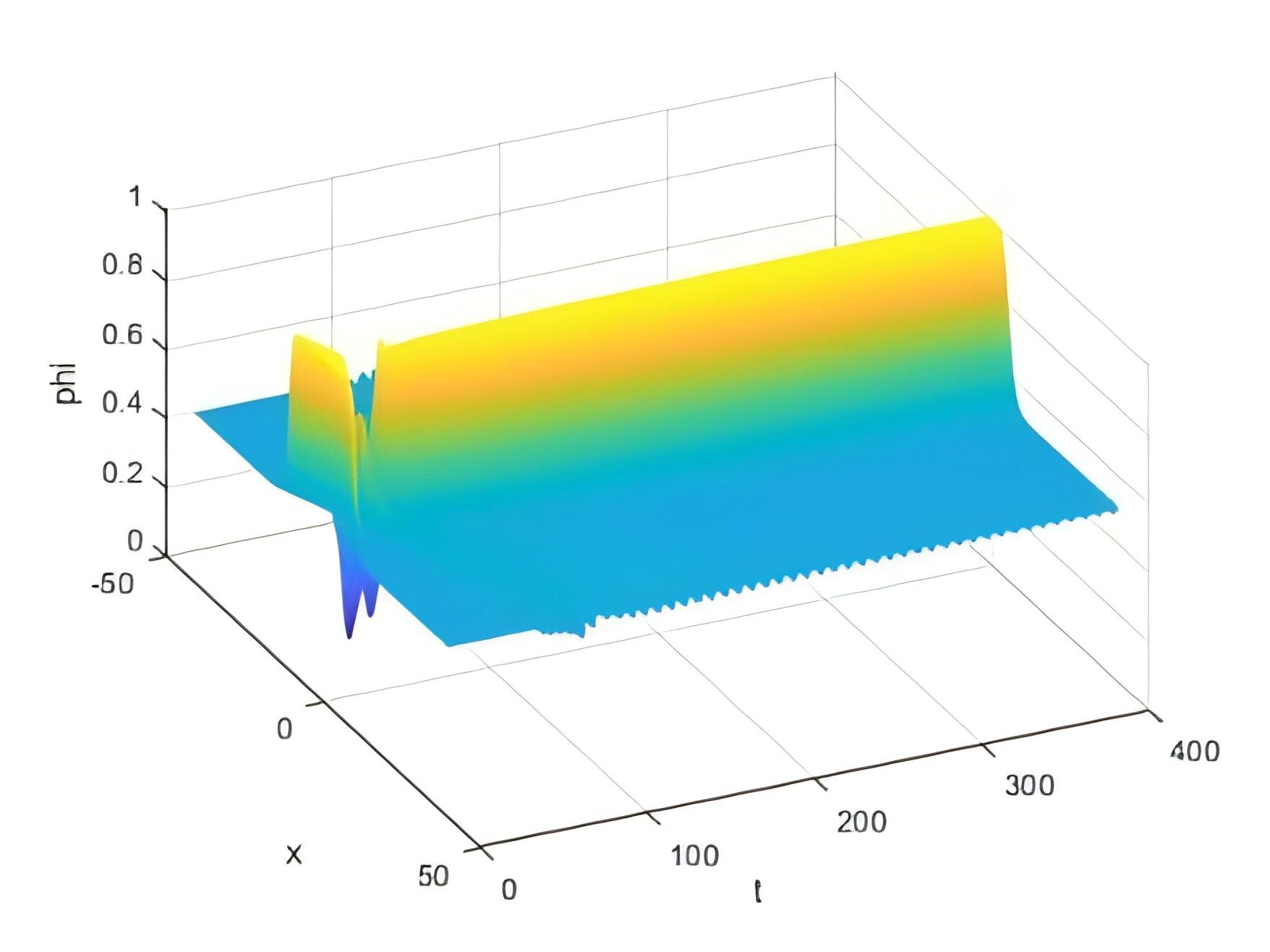}}
\subcaptionbox{$v=0.85$}{\includegraphics[width=0.3\textwidth]{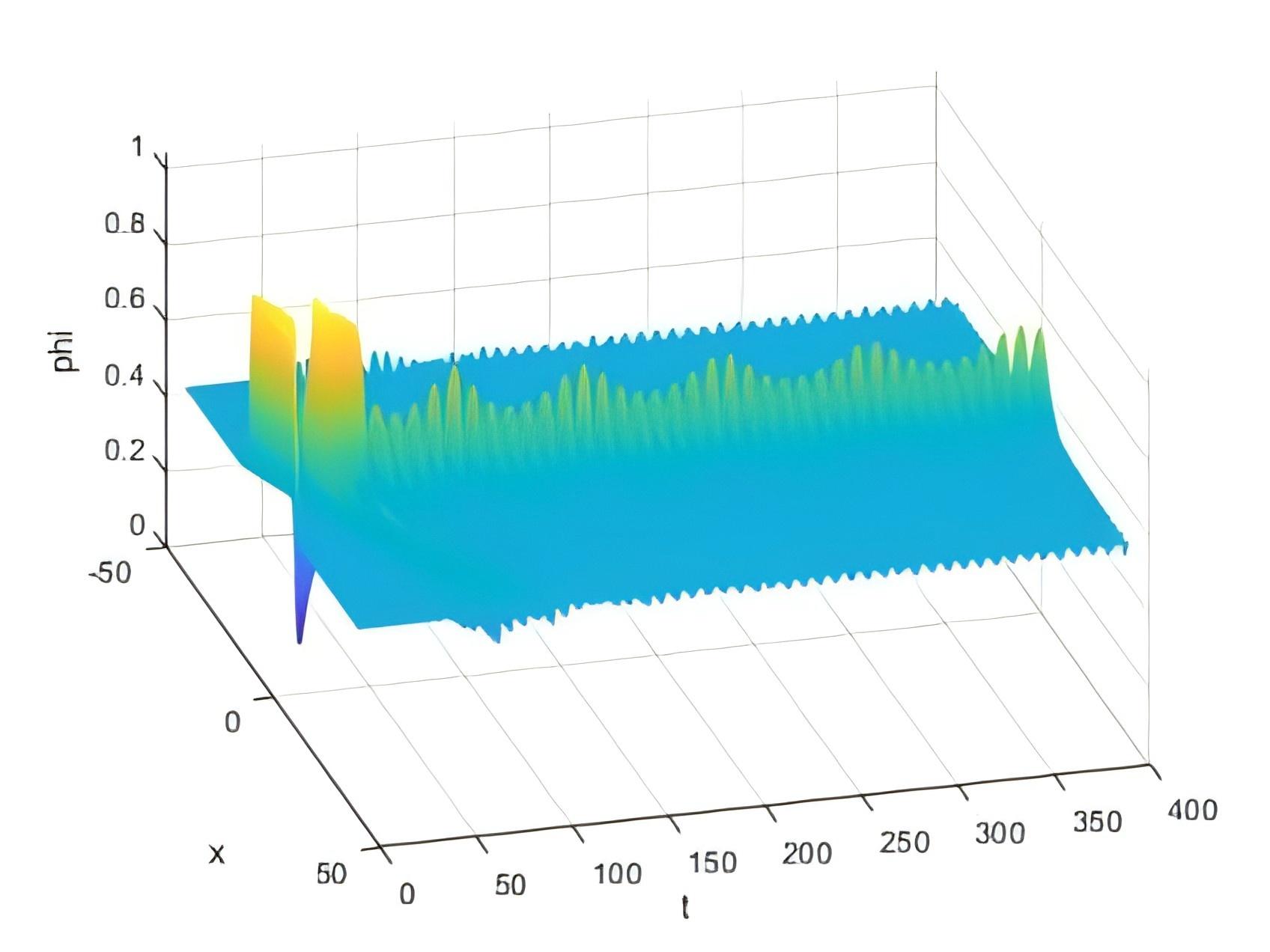}}
\caption{3D plot of $\phi_{K\bar{K}}$ with the incident velocities $v=0.1$, $v=0.75$ and $v=0.85$, respectively.}
\label{Fig:9}
\end{figure}

Figure \ref{Fig:8} and Figure \ref{Fig:9} illustrate the time evolution of $\phi_{K\bar{K}}$ over the range of $x \in [-200,200]$, revealing three distinct dynamical outcomes.
For the case of $v=0.85$, the energy transfer process is complex, which is uncommon in the kink and anti-kink collisions. After the first bounce, the system undergoes substantial energy dissipation, which rapidly depletes the pair's kinetic energy. The bion decays to the vacuum state, with some radiation escaping to the boundary. 
\paragraph{(2) $n = p_2/p_1 = 3$}\mbox{}\\

\subparagraph{(i) Topological sector $(-1, -1/3)$}\mbox{}\\
\begin{figure} [h]
\centering
\subcaptionbox{$v=0.088$}{\includegraphics[height=0.24\textwidth]{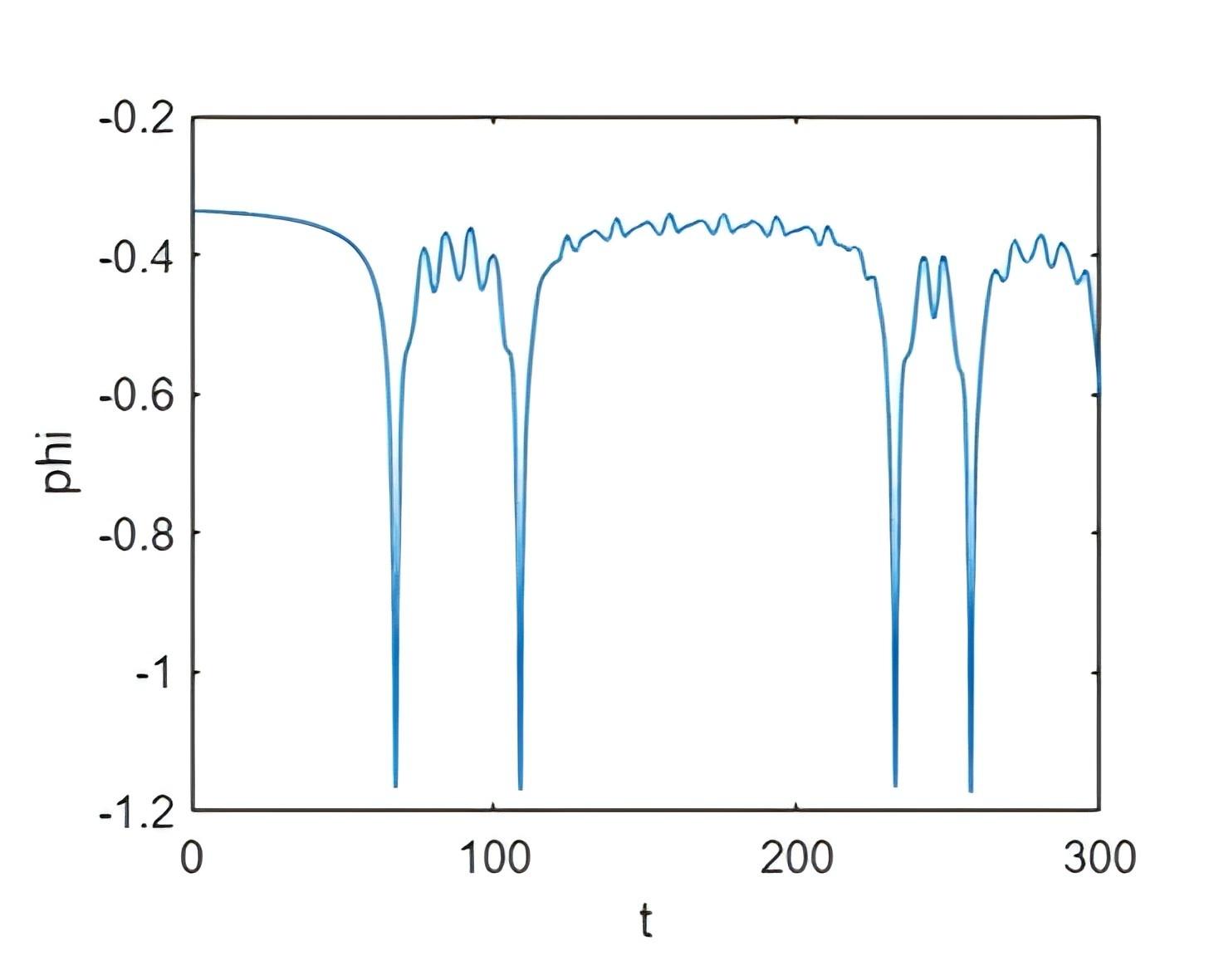}}
\subcaptionbox{$v=0.122$}{\includegraphics[height=0.24\textwidth]{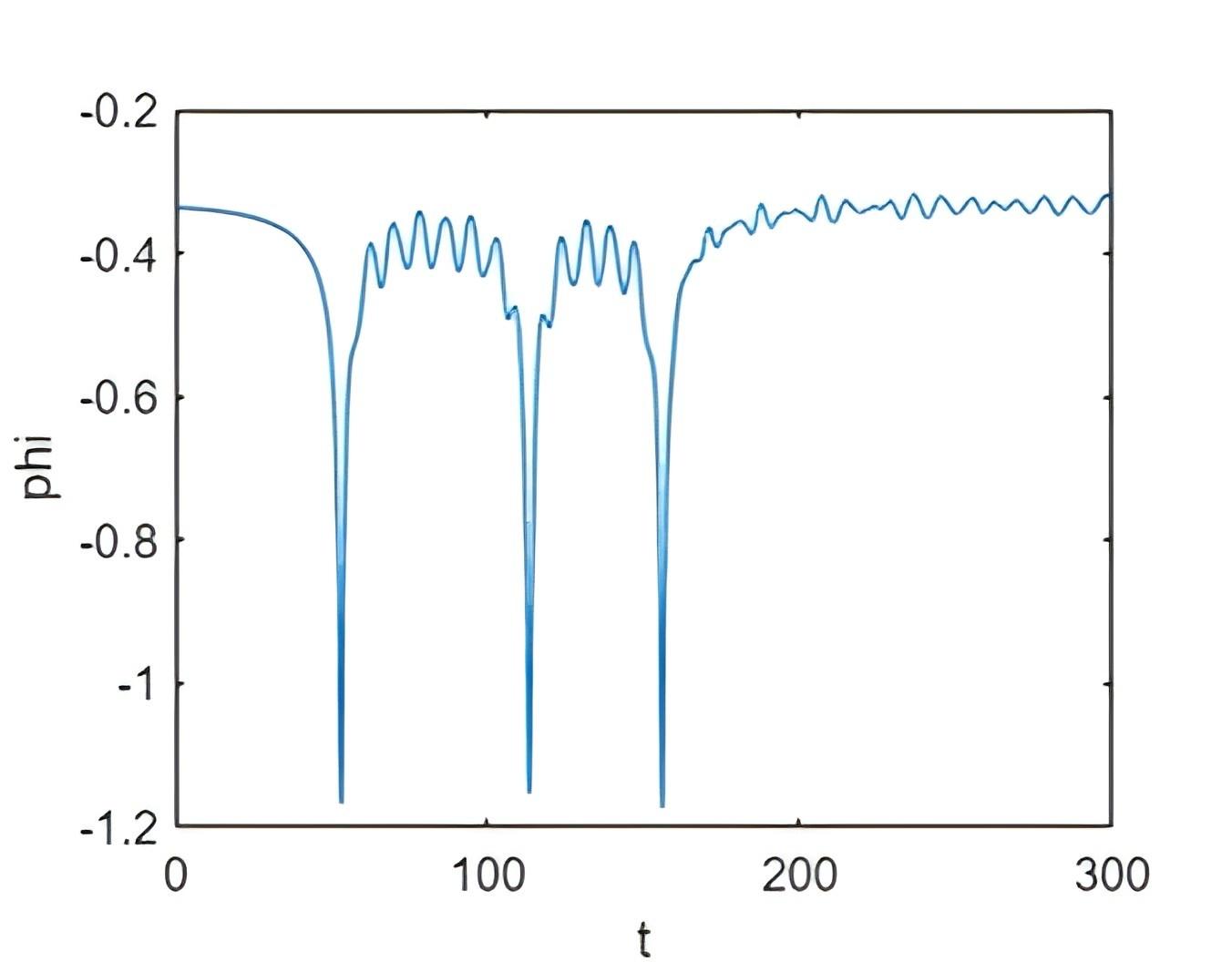}}
\subcaptionbox{$v=0.142$}{\includegraphics[height=0.24\textwidth]{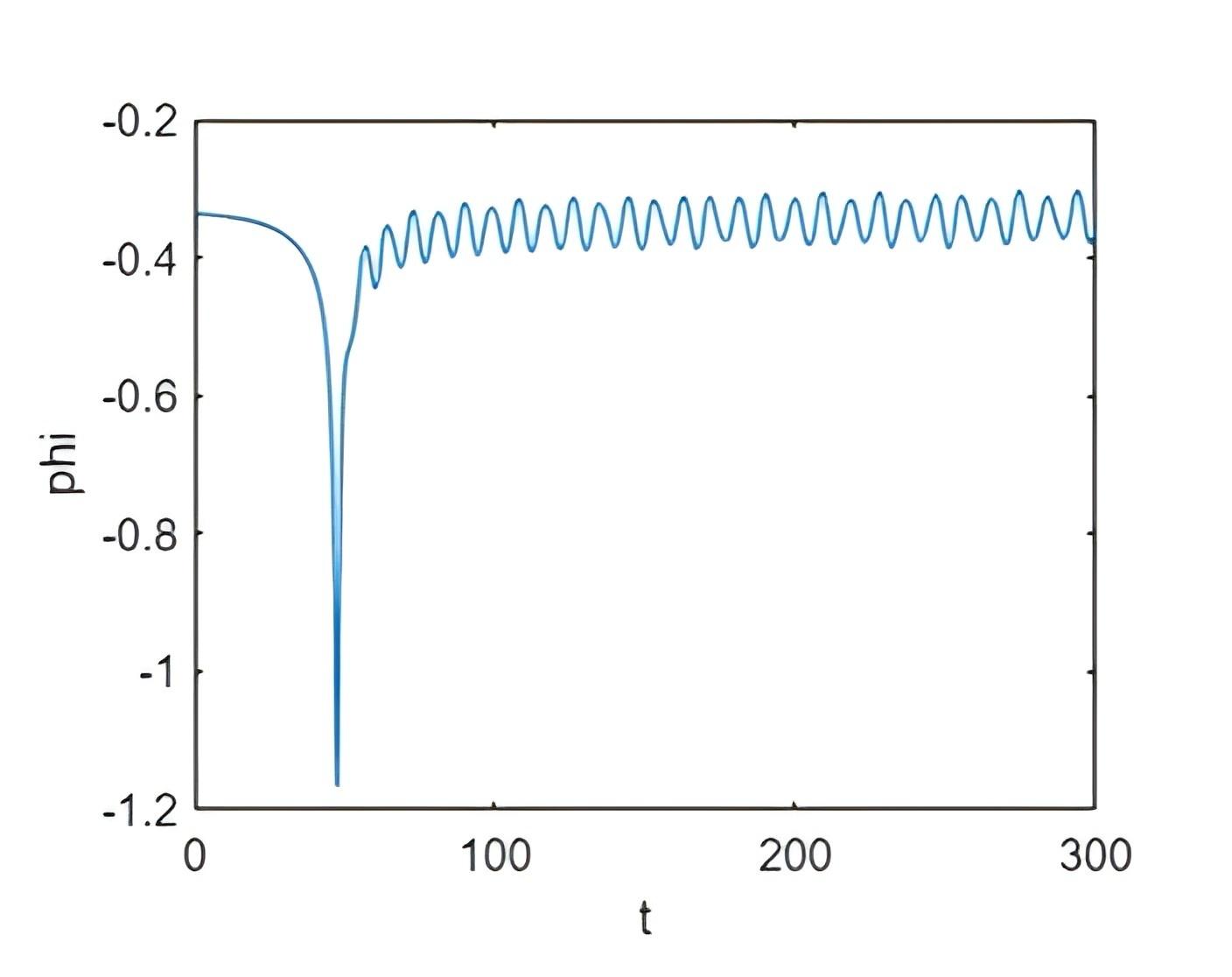}}
\caption{2D plot of the $\phi_{K\bar{K}}$ at $x=0$. From left to right panels, incident velocities are $v=0.088$, $v=0.122$ and $v=0.142$, respectively.}
\label{Fig:10}
\end{figure}

\begin{figure} [h]
\centering
\subcaptionbox{$v=0.088$}{\includegraphics[height=0.23\textwidth]{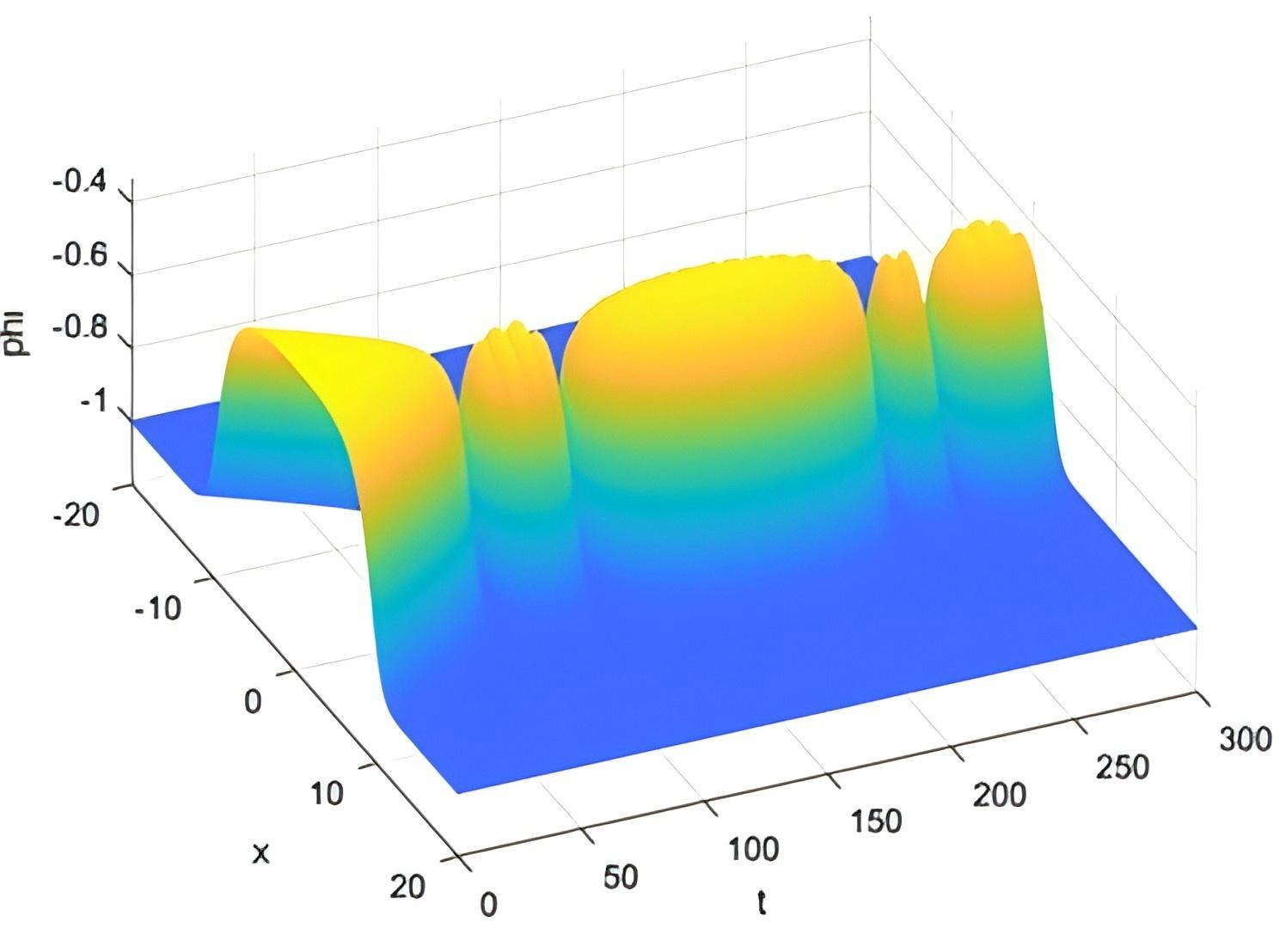}}
\subcaptionbox{$v=0.122$}{\includegraphics[height=0.23\textwidth]{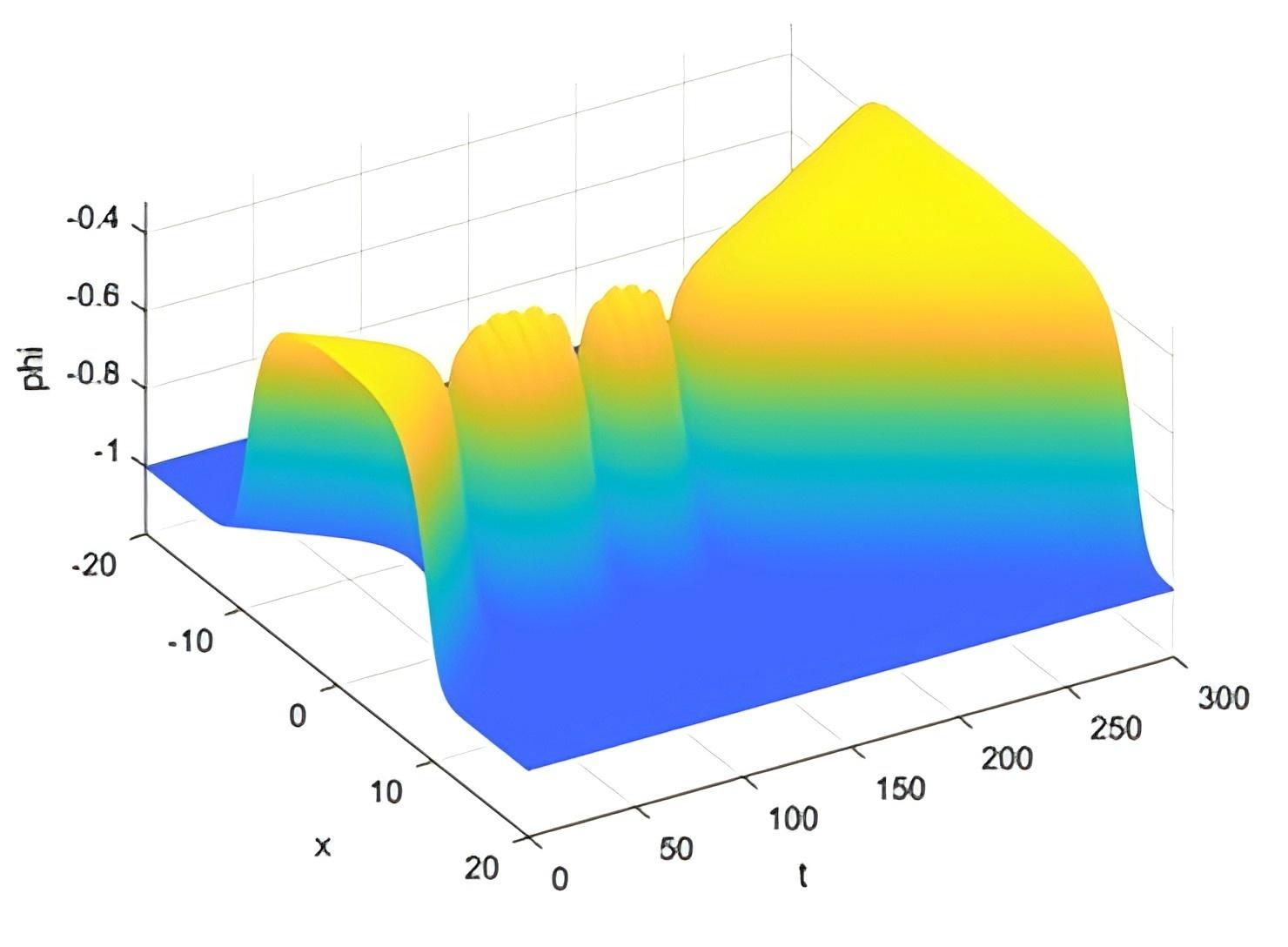}}
\subcaptionbox{$v=0.142$}{\includegraphics[height=0.23\textwidth]{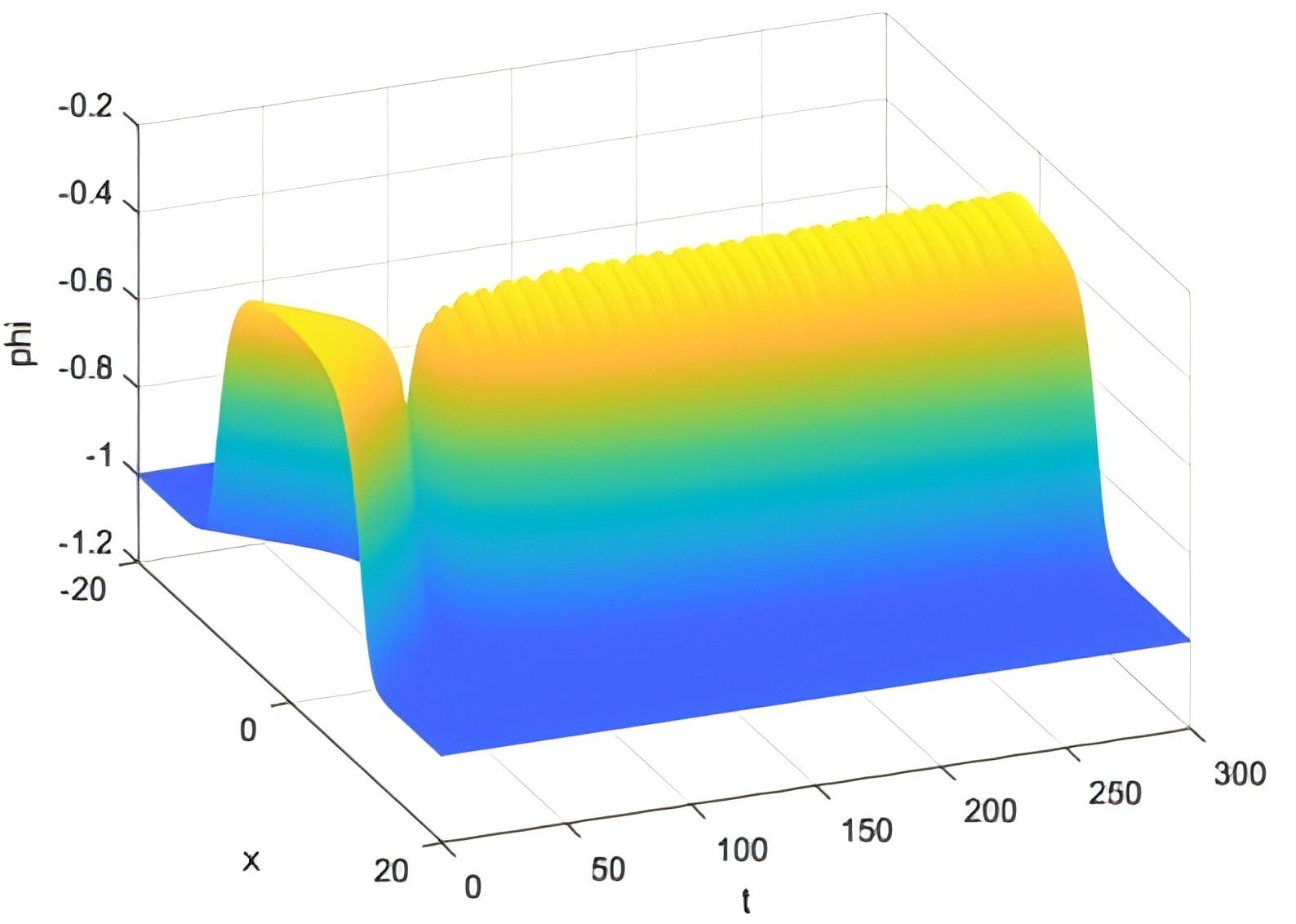}}
\caption{3D plot of $\phi_{K\bar{K}}$ with the initial velocities $v=0.088$, $v=0.122$ and $v=0.142$, respectively.}
\label{Fig:11}
\end{figure}

Figure \ref{Fig:10} and Figure \ref{Fig:11}  demonstrate the time evolution of $\phi_{K\bar{K}}$ for the topological sector $(-1,-1/3)$.  
For $v=0.088$, the kink and antikink pair undergoes four bounces before $t=300$. For $v=0.122$, the soliton pair escapes after three bounces. For a higher velocity of $v=0.142$, the pair escapes immediately after the first collision. These three cases reveal a transition from multi-bounce resonances to direct escape as the initial velocity $v$ increasing. Ignoring the incident velocities, these collision channels are similar to the collisions in topological sector $(-1,-1/2)$ of $n=2$, and the energy transfer processes are also similar which will not be repeated here. 

\subparagraph{(ii) Topological sector $(-1/3, 1/3)$}\mbox{}\\

\begin{figure} [h]
\centering
\subcaptionbox{$v=0.276$}{\includegraphics[height=0.23\textwidth]{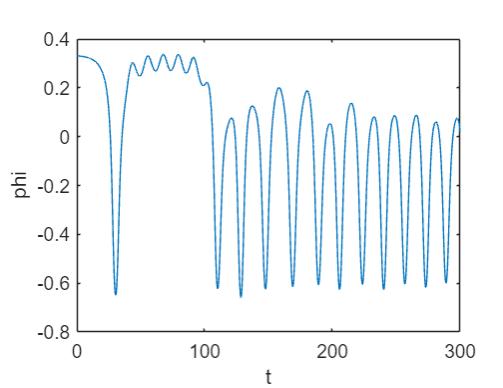}}
\subcaptionbox{$v=0.277$}{\includegraphics[height=0.23\textwidth]{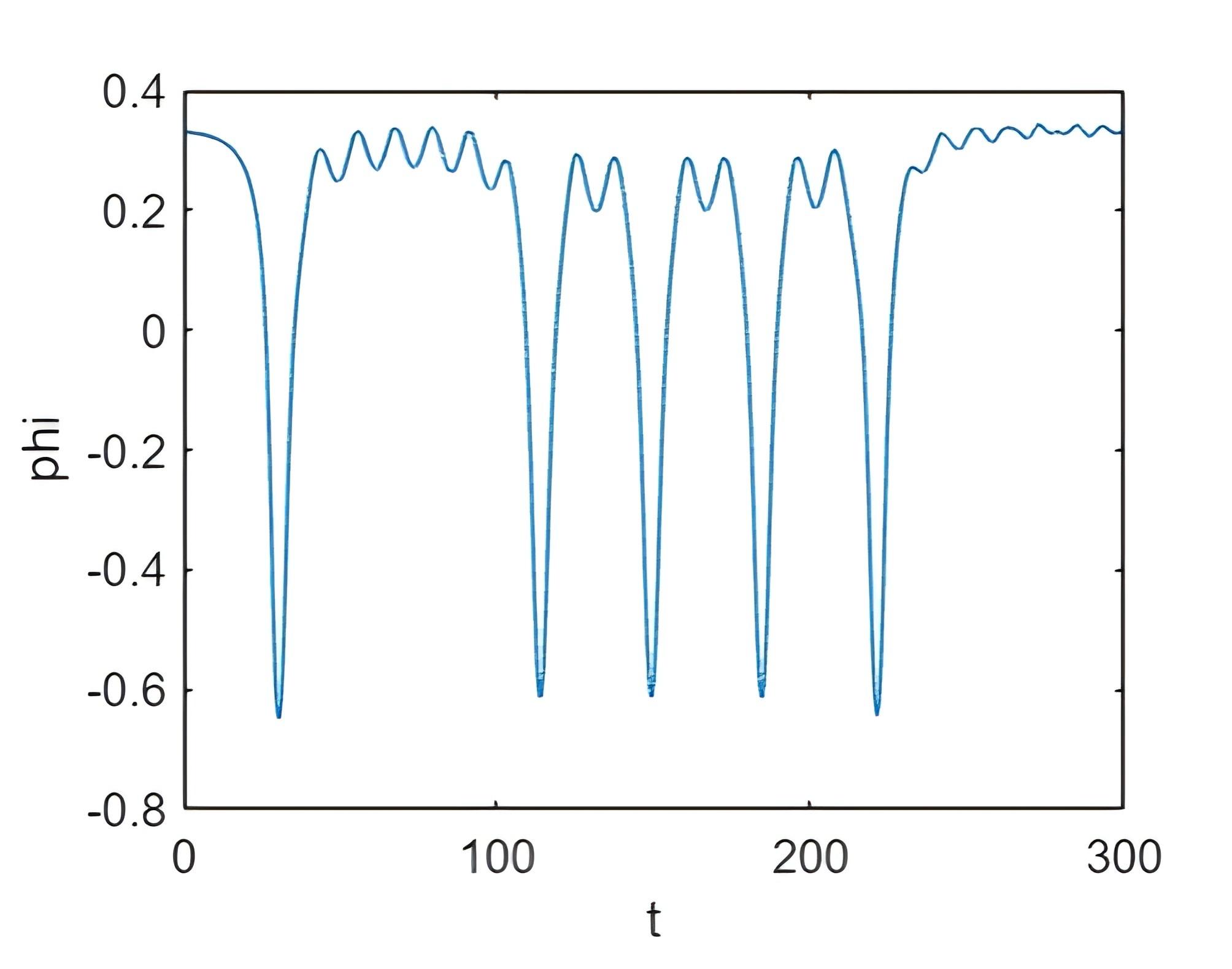}}
\subcaptionbox{$v=0.286$}{\includegraphics[height=0.23\textwidth]{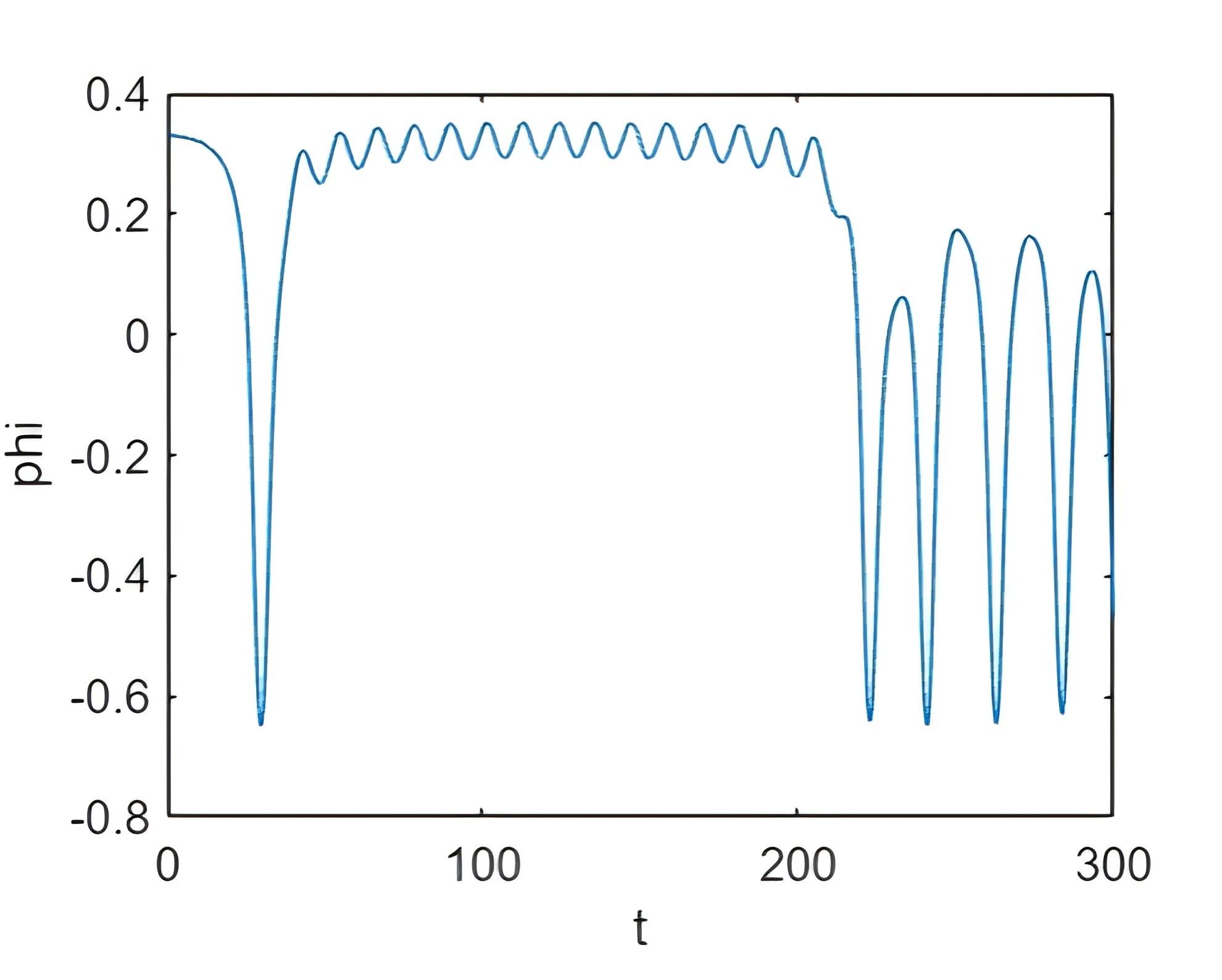}}
\caption{2D plot of the $\phi_{K\bar{K}}$ at $x=0$. Incident velocities are $v=0.276$, $v=0.277$ and $v=0.286$, respectively.}
\label{Fig:12}
\end{figure}

\begin{figure} [h]
\centering
\subcaptionbox{$v=0.276$}{\includegraphics[height=0.22\textwidth]{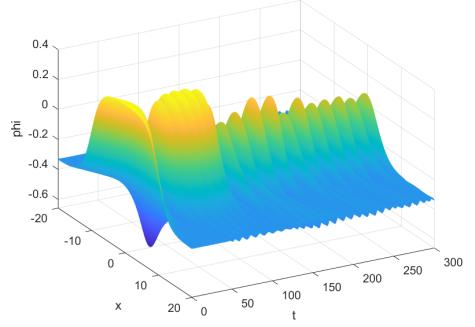}}
\subcaptionbox{$v=0.277$}{\includegraphics[height=0.22\textwidth]{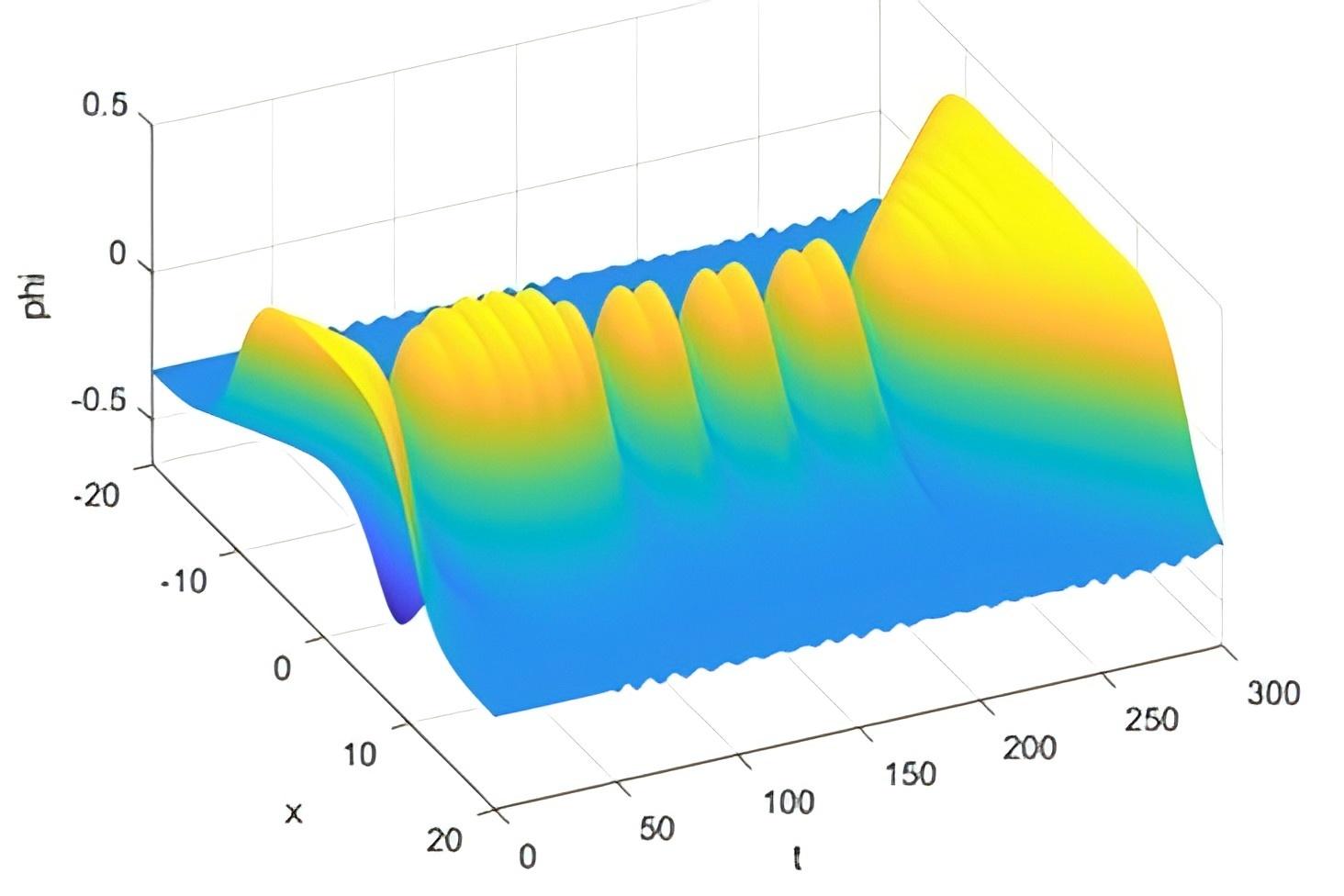}}
\subcaptionbox{$v=0.286$}{\includegraphics[height=0.22\textwidth]{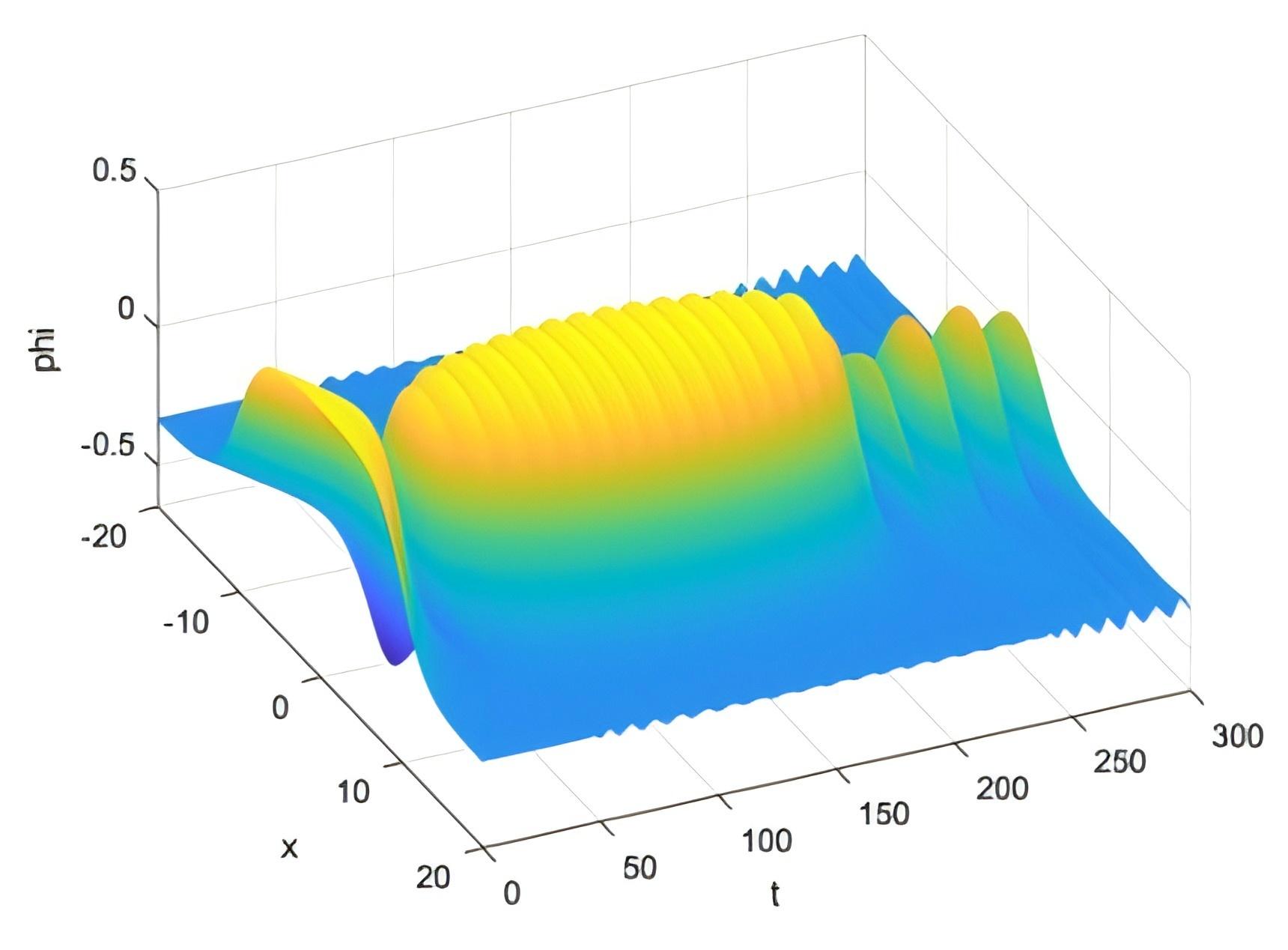}}
\caption{3D plot of $\phi_{K\bar{K}}$ with the initial velocities $v=0.276$, $v=0.277$ and $v=0.286$, respectively.}
\label{Fig:13}
\end{figure}

The evolution of the kink-antikink field configuration, $\phi_{K\bar{K}}(x,t)$, was simulated over the domain $x \in [-200,200]$ and $t \in [0,300]$. The resulting dynamics reveal a rich variety of escape scenarios dependent on the initial collision velocity $v$. Figure \ref{Fig:12} manifests the collision channels for the sector $(-1/3,1/3)$. In the left panel, the soliton pair shows a bounce first, and then forms the bion state after the second collision. In the middle panel, the pair escapes successfully after 5 bounces. In the right panel, the soliton pair separates for a long time after the first collision, with the time duration being about 200, and forms a bion state upon the second collision. 3D plots are presented below in conjunction with incident velocities in Figure \ref{Fig:13}.

For $v=0.276$, the kink-antikink pair separates for a while and then form a bion state. The time intervals between successive collisions exhibit a non-uniform pattern, approximately following a long-short one. This sequence can be understood through an energy exchange mechanism. Initially, the pair possesses sufficient kinetic energy to overcome the short-range attractive potential, resulting in a long separation. Following the first collision, energy is transferred to the internal mode excitation, leading to less energy to escape or separate for a while.  After the second collision, the internal modes disappear and  the pair form a bion state. Simultaneously, $\phi$ cannot return to $1/3$ and forms a lower bion state. The saw-toothed boundary indicates that some energy dissipates into boundary.

For $v=0.277$, a more complex five-bounce escape scenario is observed. Following an initial collision, the pair undergoes an extended separation of approximately 80 time units. Subsequently, they experience three more bounces with relatively uniform time intervals before finally escaping. The vibration modes exist at all bounce processes, but with different frequencies and amplifications. 

For $v=0.286$, the pair initially separates over an extended period of approximately 200 time units. However, unlike the previous cases, this separation does not culminate in escape. Instead, the pair re-approaches and forms a lower bion state. 

Notably, the five-bounce escape scenario observed at $v=0.277$ represents, up to our knowledge, the first reported instance of such high-order bouncing in the $\phi^8$ kink-antikink model. This result provides pivotal numerical evidence for high-order bounce resonance in this system and supports the theoretical conjecture proposed by \cite{ref13,ref22}, which stated that the number of bounces during soliton escape could exceed four.

\subparagraph{(iii) Topological sector $(1/3, 1)$}\mbox{}\\ \label{sec:131}
\begin{figure} [h]
\centering
\subcaptionbox{$v=0.05$}{\includegraphics[width=0.3\textwidth]{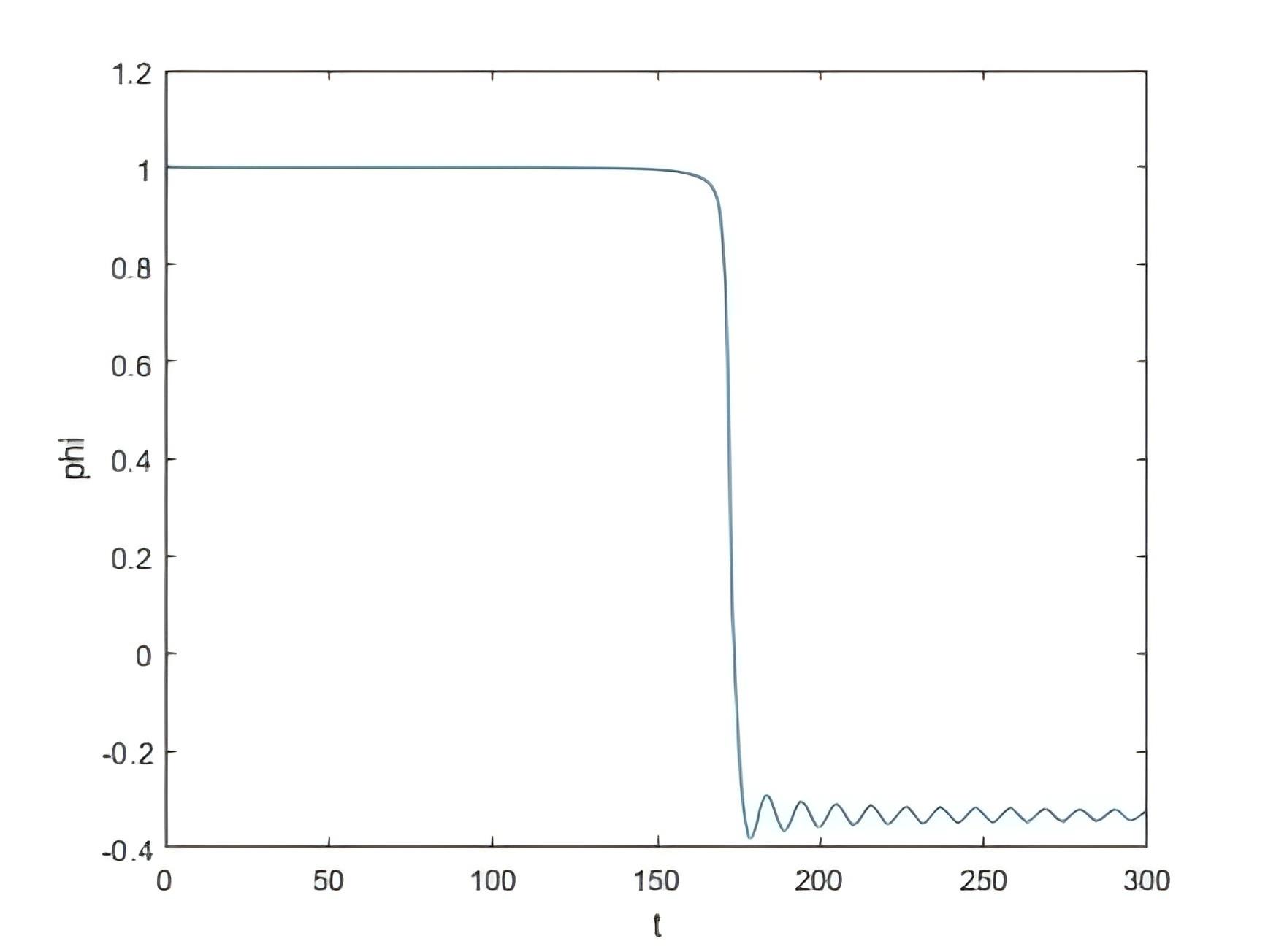}}
\subcaptionbox{$v=0.5$}{\includegraphics[width=0.3\textwidth]{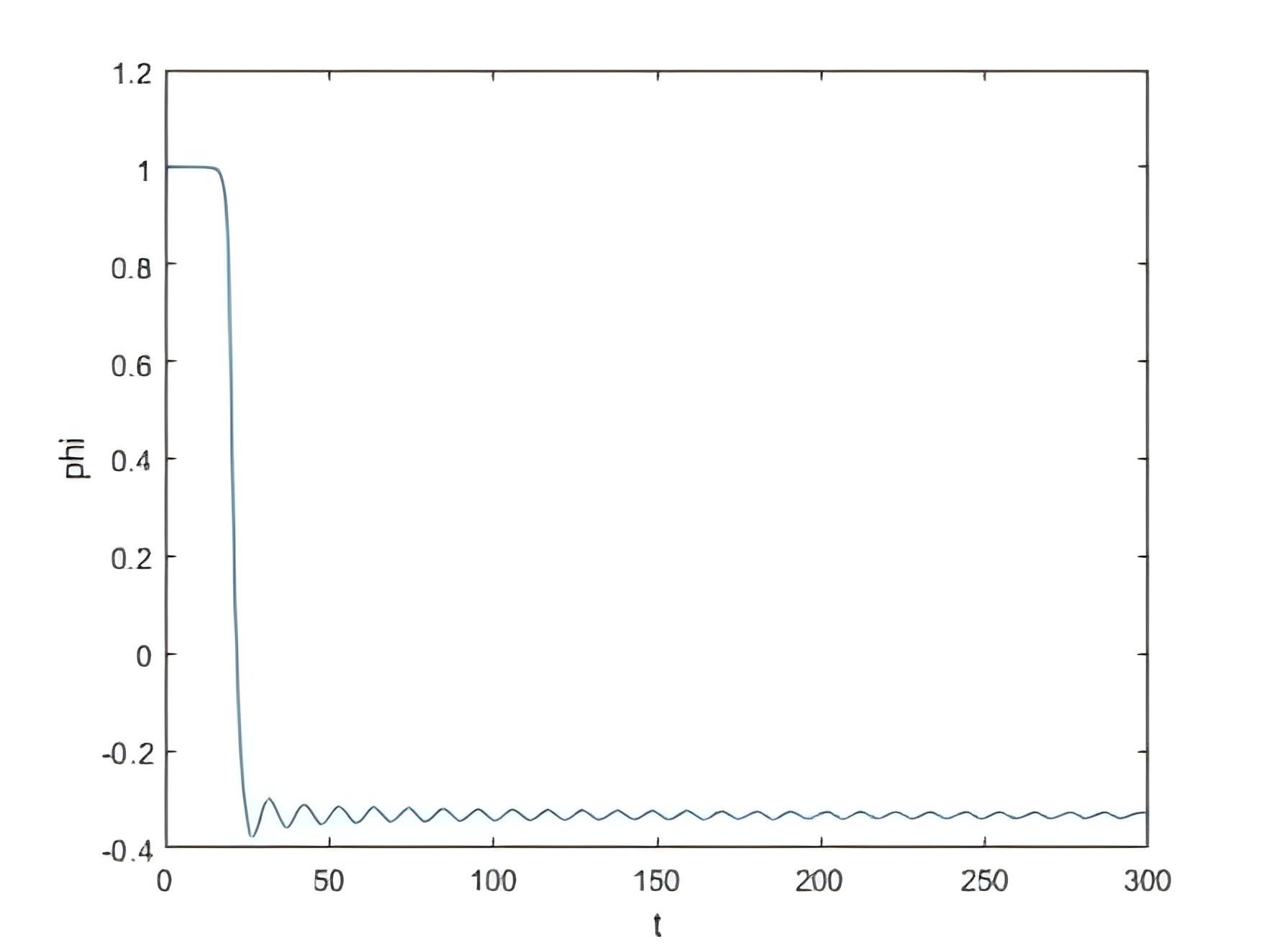}}
\subcaptionbox{$v=0.99$}{\includegraphics[width=0.3\textwidth]{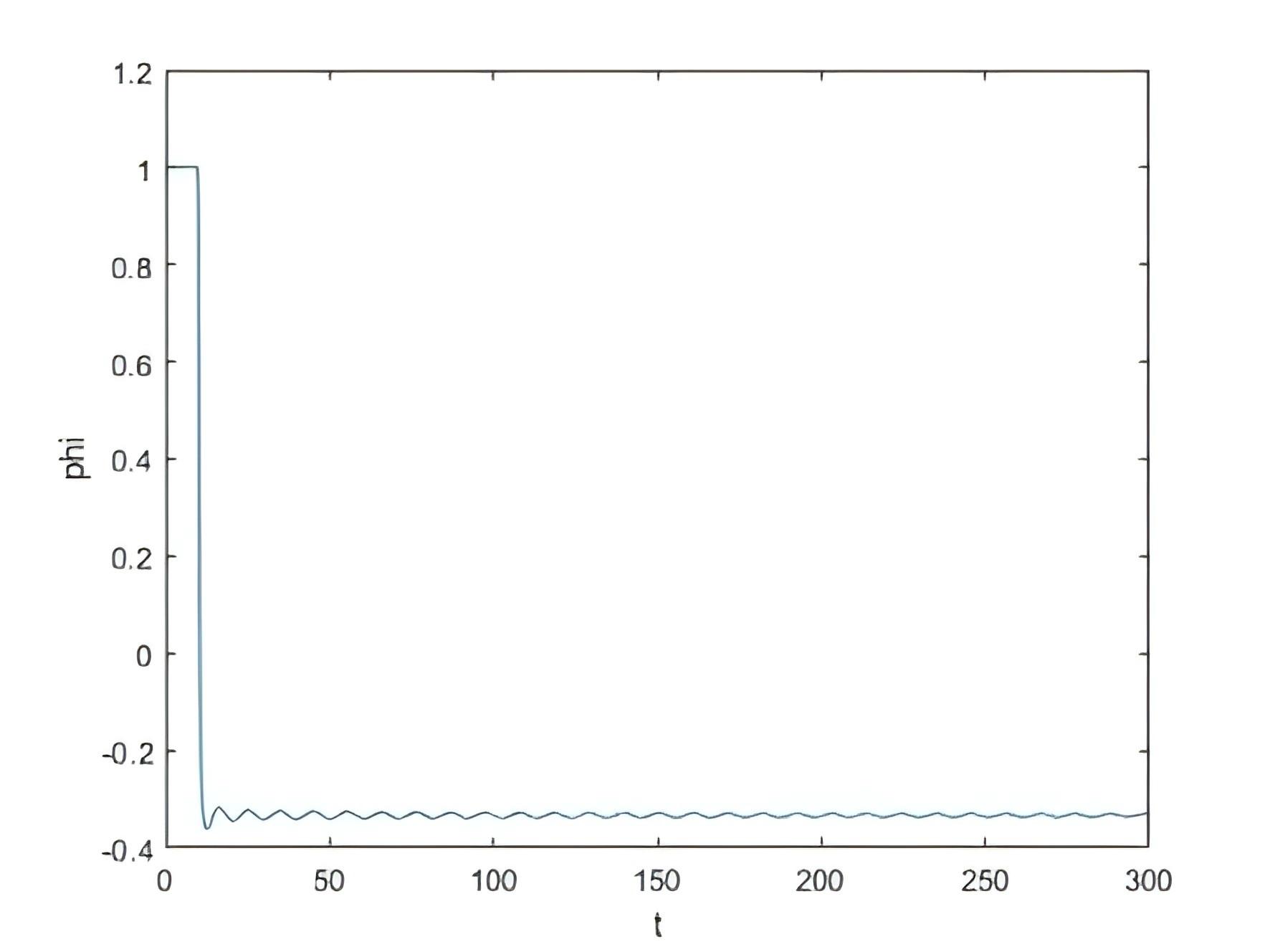}}
\caption{2D plot of the $\phi_{K\bar{K}}$ at $x=0$. Incident velocities are $v=0.05$, $v=0.5$ and $v=0.99$, respectively.}
\label{Fig:14}
\end{figure}

\begin{figure} [h]
\centering
\subcaptionbox{$v=0.05$}{\includegraphics[width=0.3\textwidth]{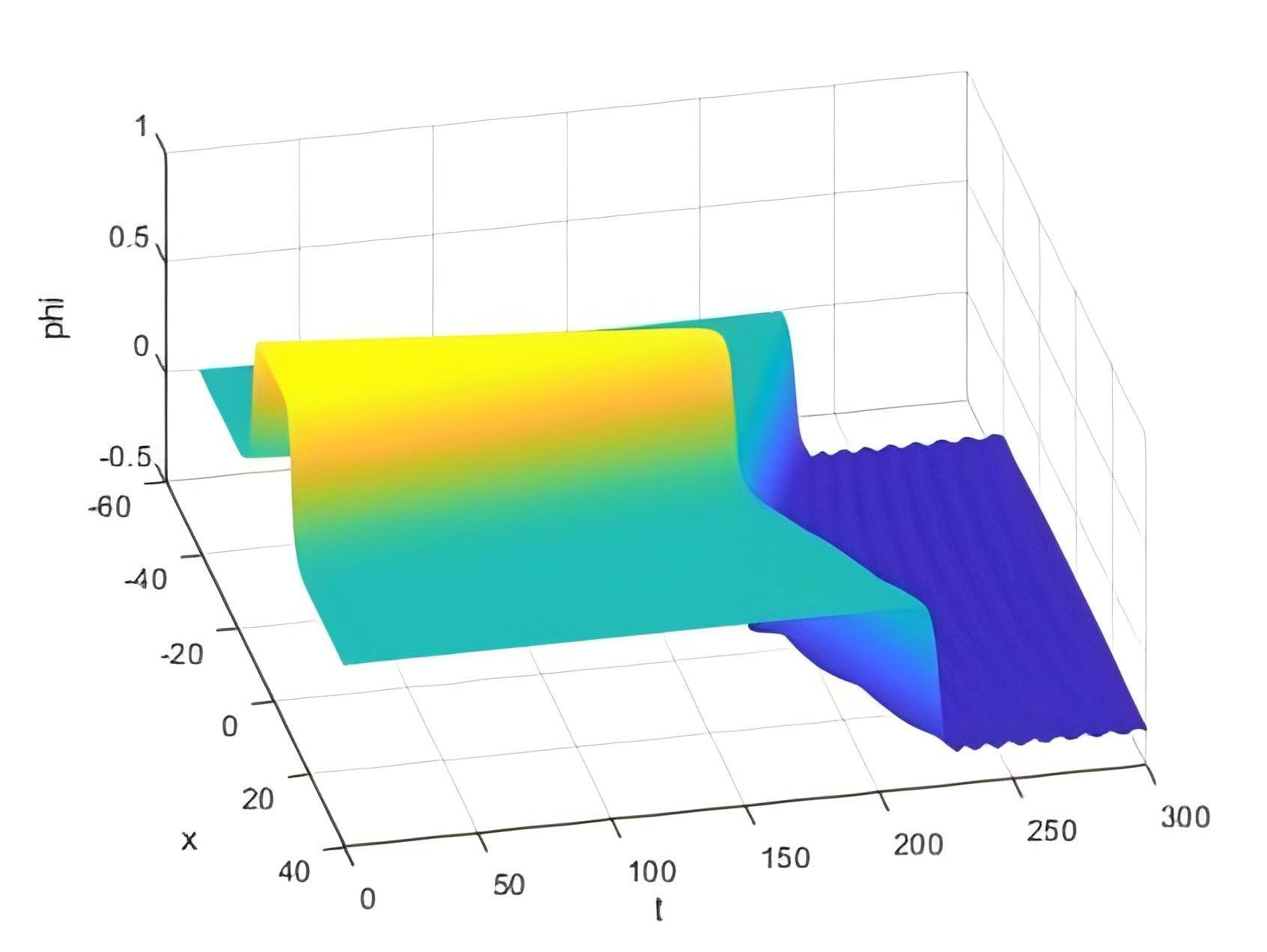}}
\subcaptionbox{$v=0.5$}{\includegraphics[width=0.3\textwidth]{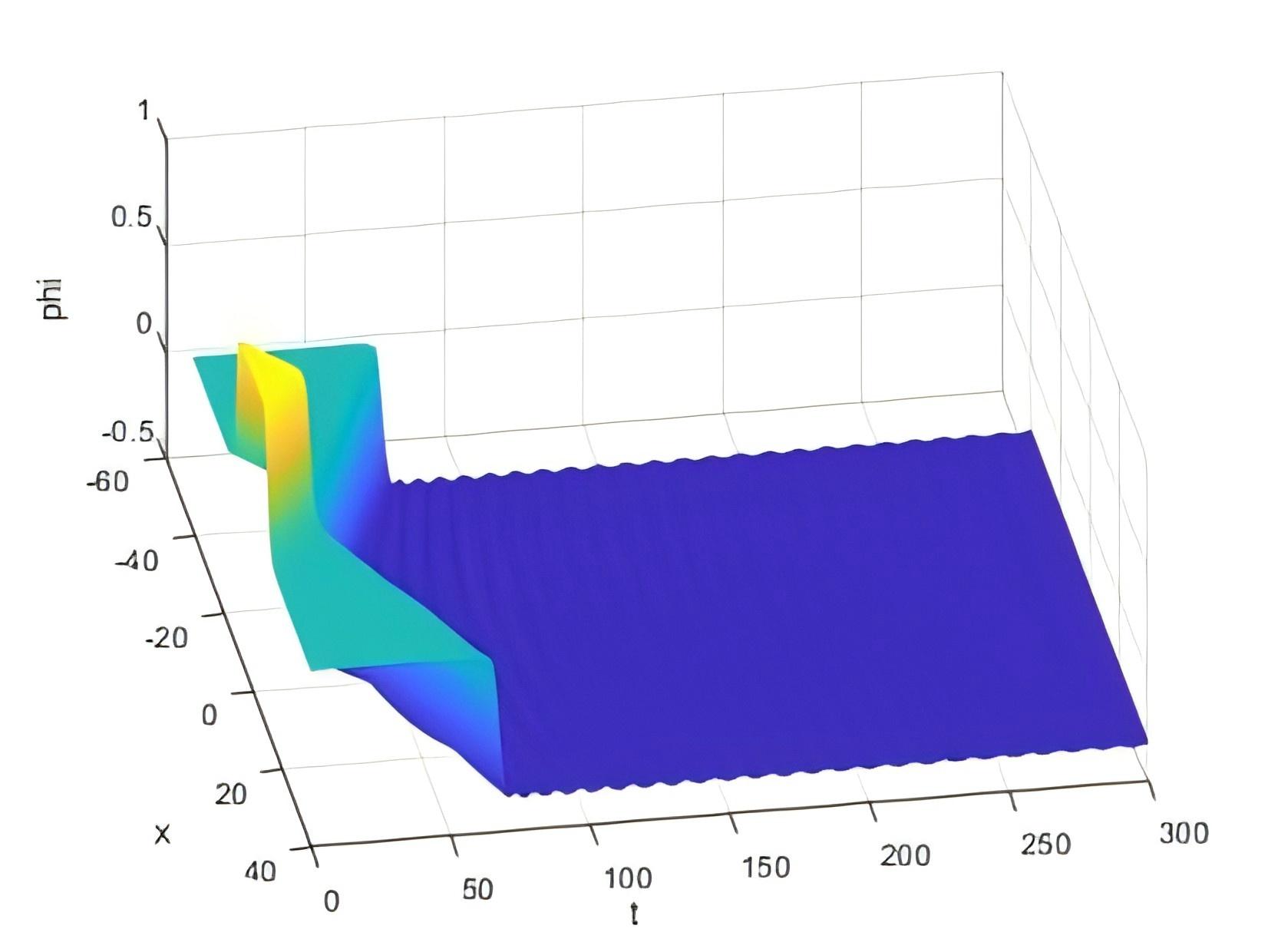}}
\subcaptionbox{$v=0.99$}{\includegraphics[width=0.3\textwidth]{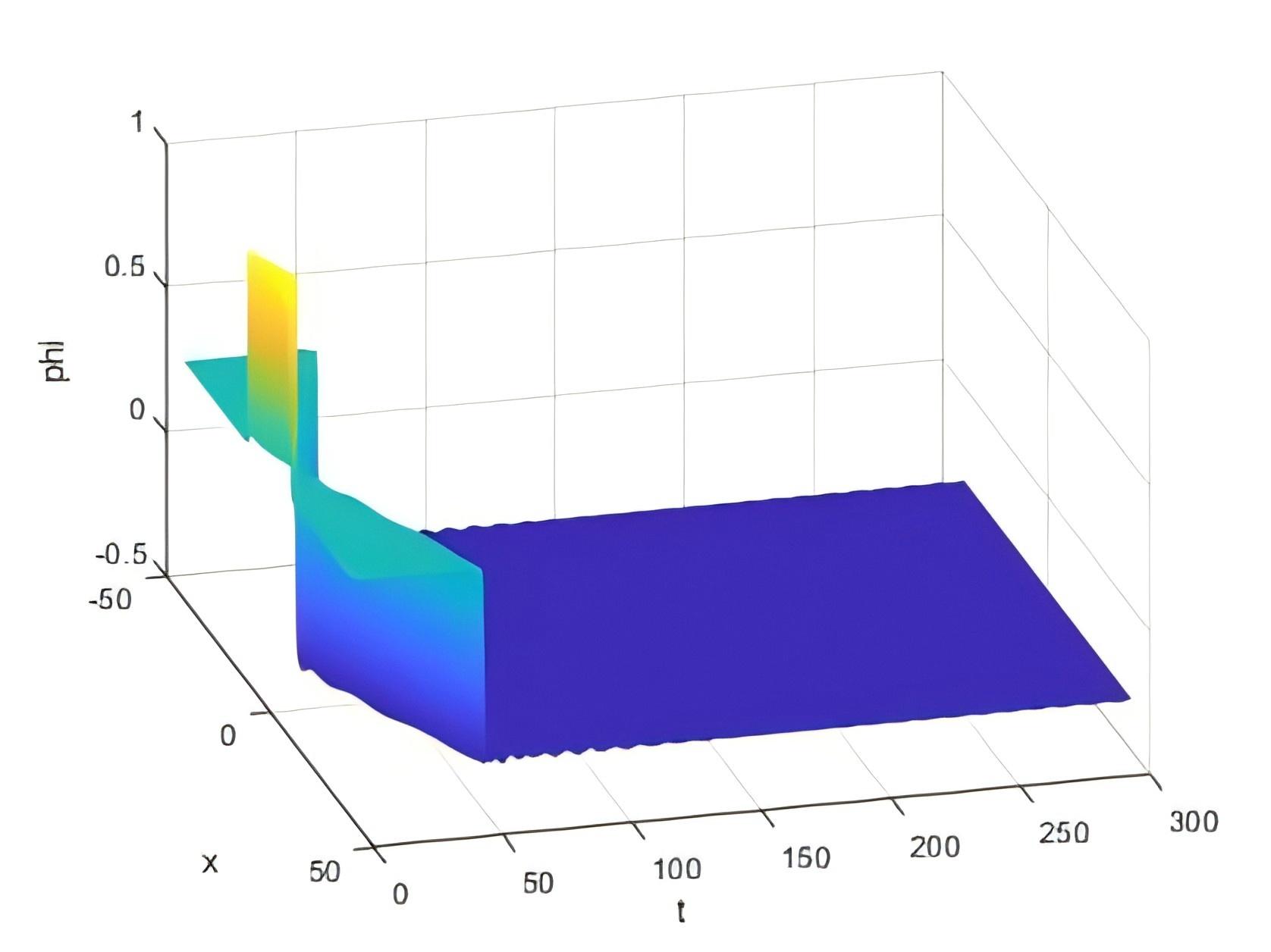}}
\caption{3D plot of $\phi_{K\bar{K}}$ with the initial velocities $v=0.05$, $v=0.5$ and $v=0.99$, respectively.}
\label{Fig:15}
\end{figure}

\begin{figure} [h]
\centering
\subcaptionbox{$v=0.05$}{\includegraphics[width=0.3\textwidth]{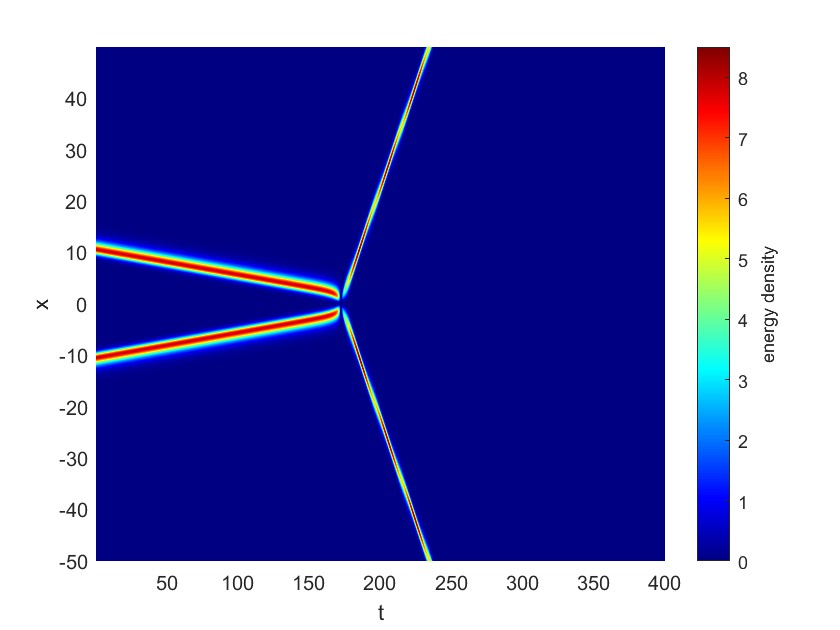}}
\subcaptionbox{$v=0.5$}{\includegraphics[width=0.3\textwidth]{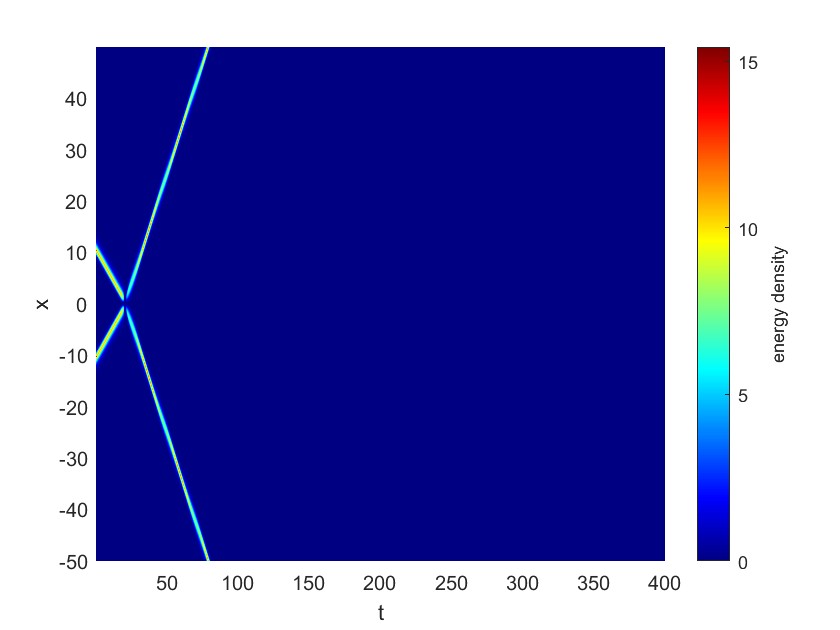}}
\subcaptionbox{$v=0.99$}{\includegraphics[width=0.3\textwidth]{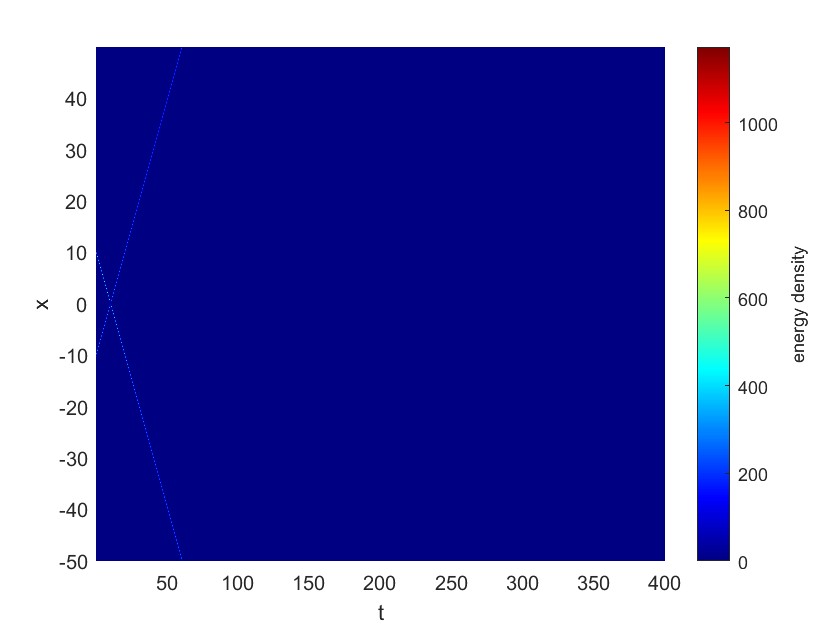}}
\caption{energy density plots of $\phi_{K\bar{K}}$ with the initial velocities $v=0.05$, $v=0.5$ and $v=0.99$, respectively.}
\label{Fig:B}
\end{figure}

Figure \ref{Fig:14} demonstrates the time evolution of $\phi_{K\bar{K}}$ at $x=0$ for the sector $(1/3, 1)$. The corresponding  3D plots are presented in Figure \ref{Fig:15}. Figure \ref{Fig:14} shows that in all three cases, the kink-antikink pair neither reflects back nor forms a bion state after the first collision. Instead, it forms a new soliton pair and changes the topological sector to $(-1/3,1/3)$. Then, the soliton pair escapes to infinity, and the vibration at vacuum $1/3$ remains at all spatial grids. This behavior is unique for the full  range of initial velocities within this topological sector. To understand this phenomenon, we also presented its energy density plot in Figure \ref{Fig:B}. Such topological sector changing is known in the sine-Gordon model \cite{ref23,ref24}, as well as in $\phi^8$ model \cite{ref16}. 

\begin{figure} [h]
\centering
\subcaptionbox{2D evolution}{\includegraphics[width=0.45\textwidth]{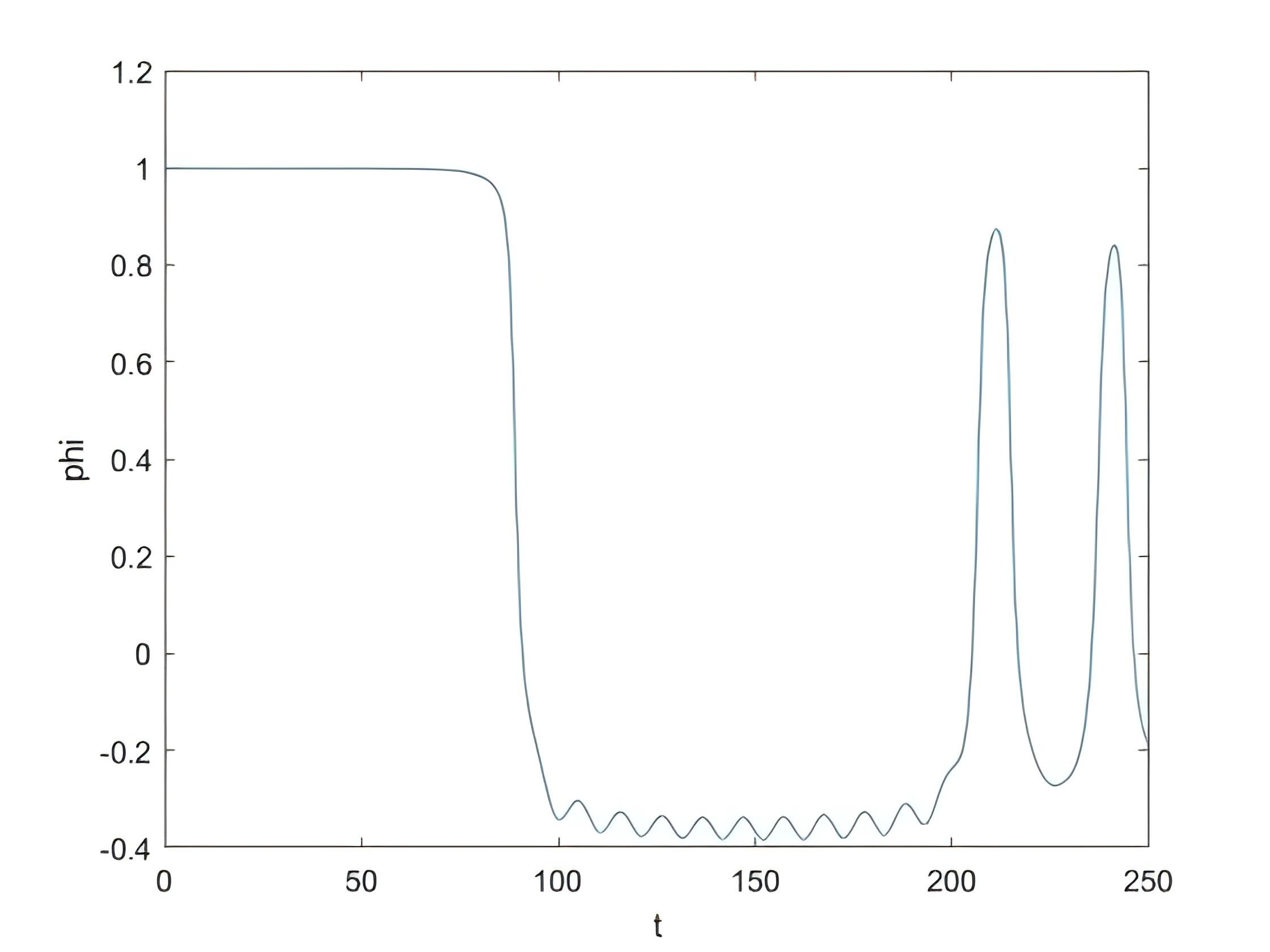}}
\subcaptionbox{3D evolution}{\includegraphics[width=0.45\textwidth]{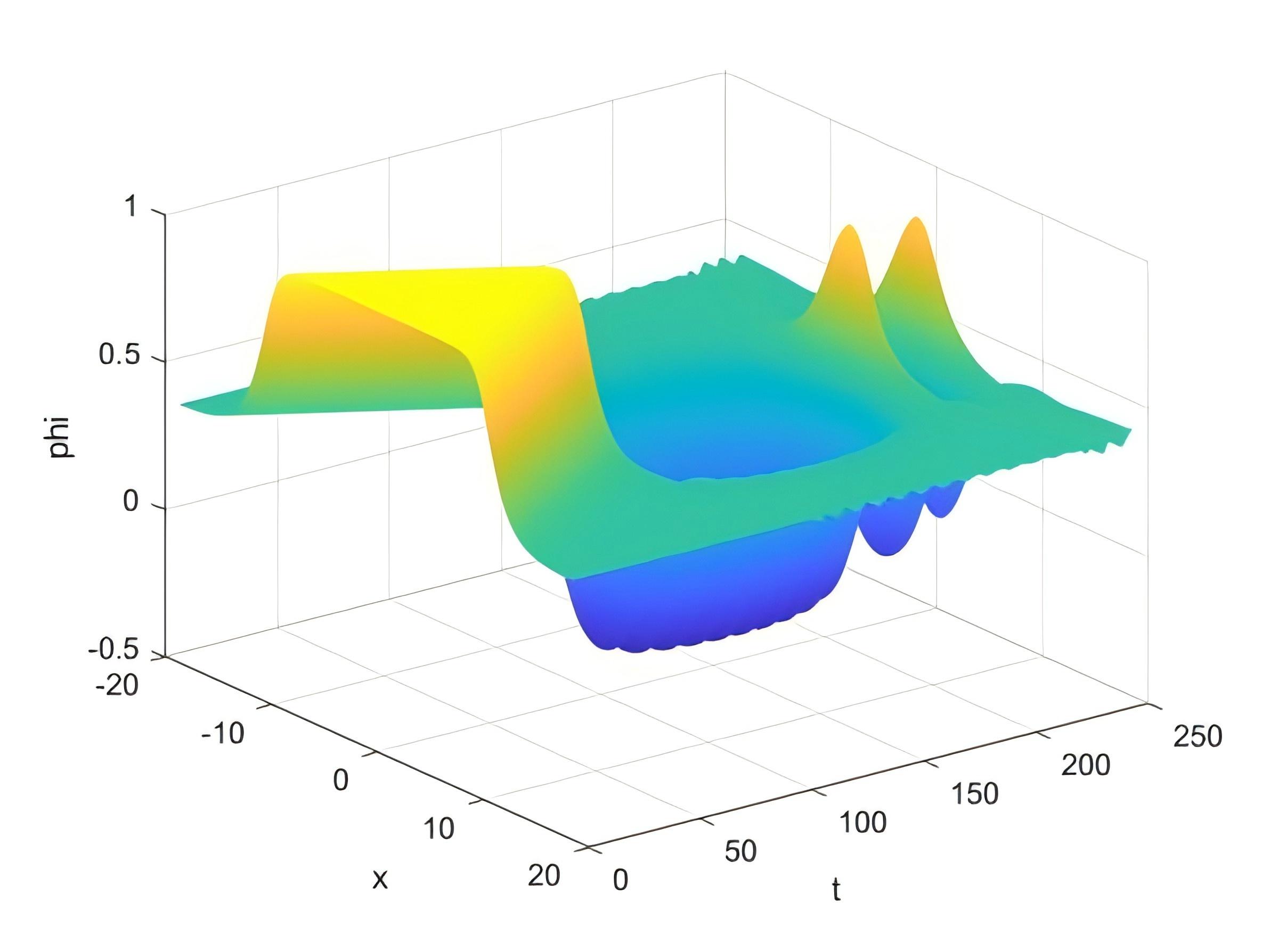}}
\caption{2D and 3D plots of kink-antikink collision in the topological sector $(0.38005, 1)$ at $v=0.1$}
\label{Fig:16}
\end{figure}

\paragraph{(3) Other $n$ values}\mbox{}\\
The  boundary conditions employed herein are derived in section \ref{Sec:bondary} by using the mathematical approximations. The numerical calculation considers methods such as the Newton-Raphson and pseudospectral methods. By combining these initial configurations with the fourth-order Runge-Kutta (RK4) method, the time evolution of the field with other $n$ values can be obtained. One specific case  with topological sector $(0.38005, 1)$ at $v=0.1$ is presented in Figure \ref{Fig:16}.

It was observed that the kink-antikink pair first changes the sector after the first collision,  from the outer sector (0.38005,1) to the inner sector $(-0.38005, 0.38005)$. Then, the system subsequently returns back to the outer sector (0.38005,1), and forms a bion state. This process, up to our knowledge, is also observed for the first time for $\phi^8$ theory. 

\begin{figure} [h]
\centering
\subcaptionbox{2D evolution}{\includegraphics[width=0.45\textwidth]{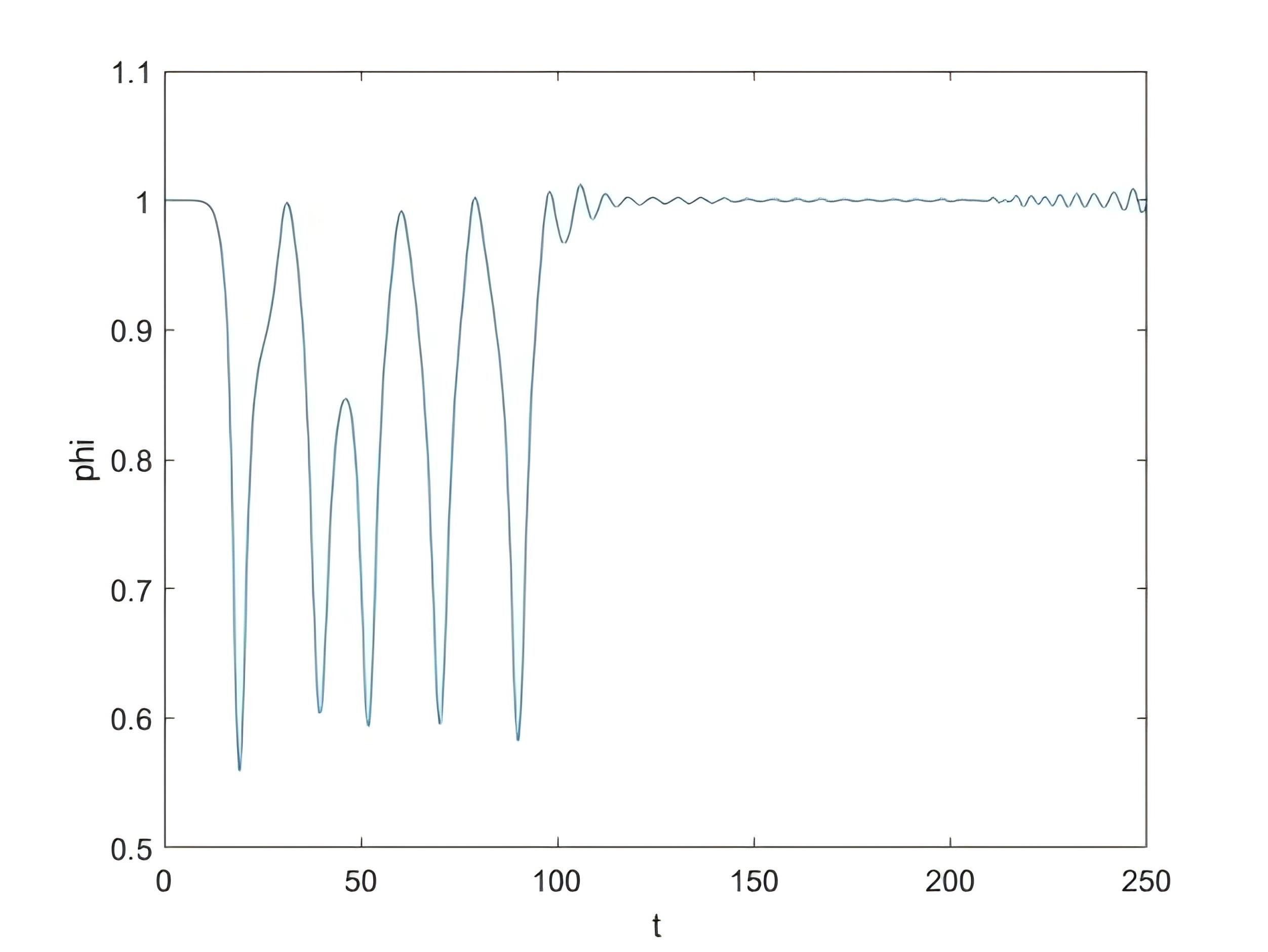}}
\subcaptionbox{3D evolution}{\includegraphics[width=0.45\textwidth]{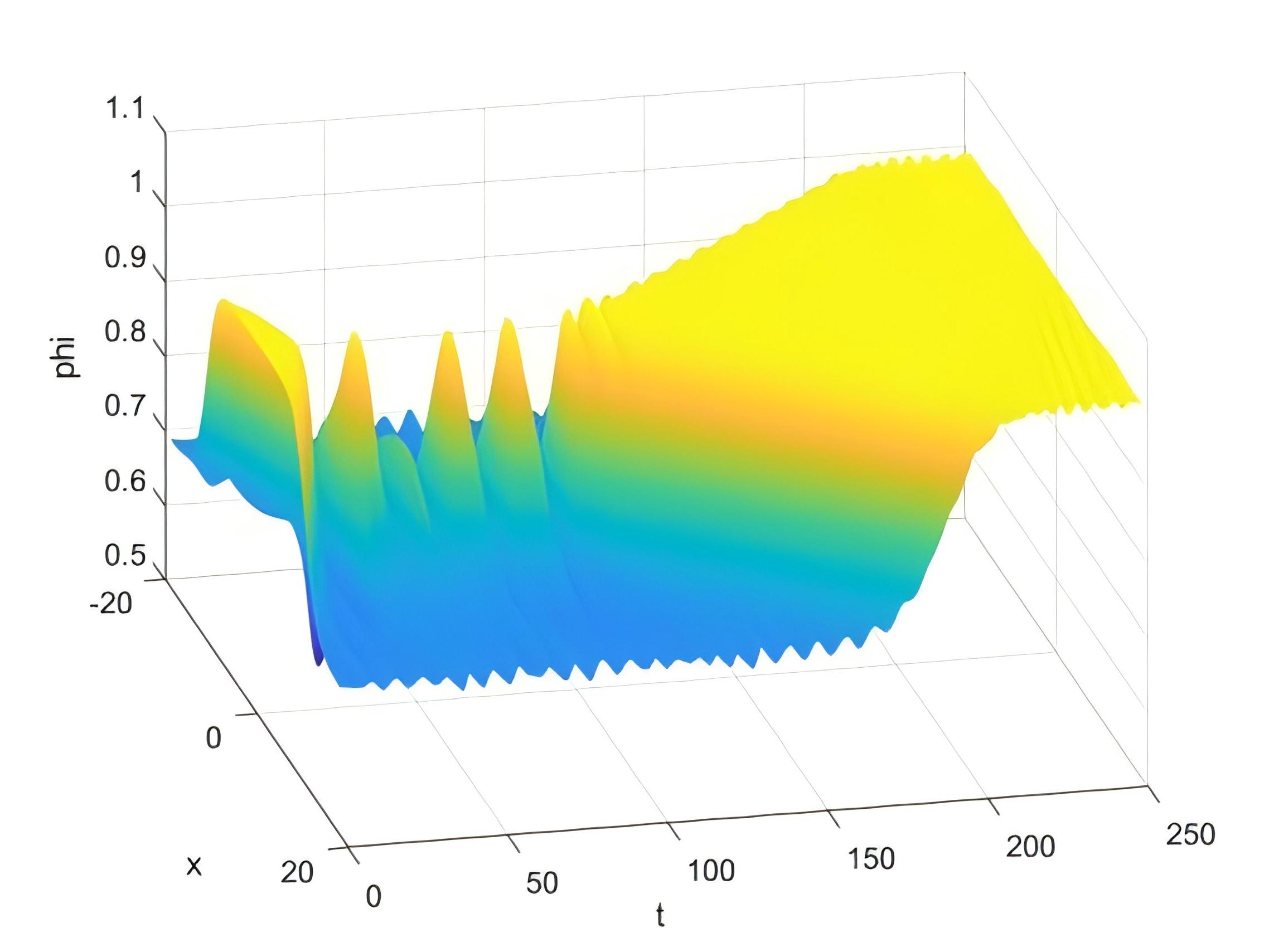}}
\caption{2D and 3D plots of kink-antikink in the Topological sector $(0.7, 1)$ at $v=0.698$}
\label{Fig:17}
\end{figure}

In Figure \ref{Fig:17}, a kink-antikink pair in the outer sector (0.7,1) with $v=0.698$ can form a bion state immediately after collision, and subsequently achieve successful escape. The post-collision bion state bounces six times. It is interesting to notice that the second bounce is a 'dwarf' one, as well as the bounces after the sixth one.  The field at $x=0$ did not return to the vacuum 1, and its duration time is smaller than other bounces. We also notice that the duration of the first bounce is slightly larger than that of bounces from the third to the sixth. At the boundary, the saw pattern indicates that some radiation leaks away. There may be more than one vibration frequency in such case. The second bounce and bounces after the seventh have shorter duration time than the first and third to sixth bounces. The energy distributes among the vibration, the radiation and the kinetic components, and some energy transits among them.

\begin{figure} [h]
\centering
\subcaptionbox{2D evolution}{\includegraphics[width=0.45\textwidth]{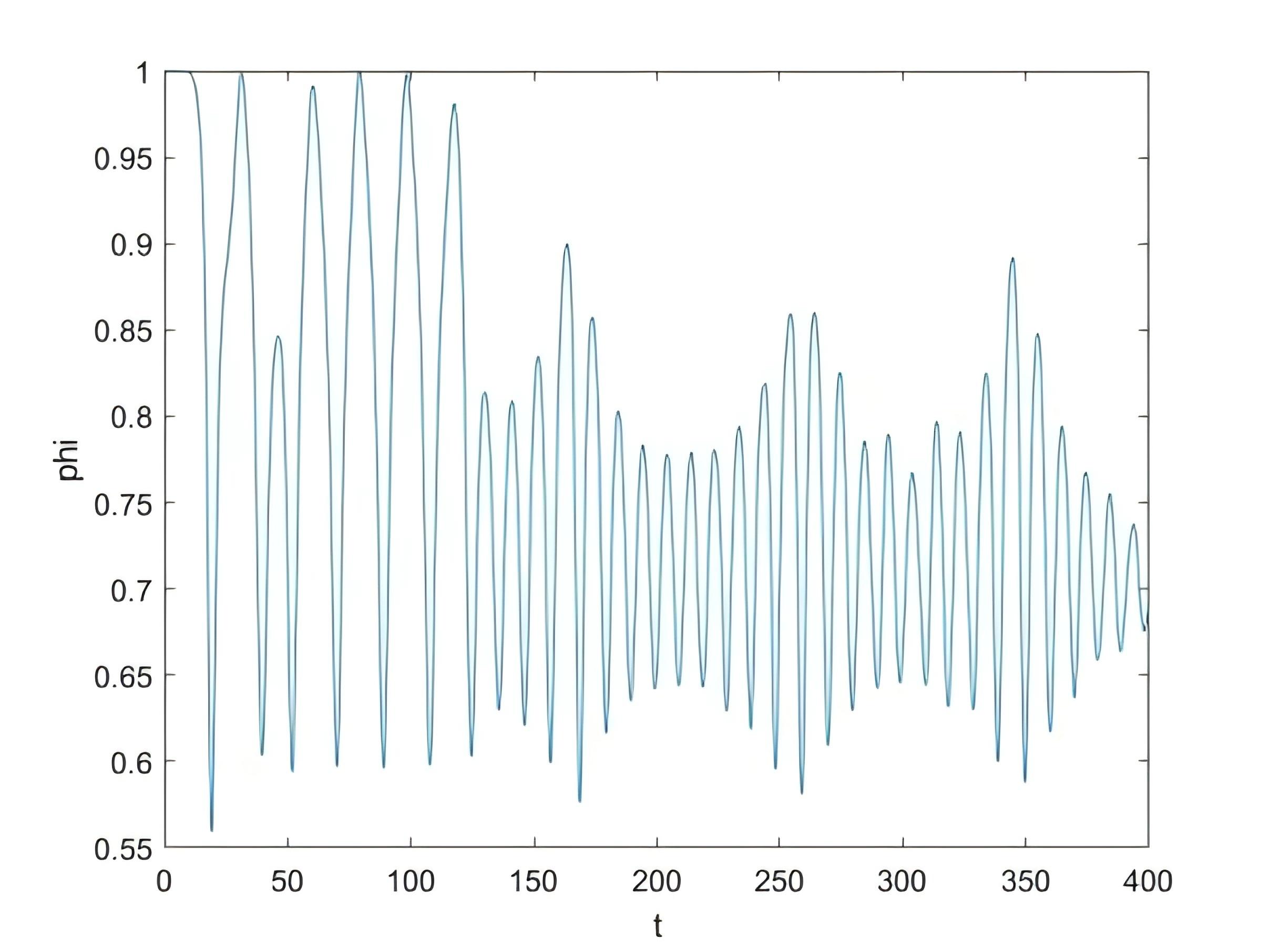}}
\subcaptionbox{3D evolution}{\includegraphics[width=0.45\textwidth]{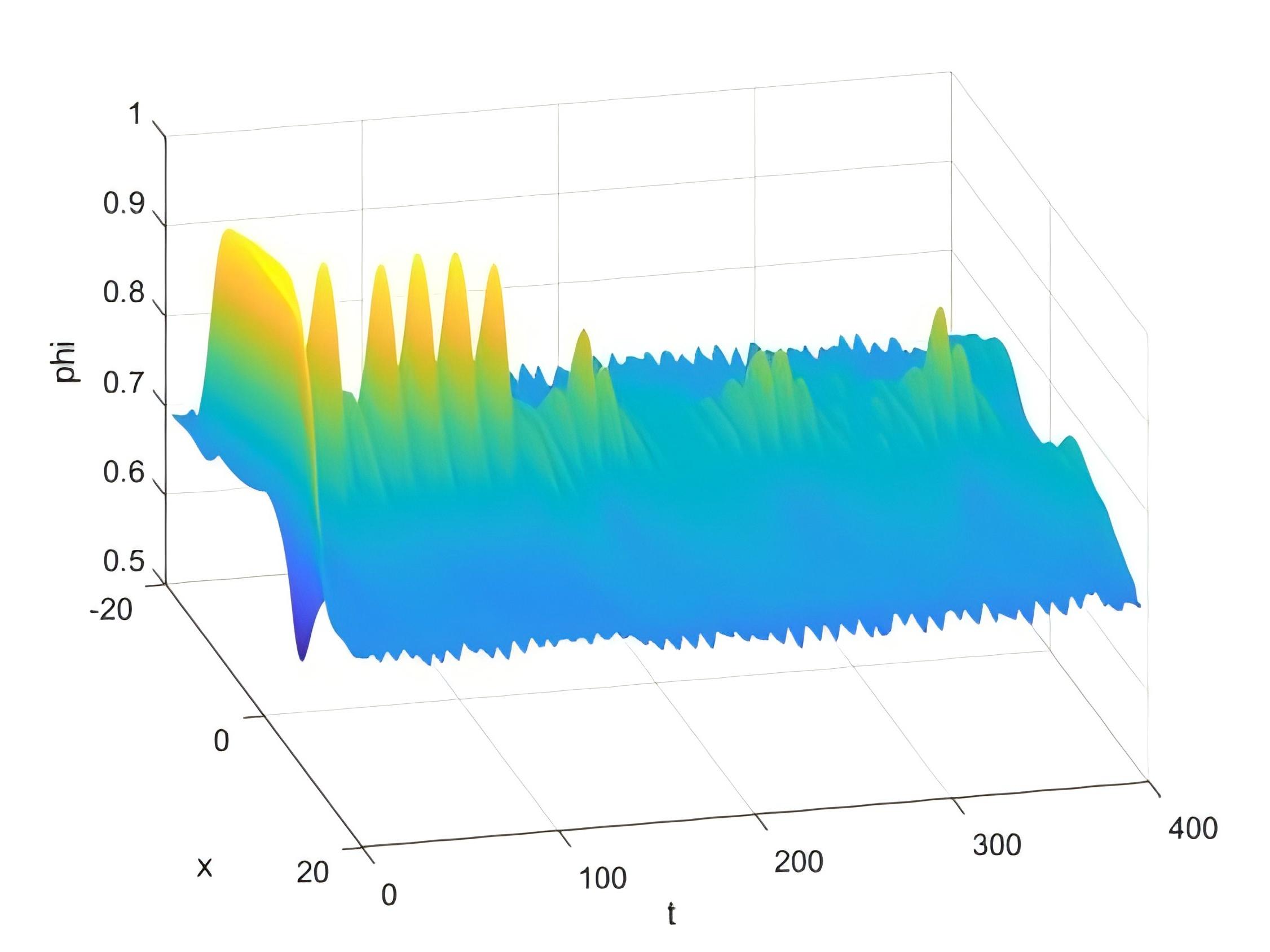}}
\caption{2D and 3D plots of kink-antikink in the Topological sector $(0.7, 1)$ at $v=0.69797$}
\label{Fig:18}
\end{figure}

A closely related dynamical process is observed at a slightly slower initial velocity  $v=0.69797$ in Figure \ref{Fig:18}. Here, the kink-antikink pair first forms a relatively high and uniform bion state after collision. Over time, this high-energy bion dissipates energy through radiation, transitioning to the 'dwarf' bion state with shorter resonance time and lower amplitude. The final state exhibits a wave packet with a profile resembling a beat frequency phenomenon.

\subsection{The rich collision channels for different topological sectors}

To show the collision channels in the whole range of $v_{in}$, we plot the $v_{in}-v_{out}$ and $v_{in}-t$ ($t$ is the outgoing time) for different topological sectors. The step size of $v_{in}$ is taken as 0.001. In numerical calculation, $v_{out}$ measures the velocity of solitons from the last bounce to the boundary. The number of bounces is counted when $\phi(x=0)$ exceeds the range of the topological sector. The setup of the numerical calculation is referred to Section \ref{sec:num}. Here potential $U=\mathrm{d^2}V/{\mathrm{d}\phi^2}$. Following \cite{ref10}, we  could also obtain the Schr\"odinger-like equation 
\begin{equation}
-\eta^{\prime\prime}+U(x)\eta=\omega^2 \eta.
\end{equation}
The numerical solution for this equation is also considered. 

\paragraph{(1) $n = p_2/p_1 = 2$}\mbox{}\\
\subparagraph{(i) Topological sector $(-1, -1/2)$}\mbox{}\\
\begin{figure}[htbp]
  \centering
  \begin{minipage}[b]{0.48\textwidth}
    \centering
    \includegraphics[width=\linewidth]{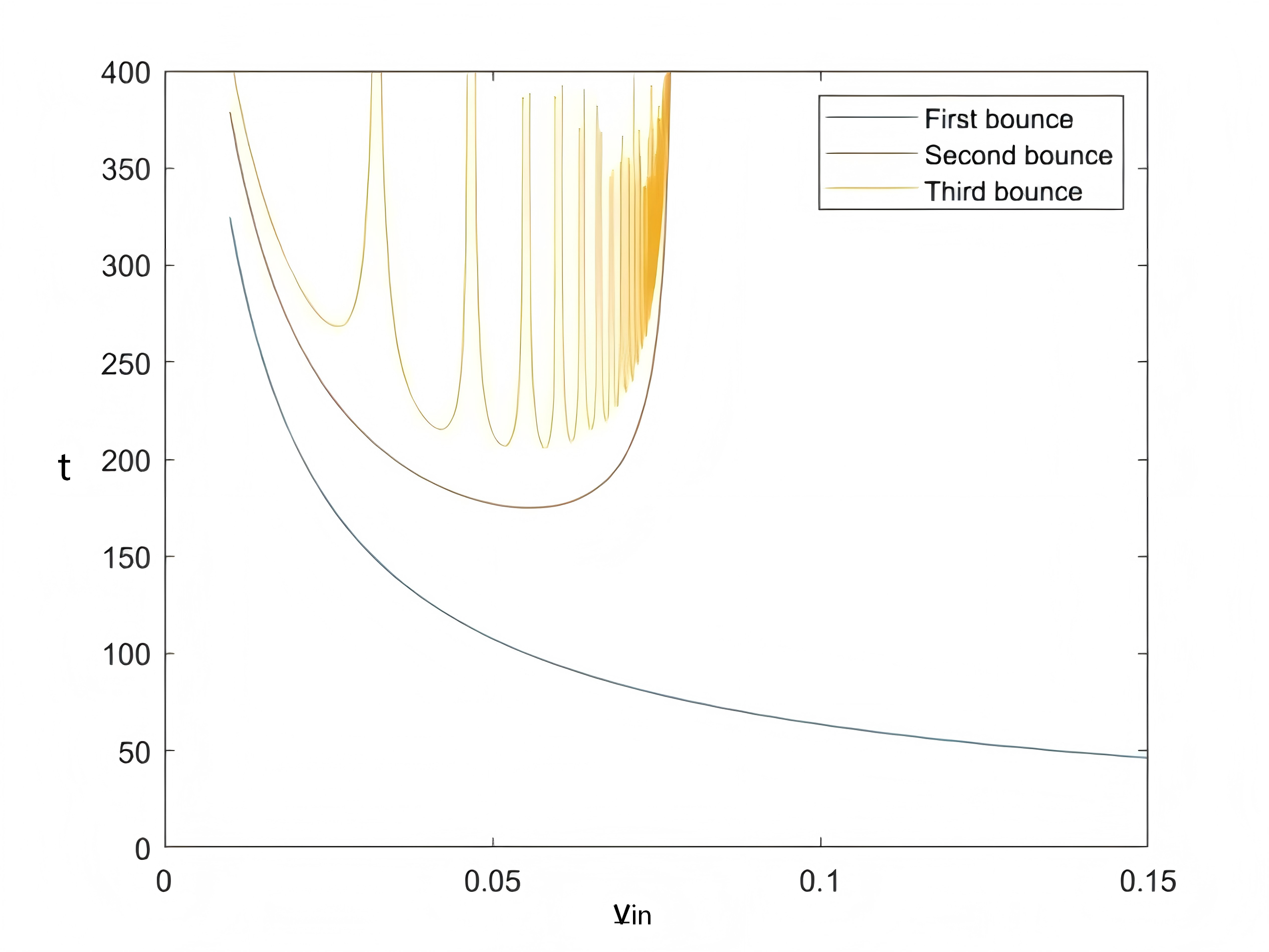}
    \subcaption{$v_{in}-t$}
    \label{Fig:22-a}
  \end{minipage}
  \hfill
  \begin{minipage}[b]{0.48\textwidth}
    \centering
    \includegraphics[width=\linewidth]{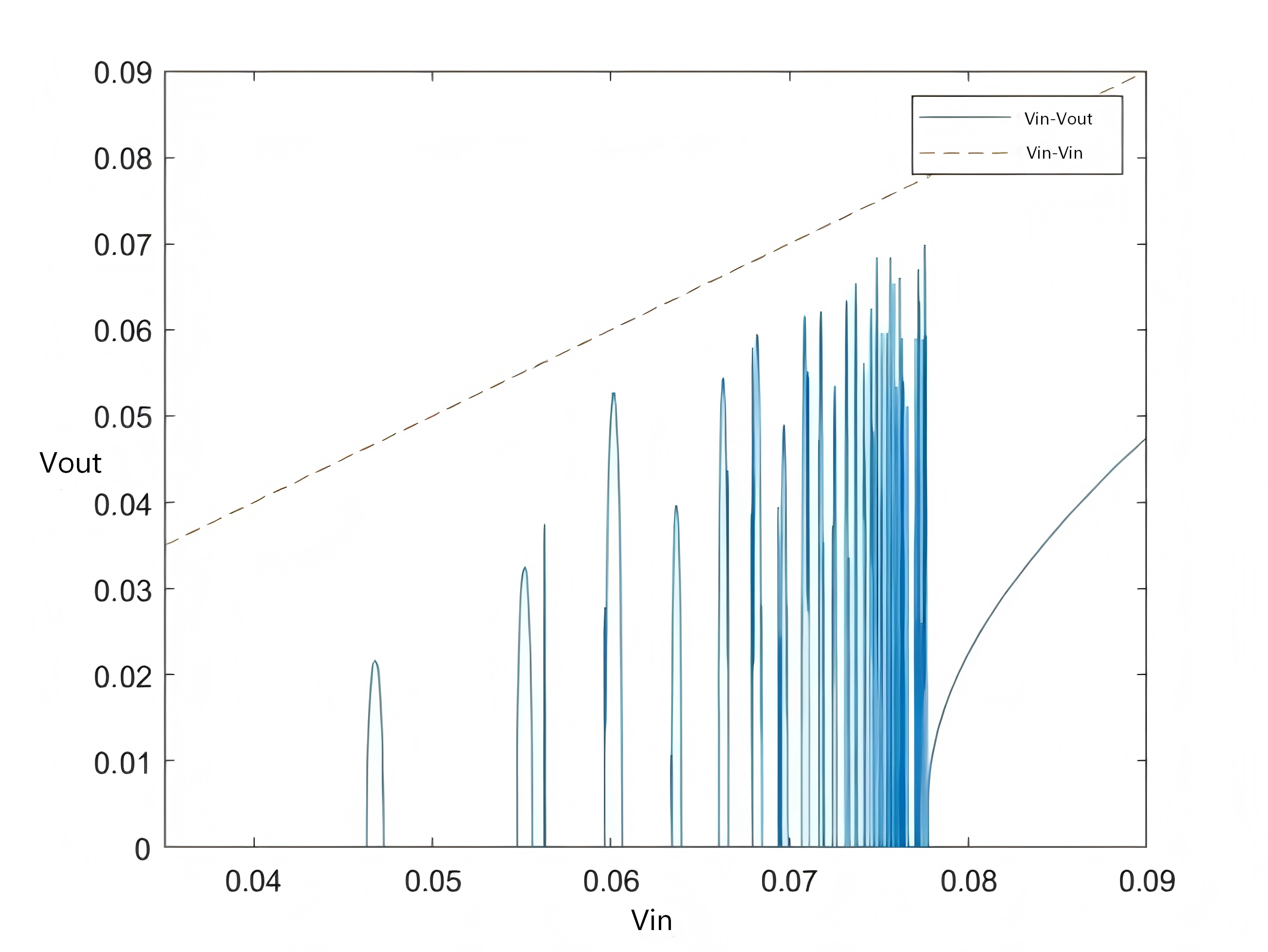}
    \subcaption{$v_{{in}} - v_{{out}}$ }
    \label{Fig:22-b}
  \end{minipage}

  \vspace{2mm} 
  \begin{minipage}[b]{0.48\textwidth}
    \centering
    \includegraphics[width=\linewidth]{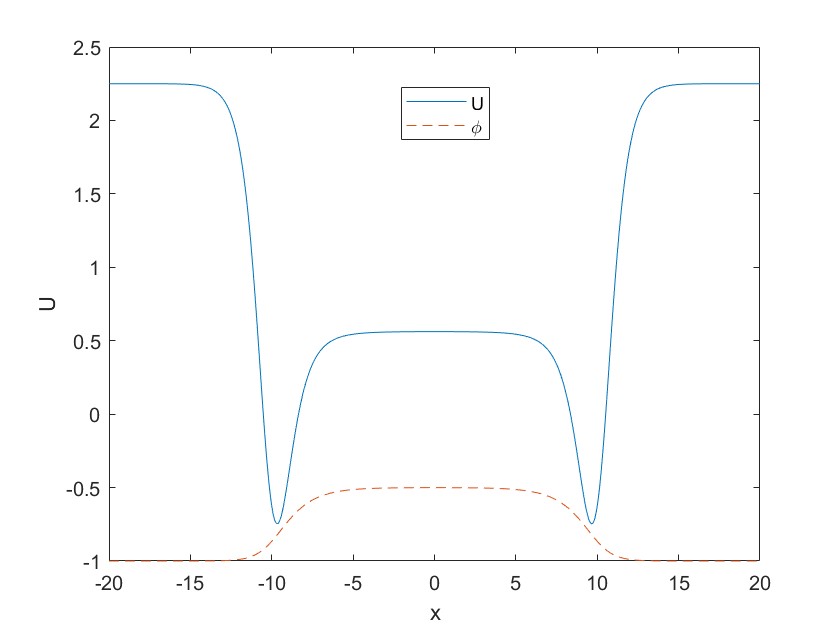}
    \subcaption{$K\bar{K}$ pair potential U(x) }
    \label{Fig:22-c}
  \end{minipage}
  \hfill
  \begin{minipage}[b]{0.48\textwidth}
    \centering
    \includegraphics[width=\linewidth]{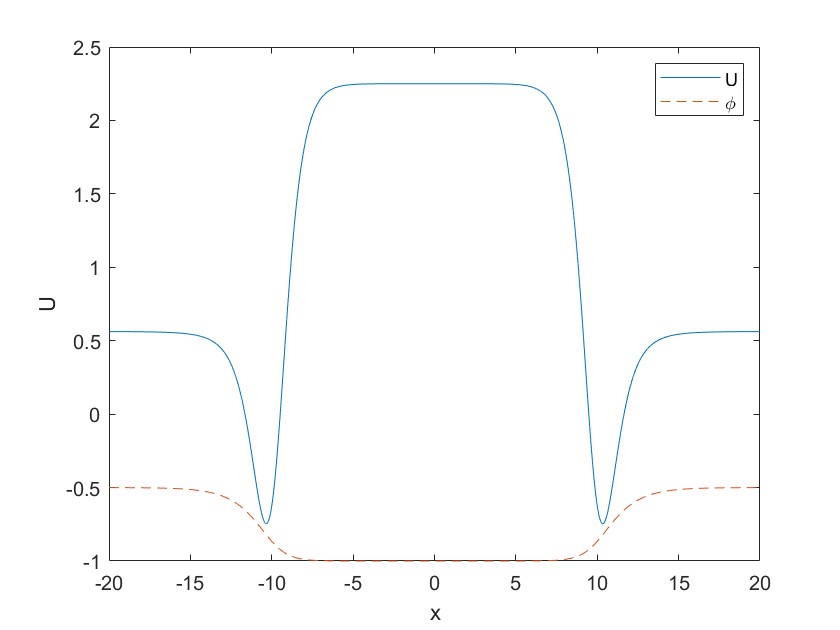}
    \subcaption{$\bar{K}K$ pair potential U(x)}
    \label{Fig:22-d}
  \end{minipage}
 \vspace{2mm} 
 \begin{minipage}[b]{0.48\textwidth}
    \centering
    \includegraphics[width=\linewidth]{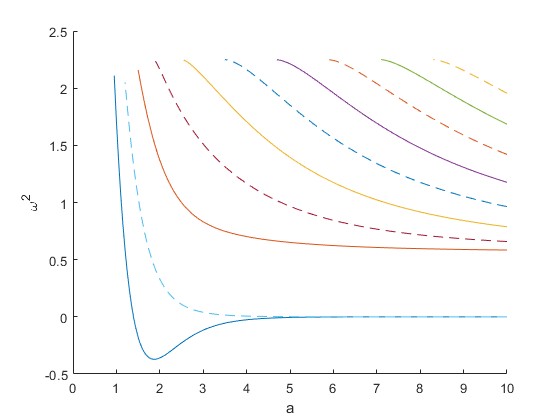}
    \subcaption{$K\bar{K}$ pair spectrum}
    \label{Fig:22-e}
  \end{minipage}
 \hfill
   \begin{minipage}[b]{0.48\textwidth}
    \centering
    \includegraphics[width=\linewidth]{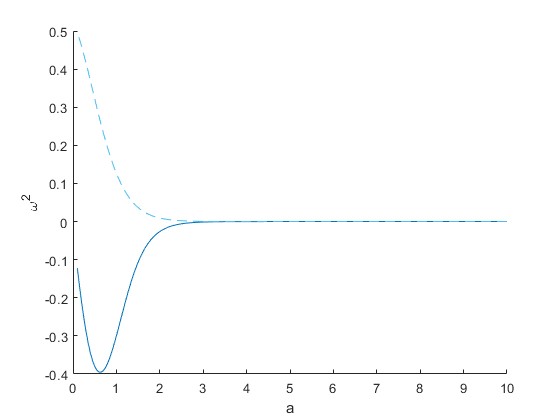}
    \subcaption{$\bar{K}K$ pair spectrum}
    \label{Fig:22-f}
  \end{minipage}

  \caption{Topological sector $(-1, -1/2)$. In panel (e), the solid and dash lines represent the even and odd bound states for $K\bar{K}$ pair. }
  \label{Fig:22}
\end{figure}
\begin{figure} [htbp]
\centering
\subcaptionbox{$v_{in} \in[0.055,0.062]$}{\includegraphics[width=0.4\textwidth]{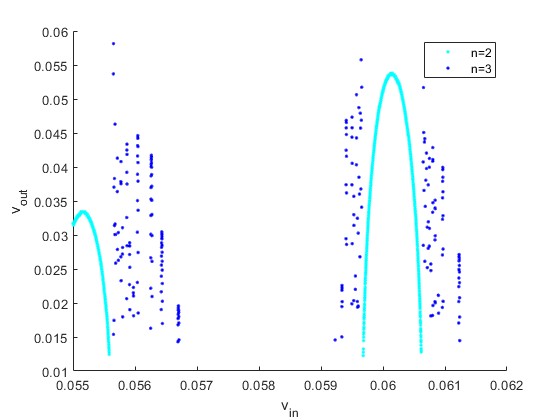}}
\subcaptionbox{$v_{in} \in[0.060695,0.060702]$}{\includegraphics[width=0.4\textwidth]{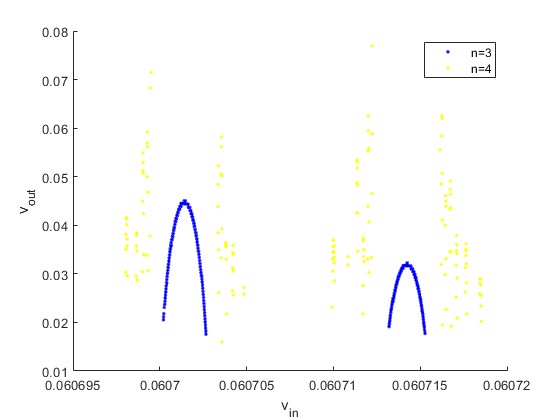}}
\caption{Fractal structure for topological sector $(-1, -1/2)$}
\label{Fig:22a}
\end{figure}
Figure \ref{Fig:22} reveals that, within this topological sector, the collision  exhibits intricate escape windows up to the critical velocity $v_c \approx 0.0777$. Beyond the critical velocity, the soliton pair escapes. In Figure \ref{Fig:22a}, the left panel shows the $v_{in} - v_{out}$ plot for the two and three bounces, and the right panel shows that for the three and four bounces. The $v_{in}$ in the right panel is a zoomed regime in the left panel. This shows the self-similar fractal structure for the soliton collisions, known in $\phi^4$ and $\phi^6$ theories \cite{ref25,ref26,ref27,ref28,ref29}.

The rich collision behaviors are very similar to the collision channels in $\phi^{6}$ theory \cite{ref10}. It was indicated that, even though there are no shape modes in kinks of $\phi^{6}$ theory, the ordered antikink and kink configuration could form a central potential well,  which induces de-localized mode responsible for the rich collision channels \cite{ref10,ref11}. 
We also plot the potential $U$ for both the $K\bar{K}$ pair and the $\bar{K}K$ pair in Figure \ref{Fig:22-c} and Figure  \ref{Fig:22-d}, respectively.
It reveals that $K\bar{K}$ pair has a central well and $\bar{K}K$ pair has a raised central plateau. The frequency $\omega^2$ as a function of half distance of soliton pair $a$ is also plotted in panel (e) and (f). From which, we could also observe that the $K\bar{K}$ pair has rich bound states, and the $\bar{K}K$ pair has no bound states. 


\subparagraph{(ii) Topological sector $(-1/2, 1/2)$}\mbox{}\\
\begin{figure}[htbp]
  \centering
\subcaptionbox{$K\bar{K}$ potential U(x)}{\includegraphics[width=0.48\textwidth]{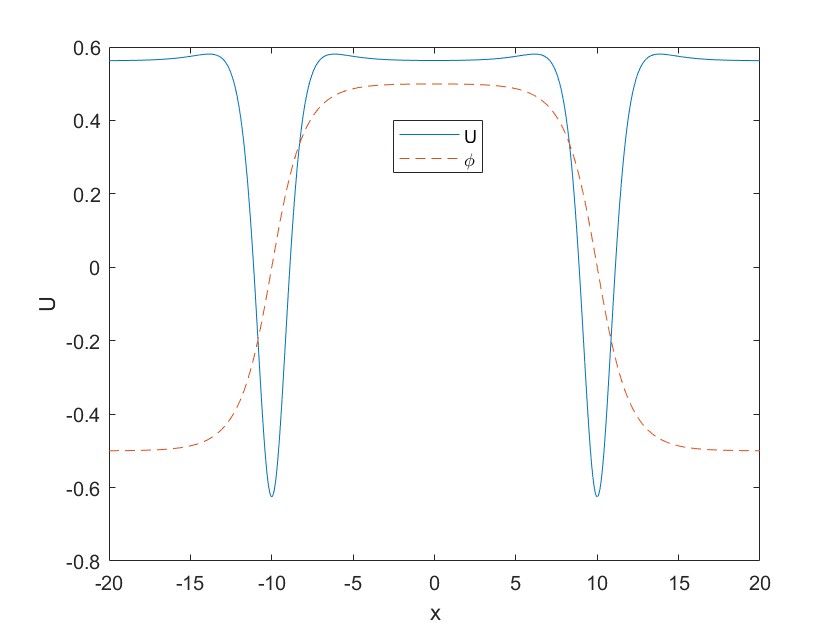}}
\subcaptionbox{$\bar{K}K$ potential U(x)}{\includegraphics[width=0.48\textwidth]{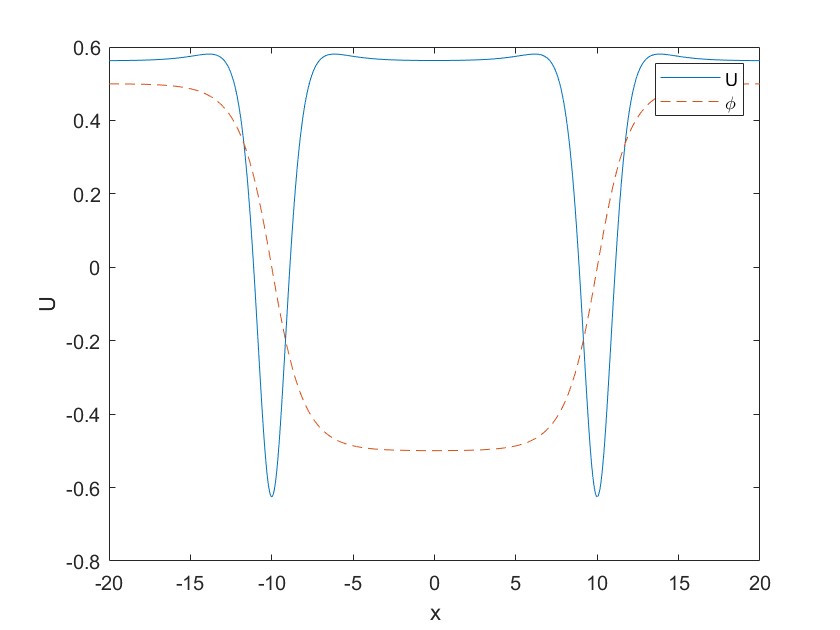}}
  \begin{minipage}[b]{0.48\textwidth}
    \centering
    \includegraphics[width=\linewidth]{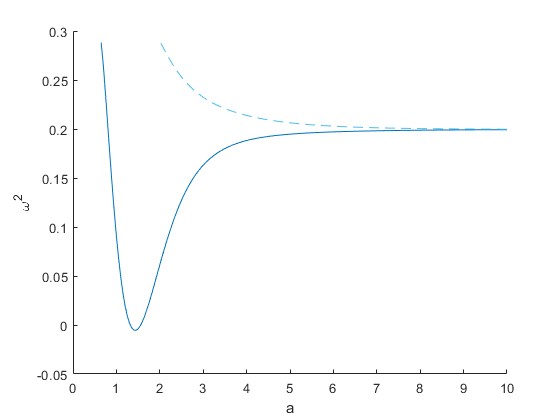}
    \subcaption{$K\bar{K}$ pair spectrum}
    \label{Fig:23-a}
  \end{minipage}
  \hfill
  \begin{minipage}[b]{0.48\textwidth}
    \centering
    \includegraphics[width=\linewidth]{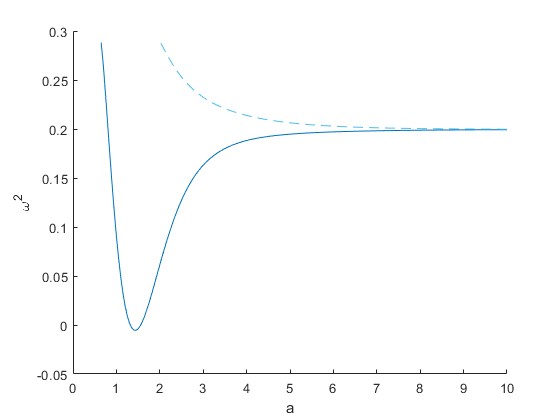}
    \subcaption{$\bar{K}K$ pair spectrum}
    \label{Fig:23-b}
  \end{minipage}
  \caption{Topological sector $(-1/2, 1/2)$}
  \label{Fig:23}
\end{figure}

There is no escape window in this topological sector, which has been indicated in Figure \ref{Fig:6} and Figure \ref{Fig:7} in Section \ref{sec:m1212}. Within this sector, the kink-antikink pair undergoes annihilation after the initial collision with any given incident velocity. Thus, there is no fractal structure. Figure \ref{Fig:23} reveals that the potential $U(x)$ of $K\bar{K}$ and $\bar{K}K$  pairs shows separated double-well, and the central plateau between wells is slightly concave. The annihilation behavior may be induced by this symmetrical potential. The spectrum shows no bound state.

\subparagraph{(iii) Topological sector $(1/2, 1)$}\mbox{}\\

\begin{figure}[htbp]
  \centering
  \begin{minipage}[b]{0.48\textwidth}
    \centering
    \includegraphics[width=\linewidth]{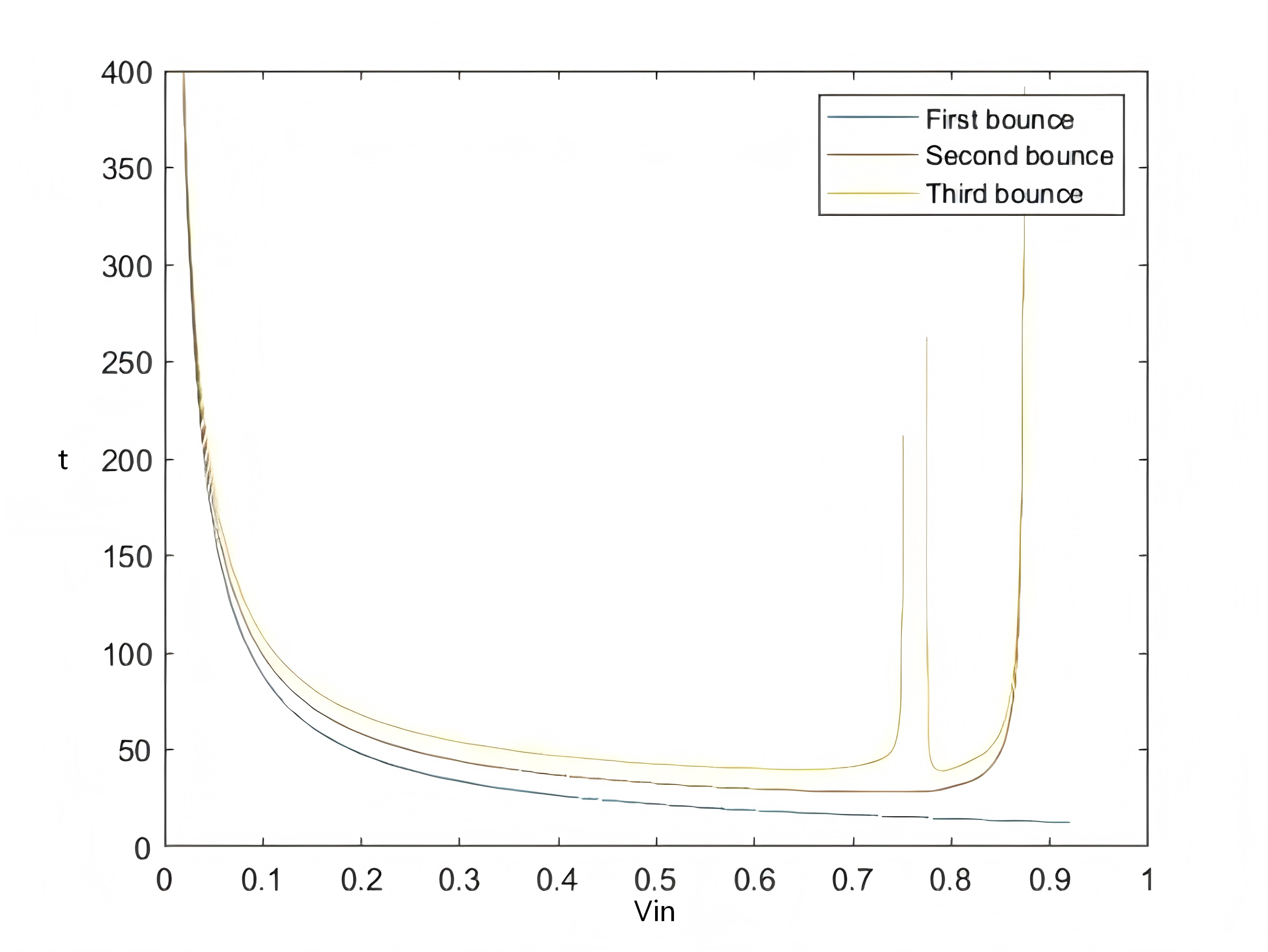}
    \subcaption{$v_{in}-t$ plot}
    \label{Fig:24-a}
  \end{minipage}
  \hfill
  \begin{minipage}[b]{0.48\textwidth}
    \centering
    \includegraphics[width=\linewidth]{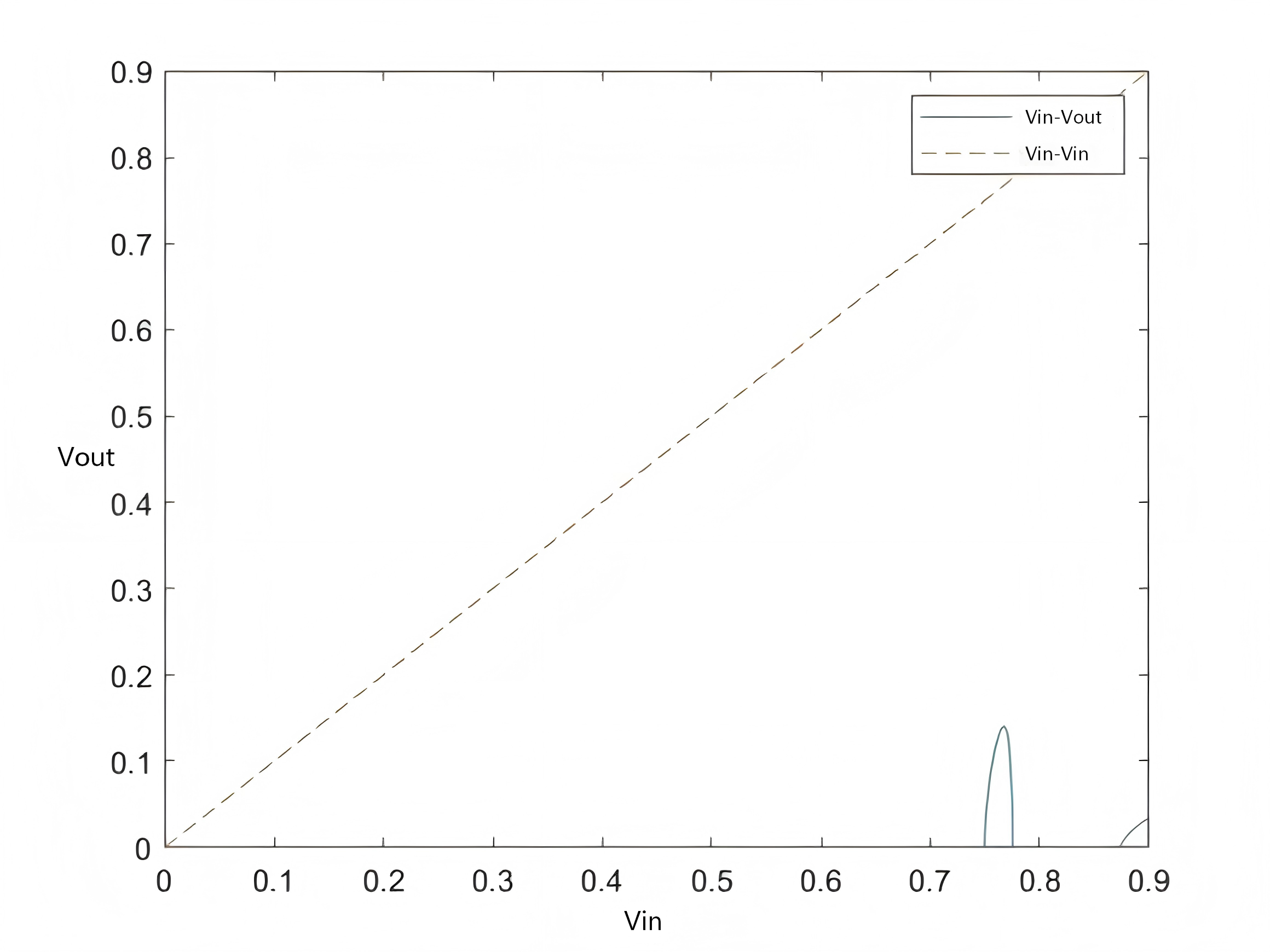}
    \subcaption{$v_{{in}} - v_{{out}}$ plot}
    \label{Fig:24-b}
  \end{minipage}

  \vspace{2mm} 
  \begin{minipage}[b]{0.48\textwidth}
    \centering
    \includegraphics[width=\linewidth]{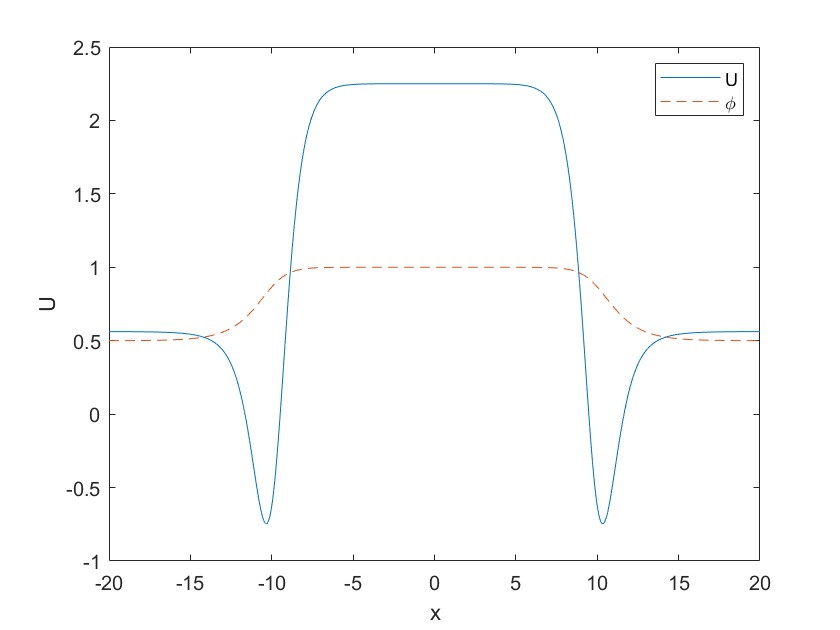}
    \subcaption{$K\bar{K}$ pair potential U(x) plot}
    \label{Fig:24-c}
  \end{minipage}
  \hfill
  \begin{minipage}[b]{0.48\textwidth}
    \centering
    \includegraphics[width=\linewidth]{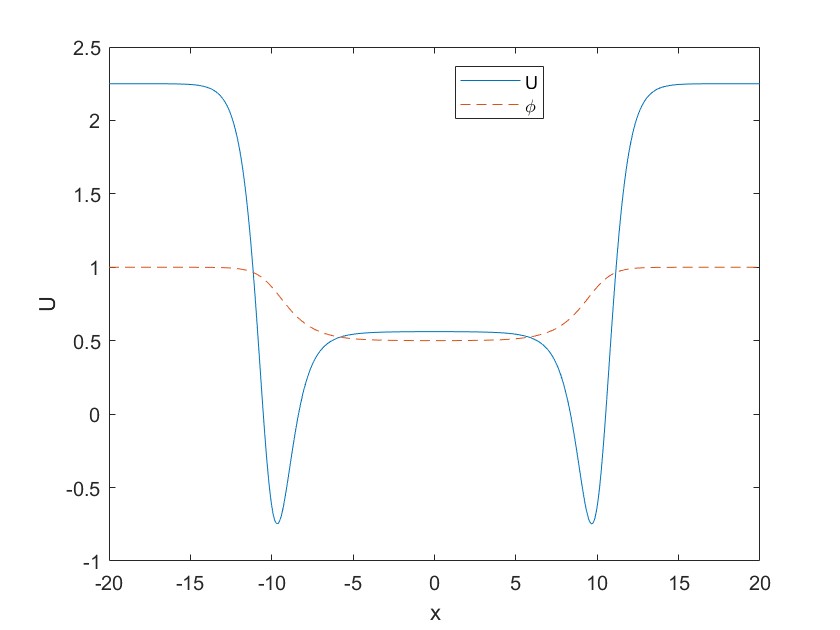}
    \subcaption{$\bar{K}K$ pair potential U(x) plot}
    \label{Fig:24-d}
  \end{minipage}

    \vspace{2mm} 
  \begin{minipage}[b]{0.48\textwidth}
    \centering
    \includegraphics[width=\linewidth]{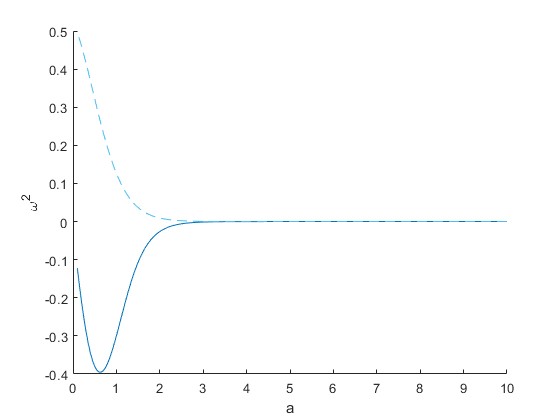}
    \subcaption{$K\bar{K}$ pair spectrum}
    \label{Fig:24-e}
  \end{minipage}
  \hfill
  \begin{minipage}[b]{0.48\textwidth}
    \centering
    \includegraphics[width=\linewidth]{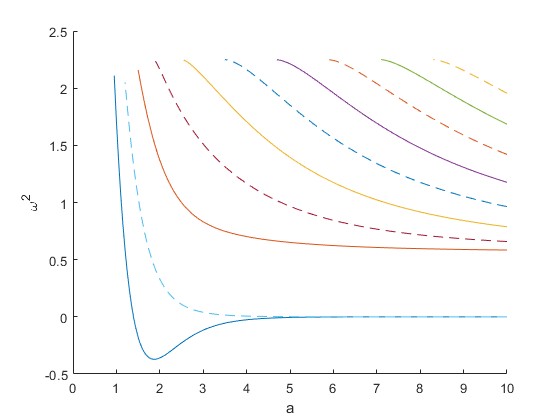}
    \subcaption{$\bar{K}K$ pair spectrum}
    \label{Fig:24-f}
  \end{minipage}

  \caption{Topological sector $(1/2, 1)$}
  \label{Fig:24}
\end{figure}

For this topological sector, Figure \ref{Fig:24}  reveals a distinct structure. Unlike the dense cluster of escape windows in the sector $(-1,-1/2)$, the escape window does not appear until the velocity $v_{in} \approx 0.87$.  In contrast with $\phi^6$ in \cite{ref10}, the pair is trapped only one time before the critical velocity. There is no fractal structure in this topological sector.
Figure \ref{Fig:24-c} reveals that $K\bar{K}$ pair has a raised central plateau. We also plot the potential $U$ for the $\bar{K}K$ pair in Figure \ref{Fig:24-d}, which shows a central well.  The collision channels for the $\bar{K}K$ pair in the sector $(1/2, 1)$ are the same as those for the $K\bar{K}$ pair in the sector $(-1, -1/2)$, which may be due to the fact that the potential $U(x)$ is the same in these two cases. The spectrum of $K\bar{K}$ shows no bound state, while the spectrum of  $\bar{K}K$ shows rich bound states.

\paragraph{(2) $n = p_2/p_1 = 3$}\mbox{}\\
\subparagraph{(i) Topological sector $(-1, -1/3)$}\mbox{}\\

\begin{figure}[htbp]
  \centering
  \begin{minipage}[b]{0.48\textwidth}
    \centering
    \includegraphics[width=\linewidth]{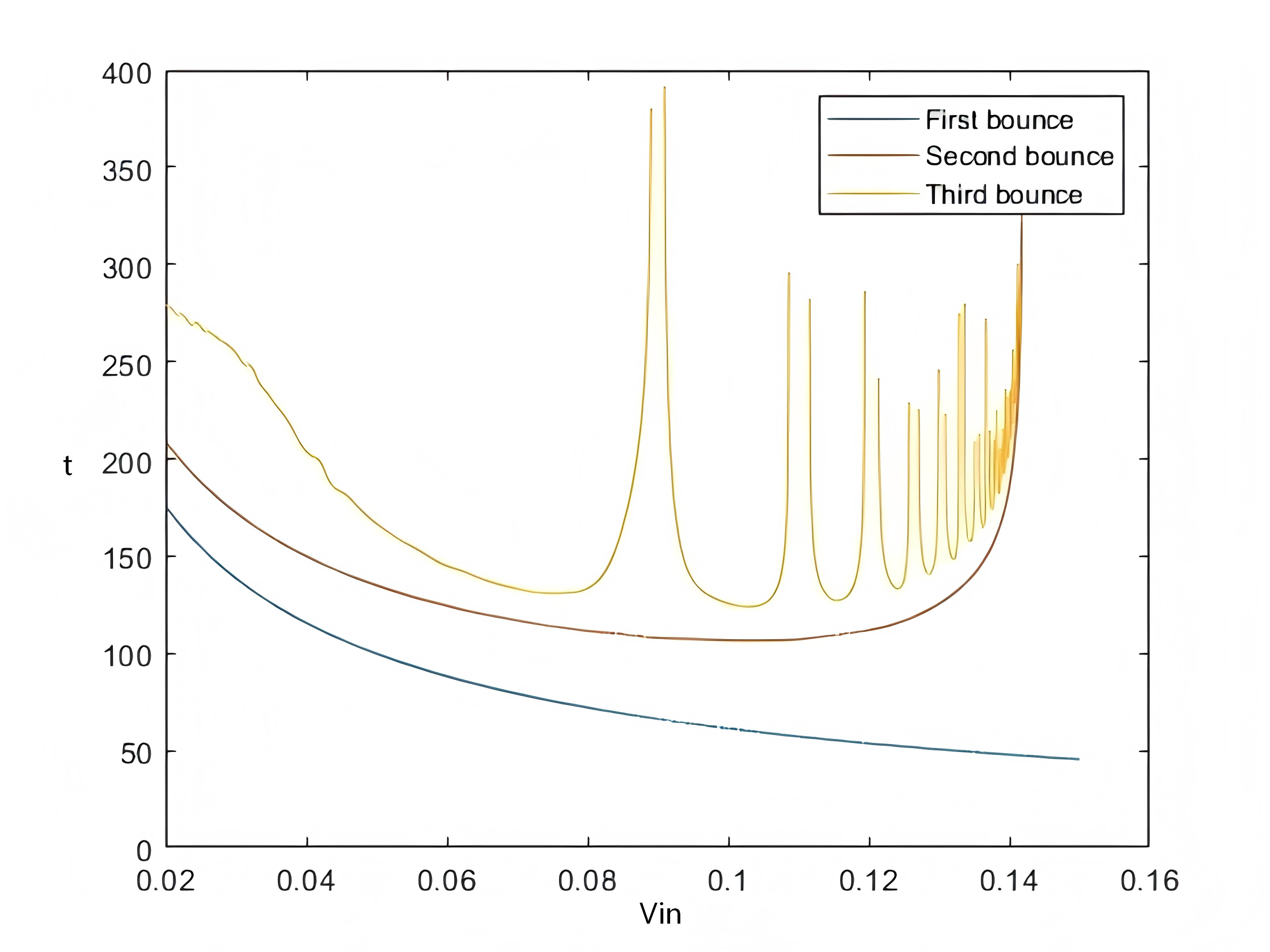}
    \subcaption{$v_{in}-t$ plot}
    \label{Fig:25-a}
  \end{minipage}
  \hfill
  \begin{minipage}[b]{0.48\textwidth}
    \centering
    \includegraphics[width=\linewidth]{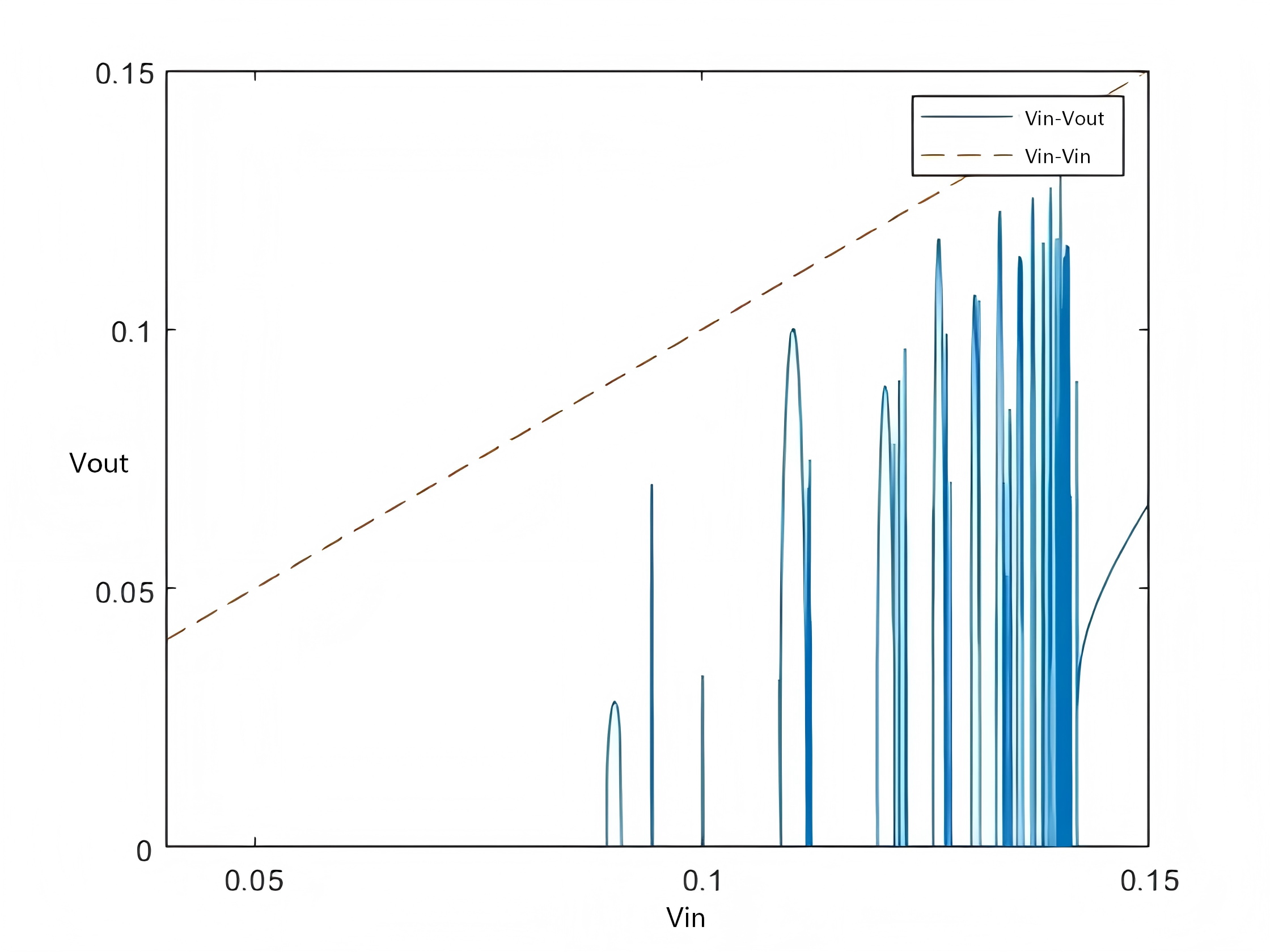}
    \subcaption{$v_{{in}} - v_{{out}}$ plot}
    \label{Fig:25-b}
  \end{minipage}

  \vspace{2mm} 
  \begin{minipage}[b]{0.48\textwidth}
    \centering
    \includegraphics[width=\linewidth]{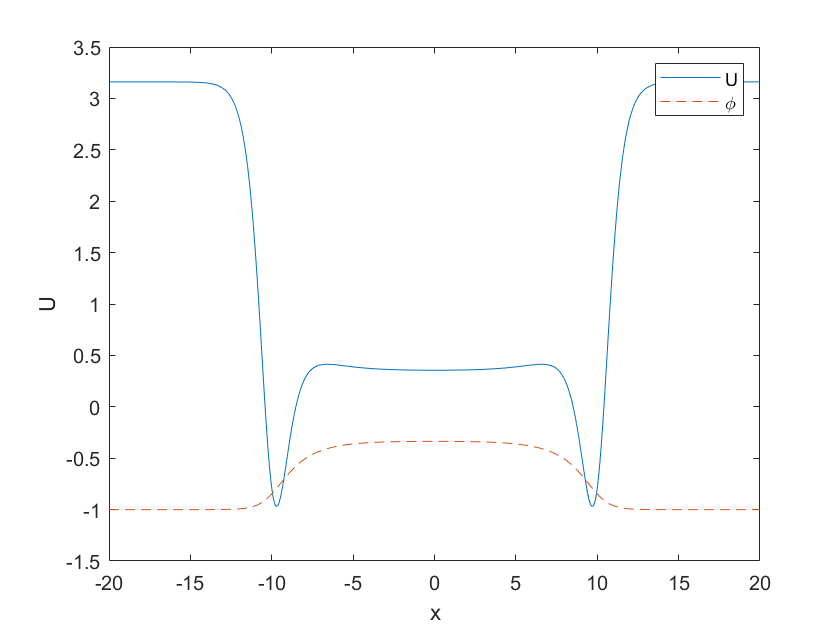}
    \subcaption{$K\bar{K}$ pair potential U(x) plot}
    \label{Fig:25-c}
  \end{minipage}
  \hfill
  \begin{minipage}[b]{0.48\textwidth}
    \centering
    \includegraphics[width=\linewidth]{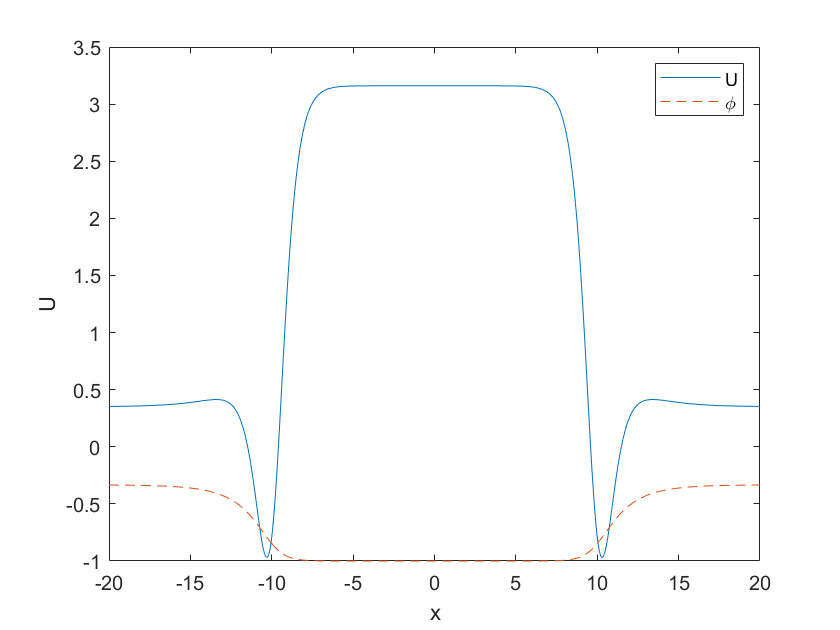}
    \subcaption{$\bar{K}K$ pair potential U(x) plot}
    \label{Fig:25-d}
  \end{minipage}
  \begin{minipage}[b]{0.48\textwidth}
    \centering
    \includegraphics[width=\linewidth]{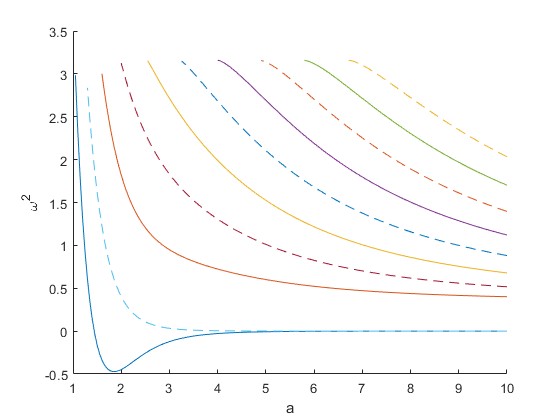}
    \subcaption{$K\bar{K}$ spectrum}
    \label{Fig:25-e}
  \end{minipage}
    \begin{minipage}[b]{0.48\textwidth}
    \centering
    \includegraphics[width=\linewidth]{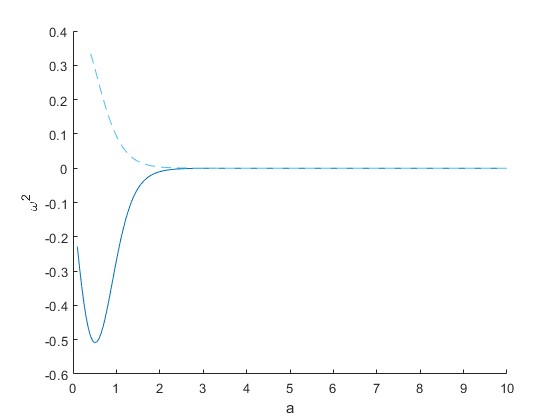}
    \subcaption{$\bar{K}K$ pair spectrum}
    \label{Fig:25-f}
  \end{minipage}
  \caption{Topological sector $(-1, -1/3)$}
  \label{Fig:25}
\end{figure}
\begin{figure} [htbp]
\centering
\subcaptionbox{$v_{in} \in[0.105,0.115]$}{\includegraphics[width=0.32\textwidth]{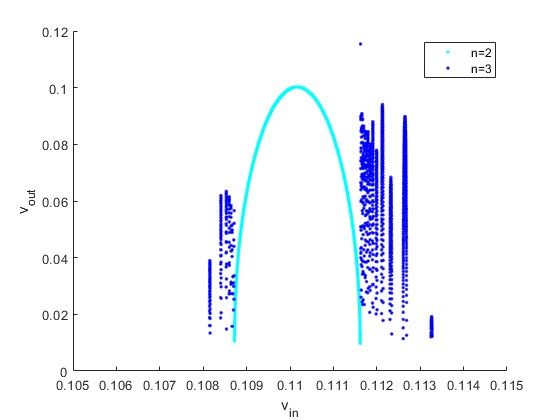}}
\subcaptionbox{$v_{in} \in[0.112,0.113]$}{\includegraphics[width=0.32\textwidth]{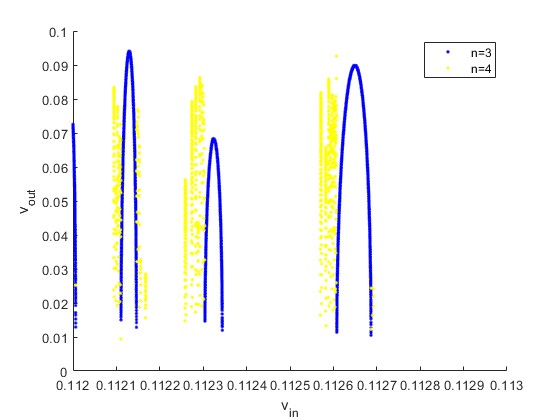}}
\subcaptionbox{$v_{in} \in[0.11255,0.1126]$}{\includegraphics[width=0.32\textwidth]{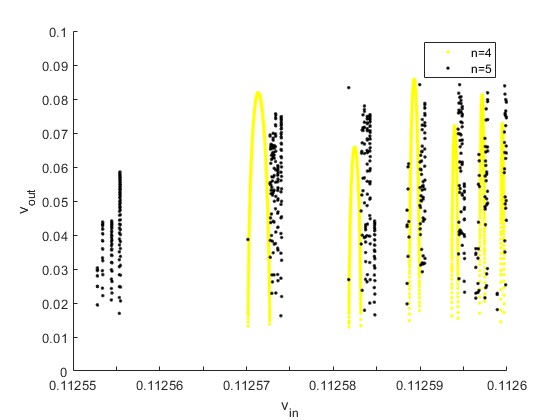}}
\caption{Fractal structure for topological sector $(-1, -1/3)$}
\label{Fig:25a}
\end{figure}

\begin{figure} [htbp] 
\centering
\subcaptionbox{$K\bar{K}$ pair potential U(x) plot}{\includegraphics[width=0.48\textwidth]{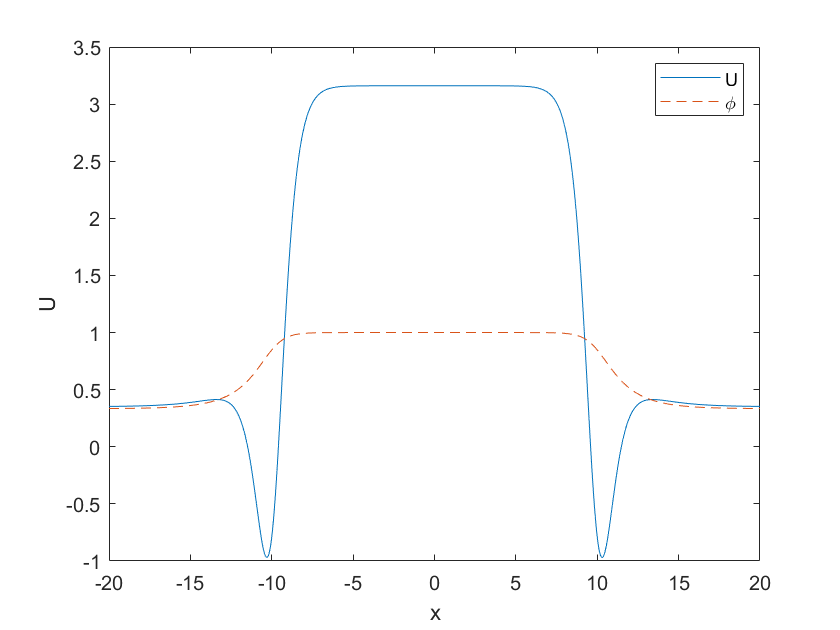}}
\subcaptionbox{$\bar{K}K$ pair potential U(x) plot}{\includegraphics[width=0.48\textwidth]{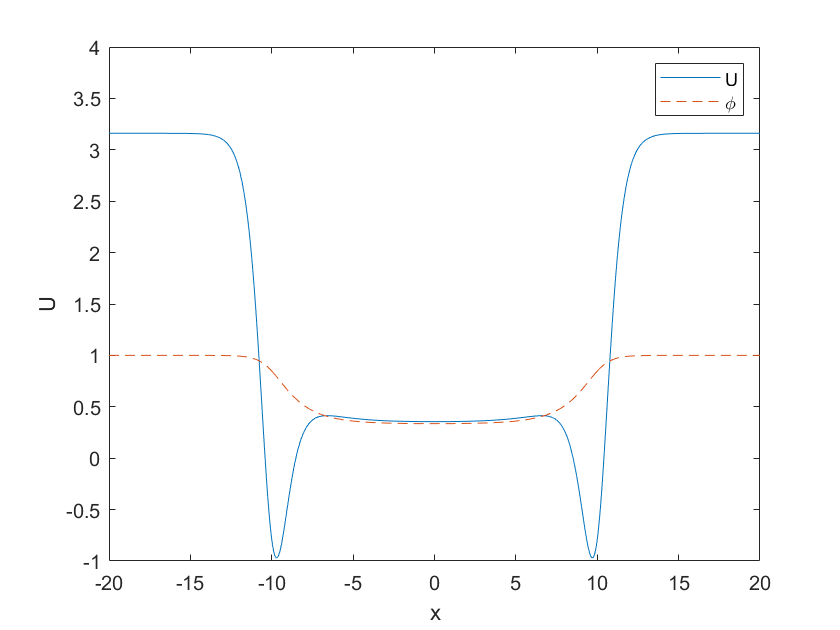}}
  \centering
  \begin{minipage}[b]{0.48\textwidth}
    \centering
    \includegraphics[width=\linewidth]{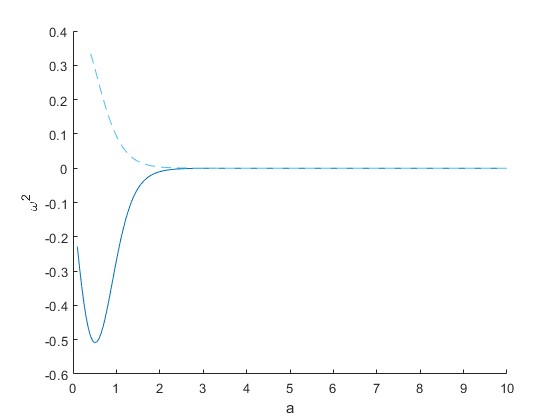}
    \subcaption{$K\bar{K}$ pair spectrum}
    \label{Fig:27-a}
  \end{minipage}
      \begin{minipage}[b]{0.48\textwidth}
    \centering
    \includegraphics[width=\linewidth]{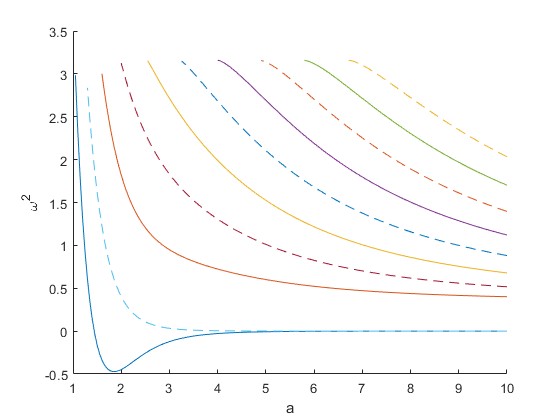}
    \subcaption{$\bar{K}K$ pair spectrum}
    \label{Fig:27-b}
  \end{minipage}
\caption{Topological sector $(1/3, 1)$}
\label{Fig:27}
\end{figure}

From Figure \ref{Fig:25-b}, we obtain that the critical velocity is $v_c \approx 0.14$. The dynamical results for this case are similar to those in Figure \ref{Fig:22}, and are not repeated here for brevity.
Figure \ref{Fig:25-c} reveals that $K\bar{K}$ pair has a central well. We also plot the potential $U(x)$ for the $\bar{K}K$ pair in Figure \ref{Fig:25-d}, which shows a raised central plateau. Here, the potential of the $\bar{K}K$ pair is similar to that of the $K\bar{K}$ pair in sector $(1/3,1)$, see Figure \ref{Fig:27}.  The $v_{in}-v_{out}$ plot of $K\bar{K}$ pair shows the fractal structure, as indicated in Figure \ref{Fig:25a}.

\subparagraph{(ii) Topological sector $(-1/3, 1/3)$}\mbox{}\\
\begin{figure}[htbp]
  \centering
  \begin{minipage}[b]{0.48\textwidth}
    \centering
    \includegraphics[width=\linewidth]{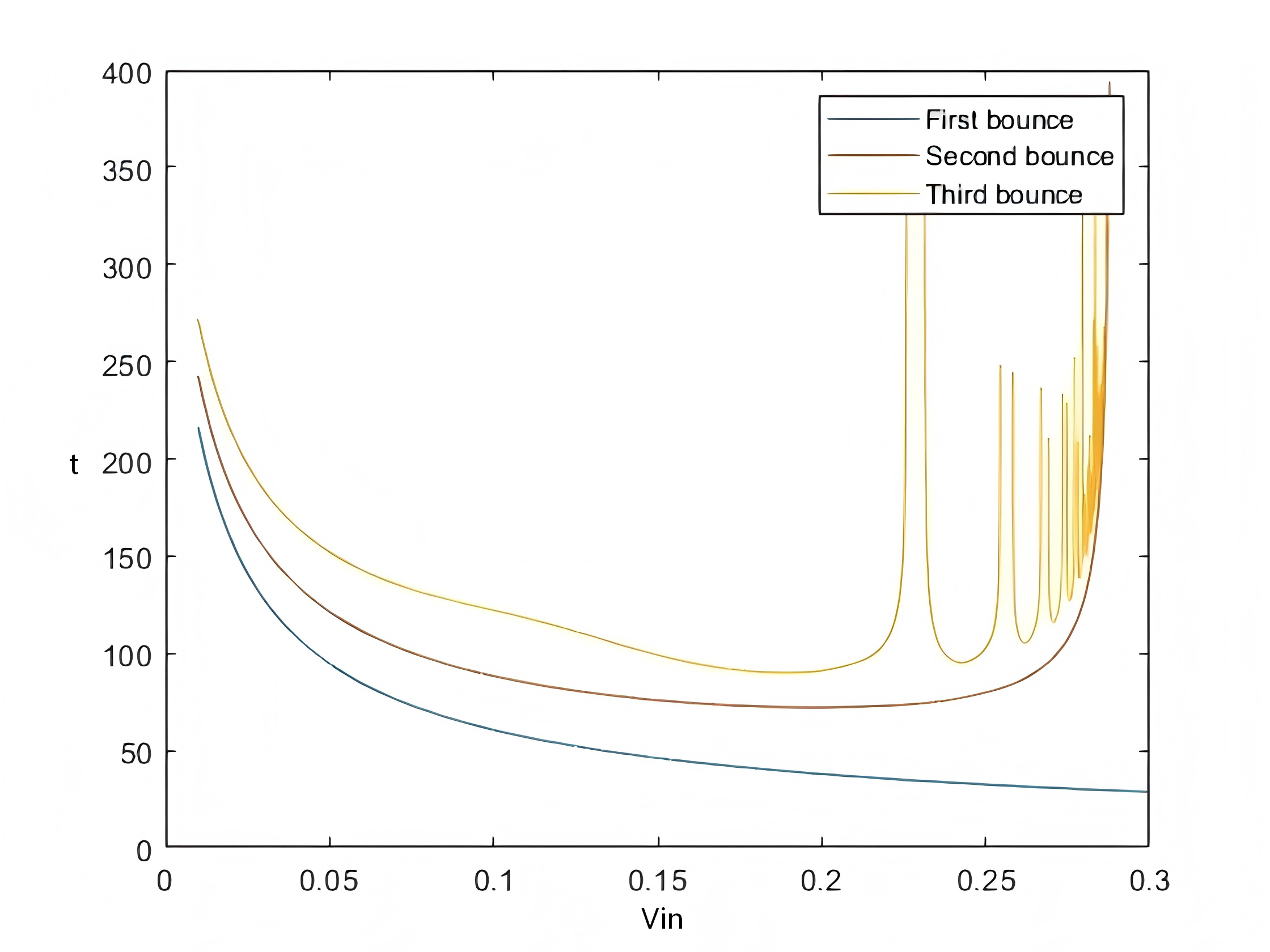}
    \subcaption{$v_{in}-t$ plot}
    \label{Fig:26-a}
  \end{minipage}
  \hfill
  \begin{minipage}[b]{0.48\textwidth}
    \centering
    \includegraphics[width=\linewidth]{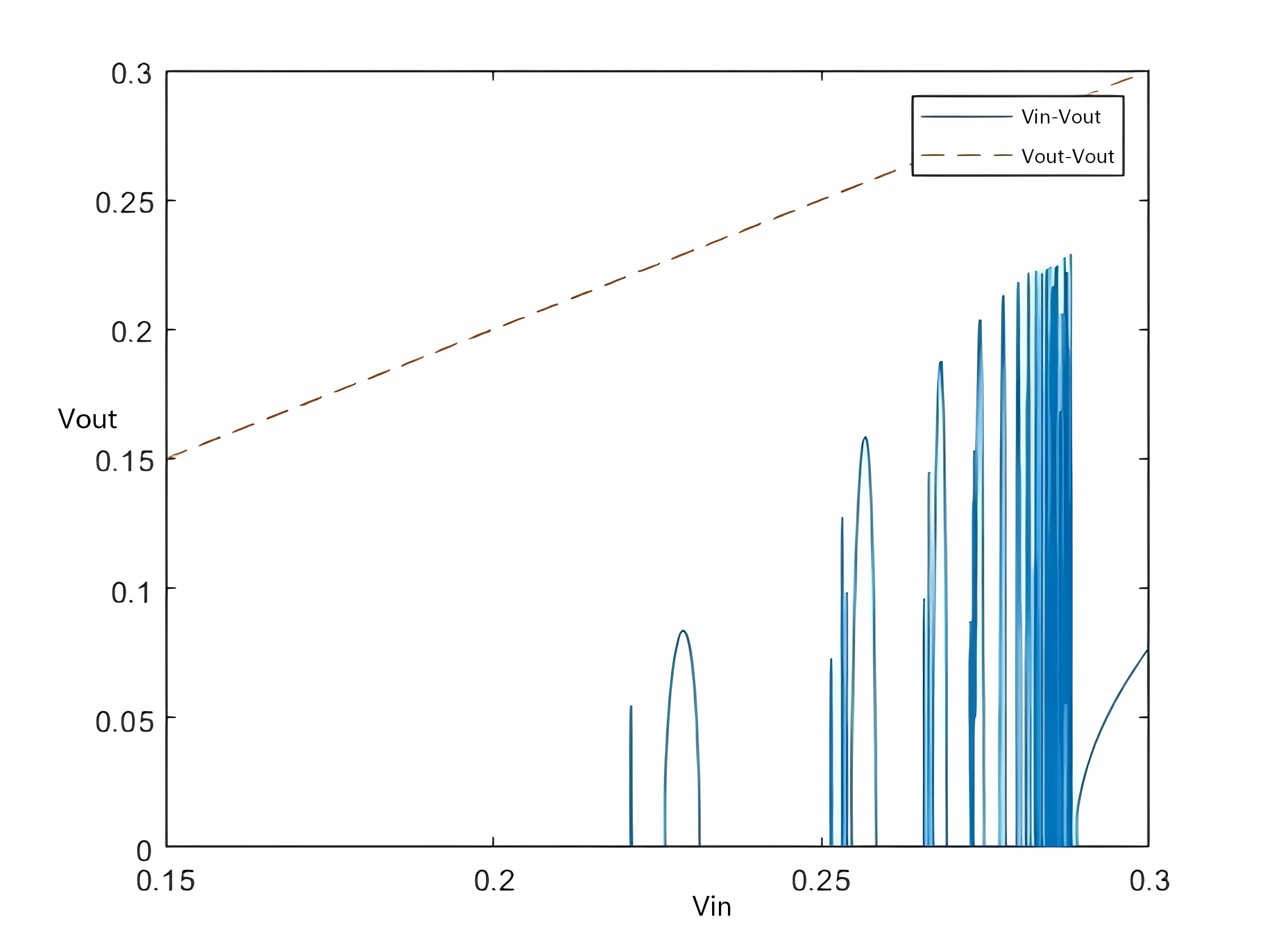}
    \subcaption{$v_{{in}} - v_{{out}}$ plot}
    \label{Fig:26-b}
  \end{minipage}

  \vspace{2mm} 
  \begin{minipage}[b]{0.48\textwidth}
    \centering
    \includegraphics[width=\linewidth]{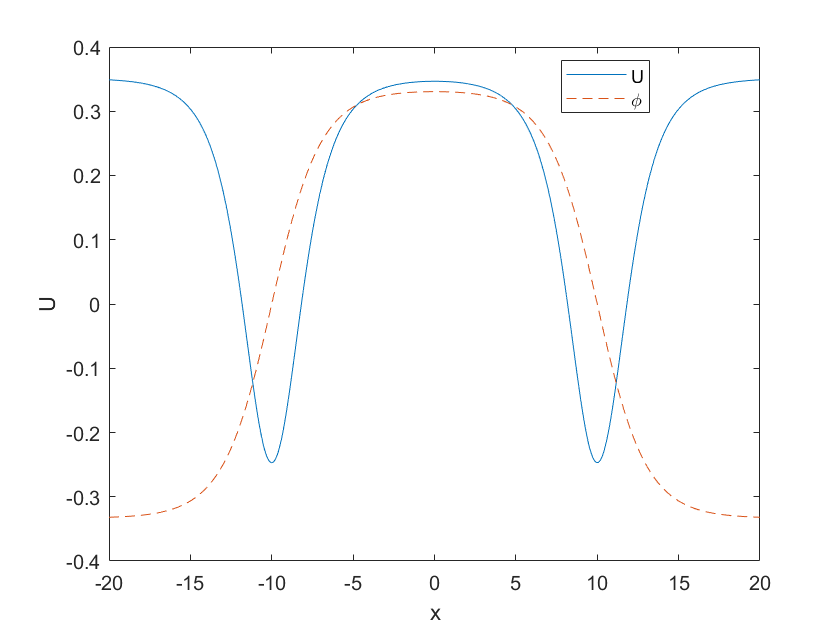}
    \subcaption{$K\bar{K}$ pair potential U(x) plot}
    \label{Fig:26-c}
  \end{minipage}
  \hfill
  \begin{minipage}[b]{0.48\textwidth}
    \centering
    \includegraphics[width=\linewidth]{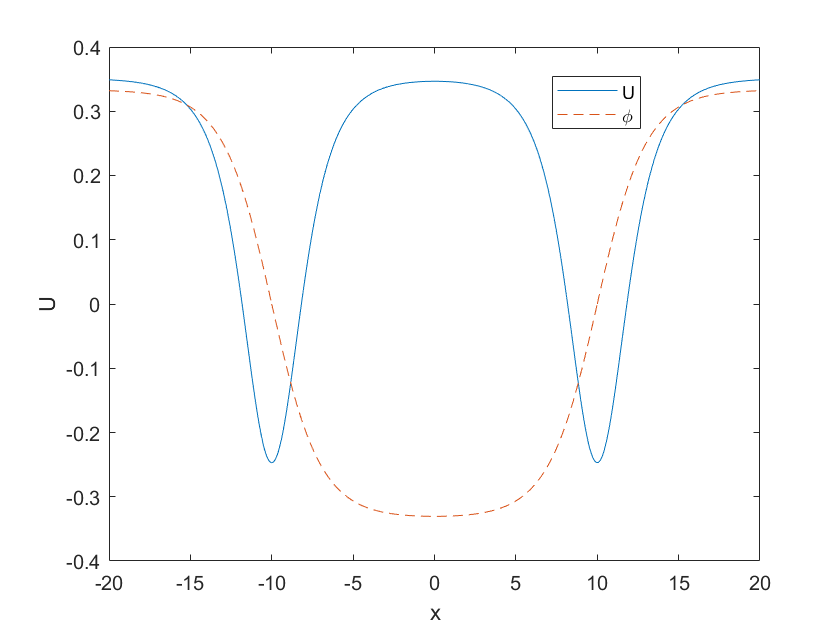}
    \subcaption{$\bar{K}K$ pair potential U(x) plot}
    \label{Fig:26-d}
  \end{minipage}
    \begin{minipage}[b]{0.48\textwidth}
    \centering
    \includegraphics[width=\linewidth]{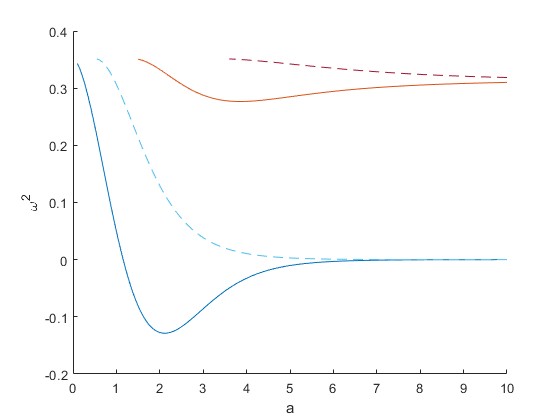}
    \subcaption{$K\bar{K}$ pair spectrum}
    \label{Fig:26-e}
  \end{minipage}
    \begin{minipage}[b]{0.48\textwidth}
    \centering
    \includegraphics[width=\linewidth]{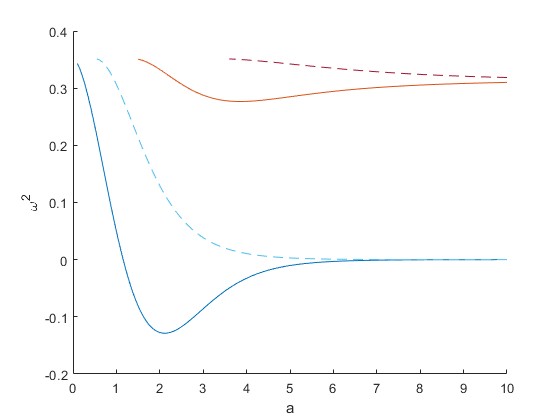}
    \subcaption{$\bar{K}K$ pair spectrum}
    \label{Fig:276-f}
  \end{minipage}
  \caption{Topological sector $(-1/3, 1/3)$}
  \label{Fig:26}
\end{figure}
\begin{figure} [htbp]
\centering
\subcaptionbox{$v_{in} \in[0.25,0.27]$}{\includegraphics[width=0.32\textwidth]{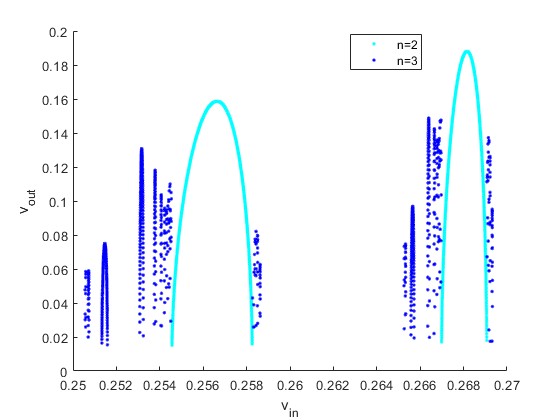}}
\subcaptionbox{$v_{in} \in[0.25,0.252]$}{\includegraphics[width=0.32\textwidth]{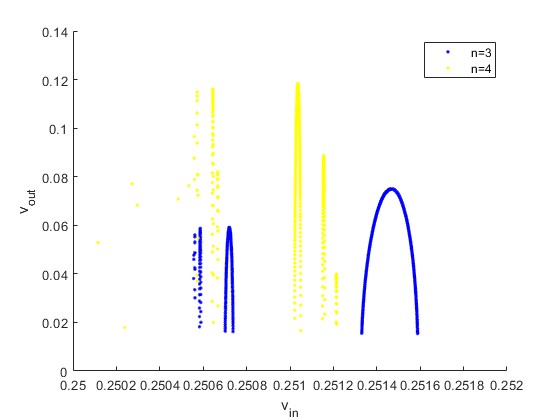}}
\subcaptionbox{$v_{in} \in[0.2510,0.2512]$}{\includegraphics[width=0.32\textwidth]{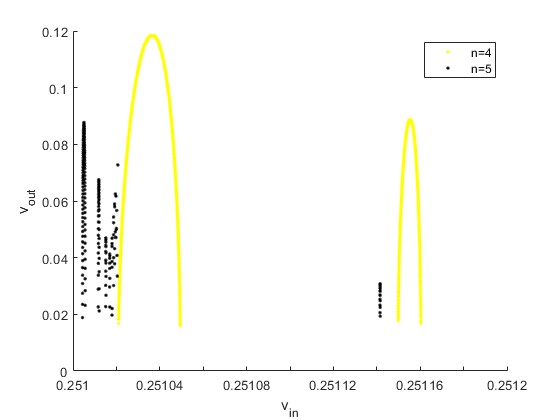}}
\caption{Fractal structure for topological sector $(-1/3, 1/3)$}
\label{Fig:144}
\end{figure}
Figure \ref{Fig:26} shows that the plots in this topological sector are similar to those in  the sector $(-1,-1/3)$. The critical velocity in this sector is $v_c \approx 0.289$. 
The potential $U(x)$ for the $K\bar{K}$ is of separated double-well,  and the raised plateau between two wells is convex. Although the double-well potentials are similar, the phenomena in sector $(-1/3,1/3)$ are quite different from those in topological sector $(-1/2,1/2)$. In Figure \ref{Fig:144}, the left panel shows the $v_{in}-v_{out}$ plot for the two and three bounces. The central panel shows the $v_{in}-v_{out}$ plot for the three and four bounces. The right panel shows the $v_{in}-v_{out}$ plot for the four and five bounces. The $v_{in}$ in the central panel is a zoomed regime in the left panel, and the $v_{in}$ in the right panel is a zoomed regime in the central panel. This shows the self-similar fractal structure for the soliton collisions.

\subparagraph{(iii) Topological sector $(1/3, 1)$}\mbox{}\\
As shown in Figure \ref{Fig:15} and Figure \ref{Fig:B} of Section \ref{sec:131}, $K\bar{K}$ pair in this sector always changes sector after collision. No fractal structure exists.
Figure \ref{Fig:27} reveals that the potential $U(x)$ of $K\bar{K}$ pair has a raised central plateau, while the potential $U(x)$ for the $\bar{K}K$ pair shows a central well. The collision channels for the  $\bar{K}K$ in the sector $(1/3, 1)$ are the same for the $K\bar{K}$ in the sector $(-1, -1/3)$, and their potentials $U(x)$ are the same. 

\clearpage
\begin{table}[htbp]
\centering
\caption{Effective-potential classification of kink-antikink collision channels}
\begin{tabular}{ccc|cccc} 
\toprule
Sector&Pair Order&Potential&Bion State&Escape&Annihilation&Sector Change\\
\midrule
$(-1,-1/2)$&$K\bar{K}$&Type I&Yes&Yes&No&No\\
$(-1,-1/2)$&$\bar{K}K$&Type IIA&Yes&Yes&Yes&No\\
$(-1/2,1/2)$&$K\bar{K}$&Type III&No&No&Yes&No\\
$(-1/2,1/2)$&$\bar{K}K$&Type III&No&No&Yes&No\\
$(1/2,1)$&$K\bar{K}$&Type IIA&Yes&Yes&Yes&No\\
$(1/2,1)$&$\bar{K}K$&Type I&Yes&Yes&No&No\\

$(-1,-1/3)$&$K\bar{K}$&Type I&Yes&Yes&No&No\\
$(-1,-1/3)$&$\bar{K}K$&Type IIB&No&No&No&Yes\\
$(-1/3,1/3)$&$K\bar{K}$&Type IV&Yes&Yes&No&No\\
$(-1/3,1/3)$&$\bar{K}K$&Type IV&Yes&Yes&No&No\\
$(1/3,1)$&$K\bar{K}$&Type IIB&No&No&No&Yes\\
$(1/3,1)$&$\bar{K}K$&Type I&Yes&Yes&No&No\\
\bottomrule
\end{tabular}
\label{table1}
\end{table}

\subsection{Effective-potential classification of collision channels}

Table \ref{table1} summarizes the relation between the effective-potential geometry and the collision channels in all topological sectors. Based on the numerical collision data, the effective-potential profiles, and the spectrum, we classify the potentials into five types: Type I, Type IIA, Type IIB, Type III, and Type IV. Type I has a central well. Type IIA has a raised central plateau with Morse-like potentials on both sides. Type IIB has a raised central plateau with a modified Morse-like structure. Type III has a separated double-well structure with a concave plateau, while Type IV has a separated double-well structure with a convex plateau.

The table shows a clear correspondence between the potential type and the final collision channel. Identical effective potentials lead to the same spectra, as expected from the Schr\"odinger-like equation. More generally, potentials belonging to the same type show similar spectral structures and lead to the same class of dynamical outcomes, even when they appear in different topological sectors or for different kink-antikink orderings. This suggests that the effective-potential geometry provides a useful classification principle for kink-antikink collisions in the $\phi^8$ model.

We also notice that, when the two solitons collide at the center $x=0$, solitons may pass through each other. At this moment, their sector and potential type could change suddenly. This could explain the annihilation and changing sector phenomena. For example, $K\bar{K}$ pair collides in the topological sector $(1/2,1)$, and then $\phi_{K\bar{K}}$ changes from $(1/2,1,1/2)$ to $(1/2,1/2,1/2)$. Thus, the field approaches a spatially uniform vacuum configuration. The localized kink-antikink structure, and hence the associated nontrivial effective-potential profile $U(x)=\mathrm{d^2}V/{\mathrm{d}\phi^2}$, disappears. There is no soliton pair solution, which leads the annihilation. In another case,  $K\bar{K}$ pair collides in the topological sector $(1/3,1)$.  After the solitons passing through each other,  $\phi_{K\bar{K}}$ changes from $(1/3,1,1/3)$ to $(1/3,-1/3,1/3)$. Thus, potential $U$ changes from Type IIB  to Type IV suddenly, leading to a new soliton pair in sector $(-1/3,1/3)$. The sector transition may occur when the instantaneous stability potential passes through a profile shared, or approximately shared, by the two soliton configurations. Such potential matching provides a possible spectral bridge between the incoming and outgoing channels. Because the new soliton pair always has outgoing velocity exceeding $0.3$, solitons always escape and there is no bion state. From a field-theoretic point of view, these two examples show how localized topological excitations can disappear or transform during real-time evolution. The annihilation channel corresponds to the decay of a soliton pair into radiation, while the sector-change channel corresponds to the collision-induced production of a new topological pair. This mechanism may be useful for understanding defect collisions in multi-vacuum scalar field theories, such as domain-wall dynamics and scalar-field phase-transition scenarios.

\section{Conclusion} \label{sec:4}

We have studied kink-antikink collisions in a $(1+1)$-dimensional $\phi^8$ scalar field theory with multiple degenerate vacua. This model provides a simple field-theoretic system for studying particle-like topological excitations. In such a system, collisions can lead not only to escape or bion formation, but also to annihilation and topological-sector change.
Building on the $\phi^8$ model introduced by Gani et al. \cite{ref13,ref15}, we derived soliton solutions for different ratios $n=p_2/p_1$. We focused on the cases $n=2$ and $n=3$, which have four distinct degenerate vacua. We also gave general boundary conditions and soliton solutions for arbitrary $n$. Numerical simulations were then performed in all topological sectors and for both $K\bar K$ and $\bar K K$ orderings.

Several dynamical features were found. First, in the $(-1/2,1/2)$ sector, the kink-antikink pair annihilates for all initial velocities. To the best of our knowledge, such a full-velocity annihilation regime has not been reported before in $\phi^8$ kink collisions. Second, we mapped fractal resonance windows in the $(-1,-1/2)$, $(-1,-1/3)$, and $(-1/3,1/3)$ sectors. Third, we found that abrupt changes of the effective-potential type can occur when soliton pairs pass through each other. This gives a possible mechanism for both annihilation and topological-sector change.

Changing the ordering of the kink and antikink provides a controlled test of this picture. For the same topological sector, different orderings can lead to different effective potentials, spectra, fractal structures, and final states. This supports the idea that the effective-potential geometry controls the collision channels.

The main result of this work is an effective-potential classification of kink-antikink collision channels. We showed that the shape of the effective potential is closely related to the spectrum and to the final dynamical outcome. This classification organizes the observed channels into escape, bion formation, annihilation, and sector change.
Our results show that the $\phi^8$ model is a useful laboratory for studying the production, annihilation, and transformation of topological excitations in scalar field theory. They also connect topological structure, spectrum and effective potential in higher-order field theories.


\nocite{*}
\bibliographystyle{elsarticle-num}
\bibliography{phi8}

\end{document}